\pdfoutput=1
\documentclass[12pt,a4paper]{article}
\usepackage{ifthen} 
\newboolean{pdflatex}
\setboolean{pdflatex}{true} 

\newboolean{articletitles}
\setboolean{articletitles}{true} 

\newboolean{uprightparticles}
\setboolean{uprightparticles}{false} 

\newboolean{inbibliography}
\setboolean{inbibliography}{false} 

\usepackage{multirow} 
\usepackage{rotating} 
\usepackage{dashrule} 
\usepackage{color} 
\usepackage{colortbl} 
\usepackage{tikz} 
\usepackage{graphpap}
\usepackage{xfrac}
\usetikzlibrary{patterns}

\definecolor{Root1}  {rgb}{0,0,0}
\definecolor{Root2}  {rgb}{1,0,0}
\definecolor{Root3}  {rgb}{0,1,0}
\definecolor{Root4}  {rgb}{0,0,1}
\definecolor{Root5}  {rgb}{1,1,0}
\definecolor{Root6}  {rgb}{1,0,1}
\definecolor{Root7}  {rgb}{0,1,1}
\definecolor{Root8}  {rgb}{0.35,0.83 ,0.33}
\definecolor{Root92} {rgb}{1   ,0.747,0   }

\usepackage[top=1in, bottom=1.25in, left=1in, right=1in]{geometry}

\usepackage{microtype}
\usepackage{lineno}  % for line numbering during review
\usepackage{xspace} % To avoid problems with missing or double spaces after
\usepackage{caption} %these three command get the figure and table captions automatically small

\usepackage{graphicx}  % to include figures (can also use other packages)
\usepackage{color}
\usepackage{colortbl}
\graphicspath{{./figs/}} % Make Latex search fig subdir for figures

\usepackage{amsmath} % Adds a large collection of math symbols
\usepackage{amssymb}
\usepackage{amsfonts}
\usepackage{upgreek} % Adds in support for greek letters in roman typeset

\newcommand*\patchAmsMathEnvironmentForLineno[1]{%
\expandafter\let\csname old#1\expandafter\endcsname\csname #1\endcsname
\expandafter\let\csname oldend#1\expandafter\endcsname\csname
end#1\endcsname
 \renewenvironment{#1}%
   {\linenomath\csname old#1\endcsname}%
   {\csname oldend#1\endcsname\endlinenomath}%
}
\newcommand*\patchBothAmsMathEnvironmentsForLineno[1]{%
  \patchAmsMathEnvironmentForLineno{#1}%
  \patchAmsMathEnvironmentForLineno{#1*}%
}
\AtBeginDocument{%
\patchBothAmsMathEnvironmentsForLineno{equation}%
\patchBothAmsMathEnvironmentsForLineno{align}%
\patchBothAmsMathEnvironmentsForLineno{flalign}%
\patchBothAmsMathEnvironmentsForLineno{alignat}%
\patchBothAmsMathEnvironmentsForLineno{gather}%
\patchBothAmsMathEnvironmentsForLineno{multline}%
\patchBothAmsMathEnvironmentsForLineno{eqnarray}%
}

\usepackage{hyperref}    % Hyperlinks in references
\usepackage[all]{hypcap} % Internal hyperlinks to floats.

\usepackage{xspace} 
\usepackage{upgreek}

\def\lhcb {\mbox{LHCb}\xspace}

\def\dzero  {\mbox{D0}\xspace}

\def\MagUp {\mbox{\em Mag\kern -0.05em Up}\xspace}

\ifthenelse{\boolean{uprightparticles}}%
{
 
 \def\Pgamma      {\ensuremath{\upgamma}\xspace}

 \def\Pmu         {\ensuremath{\upmu}\xspace}

 \def\Ppsi        {\ensuremath{\uppsi}\xspace}

 \def\PDelta      {\ensuremath{\Delta}\xspace}                 
 \def\PXi      {\ensuremath{\Xi}\xspace}                 
 \def\PLambda      {\ensuremath{\Lambda}\xspace}                 
 \def\PSigma      {\ensuremath{\Sigma}\xspace}                 
 \def\POmega      {\ensuremath{\Omega}\xspace}                 
 \def\PUpsilon      {\ensuremath{\Upsilon}\xspace}                 
 
 \def\PB      {\ensuremath{\mathrm{B}}\xspace}                 
                  
 \def\PD      {\ensuremath{\mathrm{D}}\xspace}

 \def\PJ      {\ensuremath{\mathrm{J}}\xspace}                 
 \def\PK      {\ensuremath{\mathrm{K}}\xspace}

 \def\PW      {\ensuremath{\mathrm{W}}\xspace}

 \def\PZ      {\ensuremath{\mathrm{Z}}\xspace}                 
                  
 \def\Pb      {\ensuremath{\mathrm{b}}\xspace}                 
 \def\Pc      {\ensuremath{\mathrm{c}}\xspace}

 \def\Pi      {\ensuremath{\mathrm{i}}\xspace}

 \def\Pp      {\ensuremath{\mathrm{p}}\xspace}

}
{
 
 \def\Pgamma      {\ensuremath{\gamma}\xspace}

 \def\Pmu         {\ensuremath{\mu}\xspace}

 \def\Ppsi        {\ensuremath{\psi}\xspace}                 
                  
 \mathchardef\PDelta="7101
 \mathchardef\PXi="7104
 \mathchardef\PLambda="7103
 \mathchardef\PSigma="7106
 \mathchardef\POmega="710A
 \mathchardef\PUpsilon="7107
                  
 \def\PB      {\ensuremath{B}\xspace}                 
                  
 \def\PD      {\ensuremath{D}\xspace}

 \def\PJ      {\ensuremath{J}\xspace}                 
 \def\PK      {\ensuremath{K}\xspace}

 \def\PW      {\ensuremath{W}\xspace}

 \def\PZ      {\ensuremath{Z}\xspace}                 
                  
 \def\Pb      {\ensuremath{b}\xspace}                 
 \def\Pc      {\ensuremath{c}\xspace}

 \def\Pi      {\ensuremath{i}\xspace}

 \def\Pp      {\ensuremath{p}\xspace}

}

\makeatletter
\ifcase \@ptsize \relax% 10pt
  \newcommand{\miniscule}{\@setfontsize\miniscule{4}{5}}% \tiny: 5/6
\or% 11pt
  \newcommand{\miniscule}{\@setfontsize\miniscule{5}{6}}% \tiny: 6/7
\or% 12pt
  \newcommand{\miniscule}{\@setfontsize\miniscule{5}{6}}% \tiny: 6/7
\fi
\makeatother

\DeclareRobustCommand{\optbar}[1]{\shortstack{{\miniscule (\rule[.5ex]{1.25em}{.18mm})}
  \\ [-.7ex] $#1$}}

\def\mup        {{\ensuremath{\Pmu^+}}\xspace}
\def\mun        {{\ensuremath{\Pmu^-}}\xspace} % muon negative (\mum is taken)

\def\g      {{\ensuremath{\Pgamma}}\xspace}

\def\Wp     {{\ensuremath{\PW^+}}\xspace}
\def\Wm     {{\ensuremath{\PW^-}}\xspace}

\def\Z      {{\ensuremath{\PZ}}\xspace}

\def\cquark    {{\ensuremath{\Pc}}\xspace}
\def\cquarkbar {{\ensuremath{\overline \cquark}}\xspace}

\def\bquark    {{\ensuremath{\Pb}}\xspace}
\def\bquarkbar {{\ensuremath{\overline \bquark}}\xspace}
\def\bbbar     {{\ensuremath{\bquark\bquarkbar}}\xspace}

  \def\Kbar    {{\kern 0.2em\overline{\kern -0.2em \PK}{}}\xspace}

\def\KorKbar    {\kern 0.18em\optbar{\kern -0.18em K}{}\xspace}

  \def\Dbar    {{\kern 0.2em\overline{\kern -0.2em \PD}{}}\xspace}

\def\DorDbar    {\kern 0.18em\optbar{\kern -0.18em D}{}\xspace}

\def\Bbar    {{\ensuremath{\kern 0.18em\overline{\kern -0.18em \PB}{}}}\xspace}

\def\BorBbar    {\kern 0.18em\optbar{\kern -0.18em B}{}\xspace}

\def\jpsi     {{\ensuremath{{\PJ\mskip -3mu/\mskip -2mu\Ppsi\mskip 2mu}}}\xspace}
\def\psitwos  {{\ensuremath{\Ppsi{(2S)}}}\xspace}

  \def\Y#1S{\ensuremath{\PUpsilon{(#1S)}}\xspace}% no space before {...}!

\def\proton      {{\ensuremath{\Pp}}\xspace}
\def\antiproton  {{\ensuremath{\overline \proton}}\xspace}

\def\Lbar        {{\ensuremath{\kern 0.1em\overline{\kern -0.1em\PLambda}}}\xspace}
\def\LorLbar    {\kern 0.18em\optbar{\kern -0.18em \PLambda}{}\xspace}

\def\BF         {{\ensuremath{\mathcal{B}}}\xspace}

\def\BR         {\BF}
\def\to                 {\ensuremath{\rightarrow}\xspace}

\def\AT#1     {\ensuremath{A_{\mathrm{T}}^{#1}}\xspace}           % 2

\def\C#1      {\ensuremath{\mathcal{C}_{#1}}\xspace}                       % 9
\def\Cp#1     {\ensuremath{\mathcal{C}_{#1}^{'}}\xspace}                    % 7
\def\Ceff#1   {\ensuremath{\mathcal{C}_{#1}^{\mathrm{(eff)}}}\xspace}        % 9  
\def\Cpeff#1  {\ensuremath{\mathcal{C}_{#1}^{'\mathrm{(eff)}}}\xspace}       % 7
\def\Ope#1    {\ensuremath{\mathcal{O}_{#1}}\xspace}                       % 2
\def\Opep#1   {\ensuremath{\mathcal{O}_{#1}^{'}}\xspace}                    % 7

\newcommand{\tev}{\ifthenelse{\boolean{inbibliography}}{\ensuremath{~T\kern -0.05em eV}\xspace}{\ensuremath{\mathrm{\,Te\kern -0.1em V}}}\xspace}
\newcommand{\gev}{\ensuremath{\mathrm{\,Ge\kern -0.1em V}}\xspace}
\newcommand{\mev}{\ensuremath{\mathrm{\,Me\kern -0.1em V}}\xspace}
\newcommand{\kev}{\ensuremath{\mathrm{\,ke\kern -0.1em V}}\xspace}
\newcommand{\ev}{\ensuremath{\mathrm{\,e\kern -0.1em V}}\xspace}
\newcommand{\gevc}{\ensuremath{{\mathrm{\,Ge\kern -0.1em V\!/}c}}\xspace}
\newcommand{\mevc}{\ensuremath{{\mathrm{\,Me\kern -0.1em V\!/}c}}\xspace}
\newcommand{\gevcc}{\ensuremath{{\mathrm{\,Ge\kern -0.1em V\!/}c^2}}\xspace}
\newcommand{\gevgevcccc}{\ensuremath{{\mathrm{\,Ge\kern -0.1em V^2\!/}c^4}}\xspace}
\newcommand{\mevcc}{\ensuremath{{\mathrm{\,Me\kern -0.1em V\!/}c^2}}\xspace}

\def\mum  {\ensuremath{{\,\upmu\mathrm{m}}}\xspace}

\def\mbarn{\ensuremath{\mathrm{ \,mb}}\xspace}
\def\mub{\ensuremath{{\mathrm{ \,\upmu b}}}\xspace}
\def\nb {\ensuremath{\mathrm{ \,nb}}\xspace}

\def\invpb {\ensuremath{\mbox{\,pb}^{-1}}\xspace}

\newcommand{\stat}{\ensuremath{\mathrm{\,(stat)}}\xspace}
\newcommand{\syst}{\ensuremath{\mathrm{\,(syst)}}\xspace}

\def\gsim{{~\raise.15em\hbox{$>$}\kern-.85em
          \lower.35em\hbox{$\sim$}~}\xspace}
\def\lsim{{~\raise.15em\hbox{$<$}\kern-.85em
          \lower.35em\hbox{$\sim$}~}\xspace}

\def\sPlot{\mbox{\em sPlot}\xspace}
\def\sqs   {\ensuremath{\protect\sqrt{s}}\xspace}

\def\ptot       {\mbox{$p$}\xspace}
\def\pt         {\mbox{$p_{\mathrm{ T}}$}\xspace}

\def\evtgen     {\mbox{\textsc{EvtGen}}\xspace}

\def\geant      {\mbox{\textsc{Geant4}}\xspace}

\def\photos     {\mbox{\textsc{Photos}}\xspace}

\def\pythia     {\mbox{\textsc{Pythia}}\xspace}

\def\tell1  {TELL1\xspace}
\def\ukl1   {UKL1\xspace}

\newcommand{\eg}{\mbox{\itshape e.g.}\xspace}
\newcommand{\ie}{\mbox{\itshape i.e.}\xspace}

\newcommand{\etc}{\mbox{\itshape etc.}\xspace}

\usepackage{cite} % Allows for ranges in citations
\usepackage{mciteplus}

\newcommand{\xx}{\ensuremath{\kern 0.5em }}

\def\QQ     {Q \overline Q}
\def\eff  {\varepsilon}
\def\efftot {\eff_{\rm{tot}}}
\def\effacc {\eff_{\rm{acc}}}

\def\effrecsel {\eff_{\rm{rec\&sel}}}
\def\effPID {\eff_{\rm{PID}}}
\def\efftrig {\eff_{\rm{trig}}}

\def\jpsione  {{\ensuremath{{\PJ\mskip -3mu/\mskip -2mu\Ppsi_1\mskip 2mu}}}\xspace}
\def\jpsitwo  {{\ensuremath{{\PJ\mskip -3mu/\mskip -2mu\Ppsi_2\mskip 2mu}}}\xspace}

\def\sPlot{\mbox{\em sPlot}\xspace}

\usepackage{multirow} % for complicated table
\usepackage{booktabs} % for complicated table
\usepackage{rotating}

\usepackage{longtable} 
\usepackage{subfigure}

\begin{document}

\renewcommand{\thefootnote}{\fnsymbol{footnote}}
\setcounter{footnote}{1}

\begin{titlepage}
\pagenumbering{roman}

\vspace*{-1.5cm}
\centerline{\large EUROPEAN ORGANIZATION FOR NUCLEAR RESEARCH (CERN)}
\vspace*{1.5cm}
\noindent
\begin{tabular*}{\linewidth}{lc@{\extracolsep{\fill}}r@{\extracolsep{0pt}}}
\ifthenelse{\boolean{pdflatex}}
{\vspace*{-2.7cm}\mbox{\!\!\!\includegraphics[width=.14\textwidth]{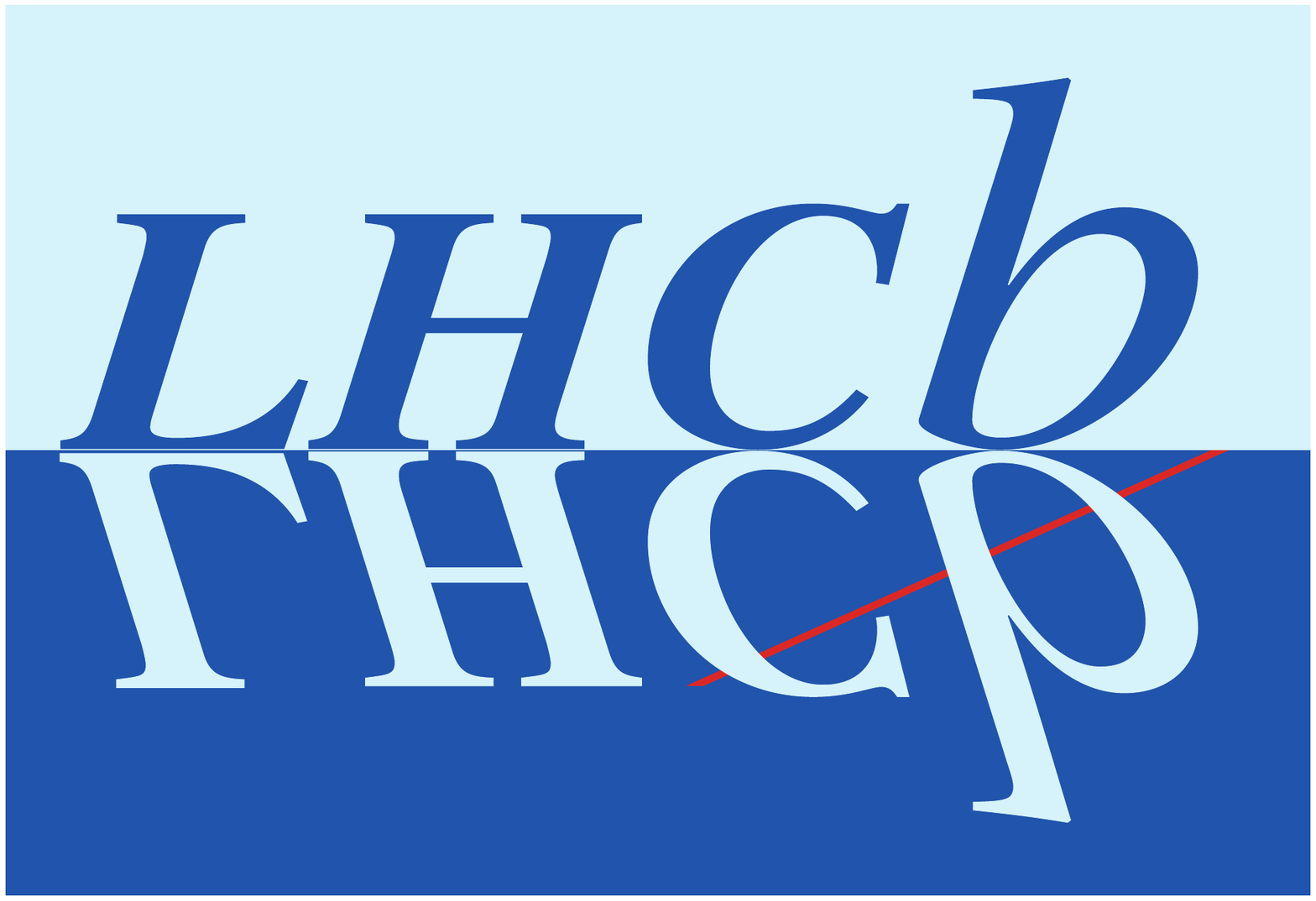}} & &}
{\vspace*{-1.2cm}\mbox{\!\!\!\includegraphics[width=.12\textwidth]{lhcb-logo.eps}} & &}
\\
 & & CERN-EP-2016-307 \\   
 & & LHCb-PAPER-2016-057 \\   
 & & Sep 28, 2017 \\ 
 & & \\
\end{tabular*}

\vspace*{4.0cm}

{\normalfont\bfseries\boldmath\huge
\begin{center}
   Measurement of the $\jpsi$ pair production cross-section in $pp$ collisions at $\sqs = 13 \tev$
\end{center}
}

\vspace*{2.0cm}

\begin{center}
The LHCb collaboration\footnote{Authors are listed at the end of this paper.}
\end{center}

\vspace{\fill}

\begin{abstract}
  \noindent
  The production cross-section of $\jpsi$ pairs is measured using a data sample of $pp$ collisions collected by the $\lhcb$ experiment at a centre-of-mass energy of \mbox{$\sqs = 13 \tev$},
  corresponding to an integrated luminosity of $279 \pm 11 \invpb$.
  The measurement is performed for $\jpsi$ mesons with a transverse momentum of less than $10\gevc$ in the rapidity range $2.0<y<4.5$.
  The production cross-section is measured to be $15.2 \pm 1.0 \pm 0.9 \nb$.
  The first uncertainty is statistical, and the second is systematic.
  The differential cross-sections as functions of several kinematic variables of the $\jpsi$ pair are measured and compared to theoretical predictions.

\end{abstract}

\vspace*{2.0cm}

\begin{center}
  Published in JHEP 06 (2017) 047
\end{center}

\vspace{\fill}

{\footnotesize 
\centerline{\copyright~CERN on behalf of the \lhcb collaboration, licence \href{http://creativecommons.org/licenses/by/4.0/}{CC-BY-4.0}.}}
\vspace*{2mm}

\end{titlepage}

\newpage
\setcounter{page}{2}
\mbox{~}

\cleardoublepage

\renewcommand{\thefootnote}{\arabic{footnote}}
\setcounter{footnote}{0}

\pagestyle{plain} 
\setcounter{page}{1}
\pagenumbering{arabic}

\section{Introduction}
\label{sec:intro}
The production mechanism of heavy quarkonia is a long-standing and intriguing problem
in quantum chromodynamics\,(QCD), 
which is not fully understood even after over forty years of study.
The colour-singlet model\,(CSM)~\cite{CSM1,CSM2,CSM3,
CSM4,CSM5,Kartvelishvili:1978id,Kartvelishvili:1980uz,
Berger:1980ni,CSM6,CSM7} assumes 
the intermediate $\QQ$ state to be colourless and to have the same $J^{PC}$ quantum 
numbers as the final quarkonium state. 
Leading-order calculations in the CSM underestimate the $\jpsi$ and $\psitwos$ production
cross-sections at high transverse momentum, $\pt$, by more than one
order of magnitude~\cite{CSMLO}.
The~gap between CSM predictions and experimental measurements is reduced when including 
next-to-leading-order corrections, but the agreement is still not 
satisfactory~\cite{CSMNLO1,CSMNLO2,CSMNLO3}.
The~non\nobreakdash-relativistic QCD\,(NRQCD) model takes into account both colour-singlet\,(CS) and 
colour-octet\,(CO) states of the~$\QQ$ pair~\cite{NRQCD1,NRQCD2,NRQCD3}.
It either describes the production cross-sections and polarisations at large $\pt$ or 
it describes the production cross-section at all \pt~values, but then fails to predict 
the polarisation~\cite{NRQCDprod1,NRQCDprod2,NRQCDprod3,NRQCDprod4,
  NRQCDprod5,NRQCDpol1,NRQCDpol2,NRQCDpol3,NRQCDpol4,NRQCDpol5,NRQCDpol6,
  NRQCDpol7,NRQCDpol8,NRQCDpol9,LHCb-PAPER-2013-008,LHCb-PAPER-2013-067}.
This puzzle can be probed via the production of pairs of quarkonia~\cite{Kartvelishvili:1984ur,
Humpert:1983yj,Humpert:1983qt,Ko:2010xy,Baranov:2015cle,Sun:2014gca}, 
where the interpretation of the~measured cross\nobreakdash-section 
could be simpler. 
In quarkonium-pair production,
the selection rules in the CS process of leading-order~(LO) NRQCD
forbid the~feed-down from cascade decays of excited $C$\nobreakdash-even states.
This~feed-down from $C$\nobreakdash-even states, \eg 
$\chi_{\cquark}\to\jpsi\g$ or 
$\chi_{\bquark}\to\PUpsilon\g$,
plays an important role in single quarkonium production. 
It significantly complicates the~precise comparison 
between data and model predictions,
and makes the~interpretation
of polarisation measurements difficult.

Besides the single parton scattering\,(SPS) process,
the process of double parton scattering\,(DPS) can also 
contribute to quarkonium pair production.
The DPS process is of great importance since it can provide information on the transverse momenta 
of the~partons and their correlations inside the proton,
and can help in understanding various backgrounds, \eg $\Z+\bbbar$, $\Wp+\Wm$, multi\nobreakdash-jets \etc, 
in searches for new physics. 
The DPS processes have been studied in several final states, 
\eg
$4{\text -}{\rm jets}$ by the~AFS~\cite{4jets1}, 
UA2~\cite{4jets2}, 
CDF~\cite{Abe:1993rv},
and ATLAS~\cite{Aaboud:2016dea} collaborations, 
$\Pgamma + 3{\text -}{\rm jets}$ by the~CDF~\cite{Gamma3jets1} and \dzero~\cite{Gamma3jets2,Abazov:2014fha} collaborations,
$2\Pgamma + 2{\text -}{\rm jets}$ by the~\dzero~\cite{Abazov:2015nnn}~collaboration,
$\PW+2{\text -}{\rm jets}$~\cite{W2jets} and 
$\PUpsilon+\PUpsilon$~\cite{DUpsilon}  by the~CMS collaboration, 
$\jpsi+\PW$~\cite{JpsiW} 
and $\jpsi+\Z$~\cite{JpsiZ} by the~ATLAS collaboration, 
and double charm~\cite{LHCb-PAPER-2012-003},
$\Z+{\rm open \ charm}$~\cite{LHCb-PAPER-2013-062} and 
$\PUpsilon+{\rm open \ charm}$~\cite{LHCb-PAPER-2015-046} 
by the~$\lhcb$ collaboration.
After having been first observed by the~NA3 collaboration
in pion-nuclear and proton\nobreakdash-nuclear interactions~\cite{NA3DJ,Badier:1985ri}, 
$\jpsi$ pair production has been measured in $\proton\proton$~collisions 
by the~LHCb~\cite{LHCB-PAPER-2011-013} and CMS~\cite{CMSDJ} experiments at $\sqs = 7 \tev$ 
and by the~ATLAS experiment~\cite{ATLASDJ} at~$\sqs = 8 \tev$. 
The~\dzero experiment~\cite{D0DJ} measured it 
using $\proton\antiproton$~collision data at~$\sqs = 1.96 \tev$.

Within the~DPS mechanism, two quarkonia are produced
independently in different partonic
interactions. Neglecting
the~parton correlations in the~proton, 
the~contribution of this mechanism 
is estimated according to the~formula~\cite{Calucci:1997ii,Calucci:1999yz,DelFabbro:2000ds}
\begin{equation}\label{eq:DPS}
  \sigma_{\mathrm{DPS}}\left(\jpsi\jpsi\right) = 
  \dfrac{1}{2} 
  \dfrac{ \sigma\left(\jpsi\right)^2}
  {\sigma_{\mathrm{eff}}},
\end{equation}
where $\sigma(\jpsi)$ is the~inclusive prompt $\jpsi$ production 
cross\nobreakdash-section,
the factor $\sfrac{1}{2}$ accounts for two identical particles in the final state,
and $\sigma_{\mathrm{eff}}$ 
is an~effective cross\nobreakdash-section, 
which provides a proper normalisation
of the~DPS cross\nobreakdash-section estimate.
The effective cross-section is related to the~transverse overlap function 
between partons in the~proton,
and is thought to be universal for all processes and energy scales.
Most of the~measured values of $\sigma_{\mathrm{eff}}$
lie in the range $12-20\mbarn$~\cite{Bansal:2014paa,LHCb-PAPER-2015-046,
Aaboud:2016dea}, which supports the expectation that
$\sigma_{\mathrm{eff}}$ is universal 
for a large range of processes with different kinematics and scales,
and for a wide spectrum of centre-of-mass energies in $\proton\proton$ and $\proton\antiproton$ collisions.

The~LHCb measurement of 
\mbox{$\sigma\left(\jpsi\jpsi\right)=5.1\pm1.0\pm1.1\nb$}
at \mbox{$\sqs = 7\tev$}
is not precise enough to distinguish
%% between
%% two contributions~\cite{Baranov:2011ch,Baranov:2012re}:
%% the~SPS contribution 
%% of $4.0\pm1.2\nb$,
%% calculated in the leading\nobreakdash-order 
%% NRQCD\,CS~\cite{Qiao:2009kg,Berezhnoy:2011xy} approach,
%% and the~DPS contribution of $3.8\pm1.3\nb$, estimated with Eq.~\eqref{eq:DPS},
%% using $\sigma\left(\jpsi\right)$ from Ref.~\cite{LHCb-PAPER-2011-003} 
%% and $\sigma_{\mathrm{eff}}=14.5\pm1.7^{+1.7}_{-2.3}\mbarn$ 
%% from Ref.~\cite{Abe:1997bp}.
between
the~SPS and DPS contributions~\cite{Baranov:2011ch,Baranov:2012re}.
The~SPS contribution is calculated to be 
$4.0\pm1.2\nb$~\cite{Qiao:2009kg,Berezhnoy:2011xy}
and
$4.6\pm1.1\nb$~\cite{Sun:2014gca}
in the~leading\nobreakdash-order 
NRQCD\,CS approach,
and
$5.4^{+2.7}_{-1.1}\nb$~\cite{Sun:2014gca} using
complete next\nobreakdash-to\nobreakdash-leading order NRQCD\,CS~approach. 
The~DPS contribution
is estimated
to be $3.8\pm1.3\nb$
with Eq.~\eqref{eq:DPS}
using $\sigma\left(\jpsi\right)$
from Ref.~\cite{LHCb-PAPER-2011-003} 
and $\sigma_{\mathrm{eff}}=14.5\pm1.7^{+1.7}_{-2.3}\mbarn$ 
from Ref.~\cite{Gamma3jets1}.
The~large number of reconstructed $\jpsi$ pair events in the CMS~data~\cite{CMSDJ}
allowed for study of \jpsi~correlations~\cite{Lansberg:2014swa}.
The~observation of events with a large separation 
in rapidity of two \jpsi~mesons indicates 
a significant DPS contribution,
leading to $\sigma_{\mathrm{eff}}=8.2\pm2.2\mbarn$~\cite{Lansberg:2014swa}, 
somewhat lower than the majority of other $\sigma_{\mathrm{eff}}$ measurements.
A similarly small value, \mbox{$\sigma_{\mathrm{eff}}=6.3\pm1.9\mbarn$},
is obtained by the ATLAS collaboration using 
a~data\nobreakdash-driven 
model\nobreakdash-independent approach~\cite{ATLASDJ}.
A small value of \mbox{$\sigma_{\mathrm{eff}}=4.8\pm2.5\mbarn$}
is also obtained by the \dzero~collaboration~\cite{D0DJ} using 
the~separation of the two \jpsi~mesons in pseudorapidity to distinguish 
SPS and DPS contributions.
Together with an even smaller value of 
\mbox{$\sigma_{\mathrm{eff}}=2.2\pm1.1\mbarn$}, 
determined by the \dzero~collaboration from the~measurement of 
the~simultaneous production of $\jpsi$ and $\PUpsilon$ mesons~\cite{Abazov:2015fbl},
and the~estimate of $\sigma_{\mathrm{eff}}=2.2-6.6\mbarn$ by the CMS collaboration
from the~production of $\PUpsilon$~pairs~\cite{DUpsilon}, 
these results question the~universality of $\sigma_{\mathrm{eff}}$.

In this paper, the $\jpsi$ pair production cross-section is measured 
using $pp$~collision data collected by the $\lhcb$ experiment in 2015 at $\sqs = 13 \tev$ 
with both $\jpsi$ mesons in the rapidity range $2.0<y<4.5$, and with 
a transverse momentum~$\pt< 10 \gevc$. 
The~polarisation of the $\jpsi$ mesons is assumed to be zero since there
is as yet no knowledge of the polarisation of $\jpsi$ pairs,
and all the LHC analyses indicate a small polarisation for 
the quarkonia~\cite{NRQCDpol7,NRQCDpol8,NRQCDpol9,LHCb-PAPER-2013-008,LHCb-PAPER-2013-067}.
The~$\jpsi$ mesons are reconstructed via the $\mup\mun$~final state.
In the following, the labels $\jpsione$ and $\jpsitwo$ are randomly assigned to the two $\jpsi$ candidates.

\section{Detector and data set}
\label{sec:Detector}
The \lhcb detector~\cite{Alves:2008zz,LHCb-DP-2014-002} is a single-arm forward
spectrometer covering the \mbox{pseudorapidity} range $2<\eta <5$,
designed for the study of particles containing \bquark or \cquark
quarks. The detector includes a high-precision tracking system
consisting of a silicon-strip vertex detector surrounding the $pp$
interaction region, a large-area silicon-strip detector~(TT) located
upstream of a dipole magnet with a bending power of about
$4{\mathrm{\,Tm}}$, and three stations of silicon-strip detectors and straw
drift tubes placed downstream of the magnet.
The tracking system provides a measurement of momentum, \ptot, of charged particles with
a relative uncertainty that varies from 0.5\% at low momentum to 1.0\% at 200\gevc.
The minimum distance of a track to a primary vertex (PV), the impact parameter, 
is measured with a resolution of $(15+29/\pt)\mum$,
where \pt is in \gevc.
Different types of charged hadrons are distinguished using information
from two ring-imaging Cherenkov detectors. 
Photons, electrons and hadrons are identified by a calorimeter system consisting of
scintillating-pad and preshower detectors, an electromagnetic
calorimeter and a hadronic calorimeter. Muons are identified by a
system composed of alternating layers of iron and multiwire
proportional chambers.

The online event selection is performed by a trigger~\cite{LHCb-DP-2012-004}, 
which consists of a hardware stage~(L0), 
based on information from the calorimeter and muon
systems, followed by a software stage, which applies a full event
reconstruction.
The L0 trigger requires two muons with $\pt(\mu_1) \times \pt(\mu_2) > (1.3 \gevc)^2$.
In the first stage of the software trigger~(HLT1), 
two muons with $\pt>330\mevc$ and $p>6\gevc$ are required 
to form a $\jpsi$ candidate with invariant mass $M(\mup\mun)>2.7\gevcc$;
alternatively, 
the event can also be accepted when it has a good quality muon 
with $\pt>4.34\gevc$ and $p>6\gevc$.
In the second stage of the software trigger~(HLT2), 
the two $\jpsi$ mesons are reconstructed from $\mup\mun$~pairs 
with good vertex-fit quality and invariant masses within $\pm 120 \mevcc$
of the~known $\jpsi$ mass~\cite{PDG2016},
using algorithms identical to the~offline reconstruction. 
In the offline selection, all four muons in the~final state are required to 
have $\pt>650\mevc$, $6<p<200\gevc$ and $2<\eta<5$.
Each track must have a good-quality track fit and be identified as a muon.
The four muon tracks are required to originate from the same PV.
This reduces to a negligible level the number of pile-up candidates, \ie $\jpsi$ pairs from two 
independent $pp$ interactions. 
The reconstructed $\jpsi$ mesons are required to have a good-quality 
vertex and an invariant mass in the range $3000<M(\mup\mun)<3200\mevcc$.
Only events explicitly triggered by one of the~\jpsi candidates
at the~L0 and the~HLT1 stages are retained.
For events with multiple candidates, 
in particular where the four muons can be combined in two different ways to form a $\jpsi$ pair, 
which account for $1.4\%$ of the total candidates,
one randomly chosen candidate pair is retained.

Simulated $\jpsi$ samples are generated to study the behaviour of the signal.
In the simulation, $pp$ collisions are generated using
$\pythia 8$~\cite{Sjostrand:2007gs,Sjostrand:2006za} 
with a specific $\lhcb$ configuration~\cite{LHCb-PROC-2010-056}.
Decays of hadronic particles are described by $\evtgen$~\cite{Lange:2001uf}, 
in which final-state radiation is generated using $\photos$~\cite{Golonka:2005pn}. 
The interaction of the generated particles with the detector, and its response,
are implemented using the $\geant$ toolkit~\cite{Allison:2006ve, *Agostinelli:2002hh} as described in Ref.~\cite{LHCb-PROC-2011-006}.

\section{Cross-section determination}
The inclusive $\jpsi$ pair production cross-section is measured as 
\begin{equation}\label{eq:crosssection}
  \sigma(\jpsi\jpsi) = \frac{N^{{\rm cor}}}{{\cal{L}} \times \BF(\jpsi \to \mup \mun)^{2}},
\end{equation}
where $N^{{\rm cor}}$ is the number of signal candidates after the efficiency correction, 
\mbox{$\BR(\jpsi \to \mup \mun) = (5.961\pm0.033)\%$} is the branching fraction of
the $\jpsi \to \mup \mun$ decay~\cite{PDG2016},
and \mbox{${\cal{L}}=279 \pm 11 \invpb$} is the integrated luminosity, determined using 
the beam-gas imaging and van der Meer scan methods~\cite{LHCb-PAPER-2014-047}.

The total detection efficiency of the $\jpsi$ pair is estimated as 
\begin{equation}
  \efftot = \effacc \times \effrecsel \times \effPID \times \efftrig,
\end{equation}
where 
$\effacc$ is the geometrical acceptance, 
$\effrecsel$ is the reconstruction and selection efficiency for candidates with all final-state muons inside the geometrical acceptance,
$\effPID$ is the muon particle identification~(PID) efficiency for selected candidates,
and $\efftrig$ is the trigger efficiency for selected candidates satisfying the PID requirement.
The~first three efficiencies of the $\jpsi$ pair, $\effacc$, $\effrecsel$ and $\effPID$, are factorized as
\begin{equation}
  \varepsilon\left(\jpsi\jpsi\right) = \varepsilon\left(\jpsione\right) \times \varepsilon\left(\jpsitwo\right).
\end{equation}
Since the HLT2 trigger selection is performed using the same reconstruction 
algorithm as the~offline selection and the selection criteria of the HLT2 trigger are a subset of those used in the final 
selection, the corresponding trigger efficiency for the reconstructed and selected events is 100\%.  
Since at least one of the two $\jpsi$ meson candidates is required to have passed the L0 and HLT1 trigger, the efficiency $\efftrig$ of the $\jpsi$ pair can be expressed as
\begin{equation}
  \varepsilon_{\mathrm{trig}}\left(\jpsi\jpsi\right)
  = 1 - \left(1-\varepsilon_{\mathrm{trig}}\left(\jpsione\right)\right) \times \left(1-\varepsilon_{\mathrm{trig}}\left(\jpsitwo\right)\right).
\end{equation}

All terms in the single $\jpsi$ efficiency are estimated in bins of $\pt$ and $y$ of the $\jpsi$ mesons using the simulation.
The track reconstruction and muon PID efficiency are corrected using data-driven techniques, as described in Sec.~\ref{sec:sys},
and the trigger efficiency measurement is validated on data. 

\begin{figure}[tb]
\begin{center}
\includegraphics[width=0.495\linewidth]{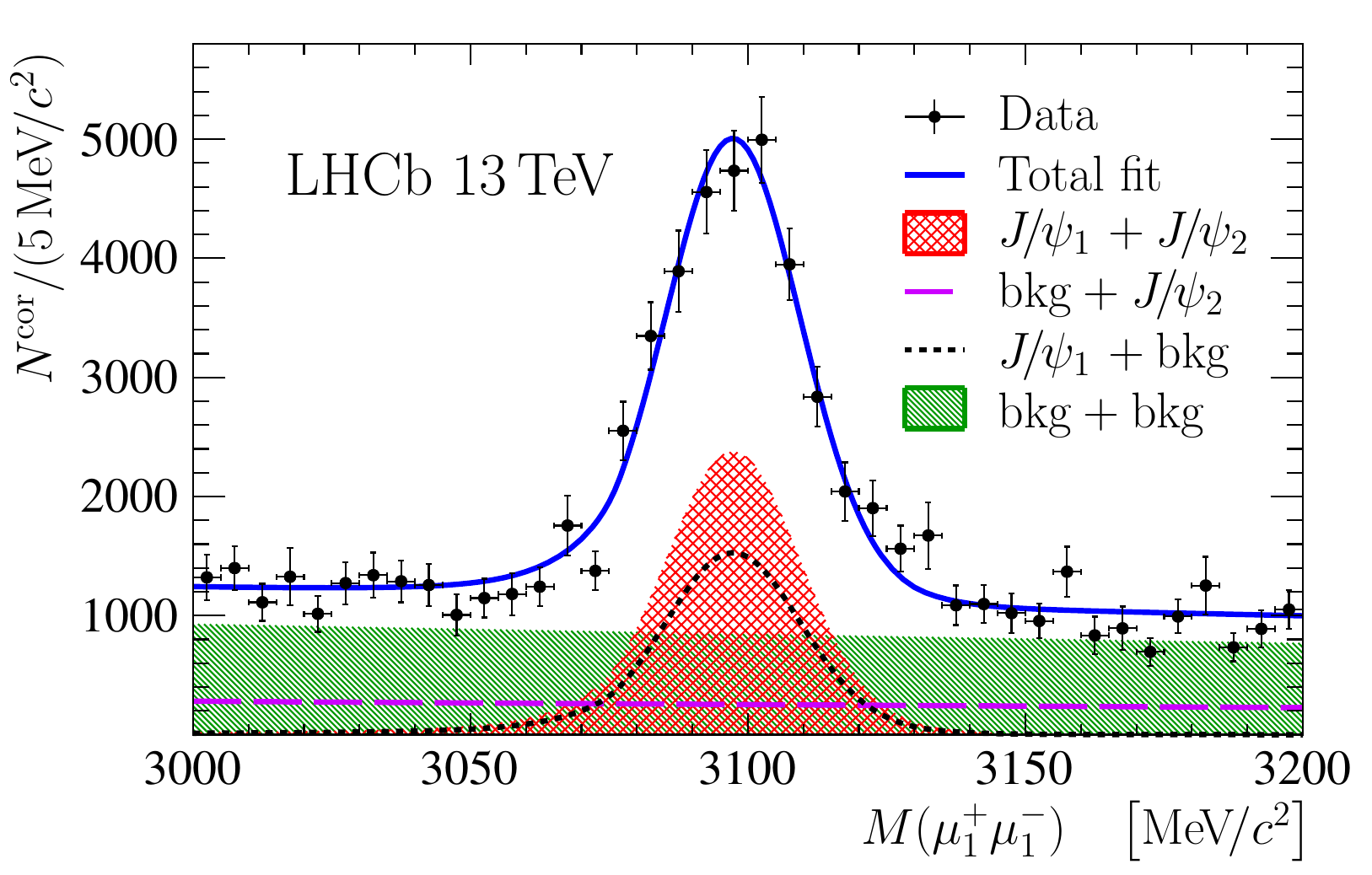}
\includegraphics[width=0.495\linewidth]{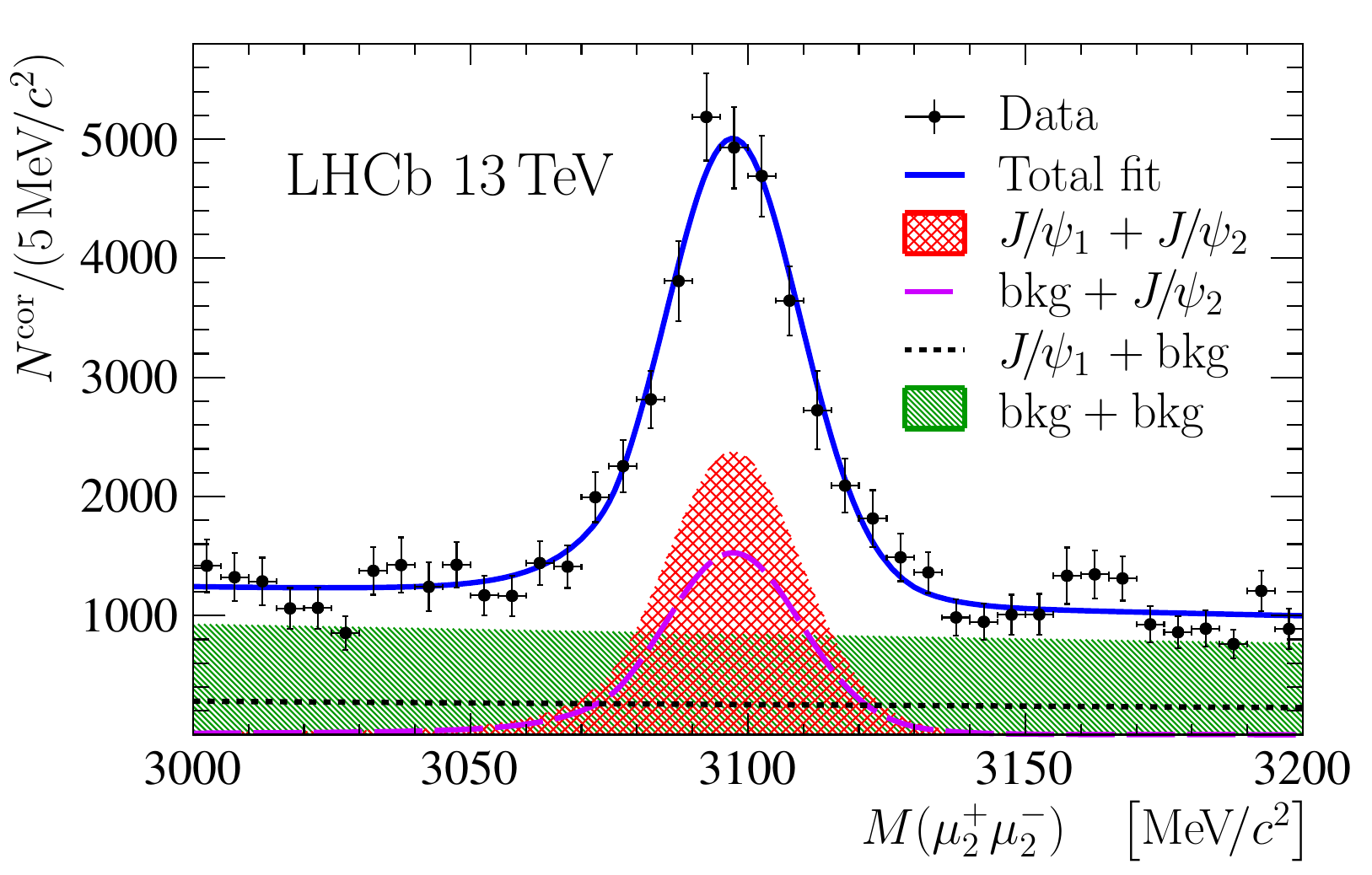}
\vspace*{-0.1cm}
\end{center}
\caption{\small Projections of the fit to the efficiency-corrected distribution of the reconstructed $\jpsi$ mass for (left)~$M(\mu^+_1\mu^-_1)$ and (right)~$M(\mu^+_2\mu^-_2)$. The (black) points with error bars represent the data. The (blue) solid line is the total fit function. The (red) cross-hatched area shows the signal distribution. The (black and magenta) dashed lines represent the background components due to the combination of a real $\jpsi$ with a combinatorial candidate. The (green) shaded area shows the purely combinatorial background.}\label{fig:fitcor}
\end{figure}

The signal yield is determined by performing an extended unbinned maximum 
likelihood fit to the efficiency\nobreakdash-corrected two\nobreakdash-dimensional 
$(M(\mu^+_1\mu^-_1),M(\mu^+_2\mu^-_2))$~mass distribution.
The total detection efficiency is applied individually on an event\nobreakdash-by\nobreakdash-event basis.
The signal is modelled by the sum of a double-sided Crystal Ball\,(DSCB) function~\cite{Skwarnicki:1986xj} and a Gaussian function, 
which share the same mean value. The~power law tail parameters of the DSCB, the relative fraction and the difference between the widths 
of the DSCB and the Gaussian function are fixed to the values obtained from simulation, leaving the peak value and the core width of the DSCB as free parameters.
The combinatorial background is described by an exponential function.
Since the labels $\jpsione$ and $\jpsitwo$ are assigned randomly,
the fit function is symmetric under the exchange of the $\jpsione$ and $\jpsitwo$ masses. 
The fit projections on $M(\mu^+_1\mu^-_1)$ and $M(\mu^+_2\mu^-_2)$ are shown in Fig.~\ref{fig:fitcor}.
The corrected yield~\footnote{The corresponding fit of the efficiency-uncorrected sample gives $(1.05 \pm 0.05) \times 10^3$ signal events.} of $\jpsi$ pairs is determined to be $N^{{\rm cor}} = (15.8 \pm 1.1) \times 10^3$.

After the fit, the residual contamination, where either one or both $\jpsi$ mesons come from $\bquark$-hadron decays, must be corrected for.
The fraction of background is evaluated with the help of simulation validated with data and normalized using the measured prompt $\jpsi$ and inclusive $\bbbar\to\jpsi$ production cross-sections
within the LHCb acceptance at $\sqs = 13 \tev$~\cite{LHCb-PAPER-2015-037}.
The fraction of candidates with $\jpsi$ mesons from $\bquark$-hadron decays is determined to be $4.5\%$. 

\section{Systematic uncertainties}
\label{sec:sys}
Several sources of systematic uncertainties on the $\jpsi$ pair production cross-section are studied and summarized in Table~\ref{tab:sys}. 
The uncertainty due to the signal shape description is estimated by replacing the nominal model with two alternative models, the Hypatia function~\cite{Hypatia} and 
a kernel estimate for the underlying probability distribution function of the simulated sample convolved with a Gaussian function~\cite{kernel}. 
The relative difference of $1.6\%$ with respect to the nominal result is taken as a systematic uncertainty.

A difference between simulation and data,
in particular in the fit quality of the candidates when constraining the muons to the PV,
can lead to a bias in the efficiency determination.
This is estimated by comparing the vertex-fit quality of the reconstructed $\jpsi$ candidates
between the simulated and the data samples,
where the background is subtracted using the \sPlot technique~\cite{Pivk:2004ty}. 
Data and simulation agree within $1.0\%$, which is taken as a systematic uncertainty.

The track reconstruction efficiency is studied in data using a tag-and-probe technique~\cite{LHCb-DP-2013-002}.
In this method, one of the muons from the $\jpsi$ is fully reconstructed as the tag track, 
and the other muon track, the probe track, is reconstructed using only information from the TT detector and the muon stations. 
The tracking efficiency is taken as the fraction of $\jpsi$ candidates whose probe tracks match fully reconstructed tracks.
The simulated sample is corrected to match the track multiplicity of events in the data.
The ratio of tracking efficiencies between data and simulation is taken as the correction factor.
A systematic uncertainty of $0.8\%$ per track is assigned for the difference in event multiplicity between data and simulation.

The muon PID efficiency is also determined using a tag-and-probe method~\cite{LHCb-DP-2013-001}, where only one track of the $\jpsi$ is identified as a muon, \ie the tag track.
The single muon PID efficiency, defined as the fraction of $\jpsi$ candidates with the other track~(probe track) identified as a muon, is determined in bins of $p$ and $\eta$ of the probe track. 
Systematic effects arising from the choice of the binning scheme and for the difference in event multiplicity between data and simulation are studied.
In total, the muon PID efficiency uncertainty is determined to be $2.3\%$.

The trigger efficiency $\varepsilon_{\mathrm{trig}}\left(\jpsi\right)$ measured with simulation is compared with the result obtained in data for inclusive $\jpsi$ events using a tag-and-probe method~\cite{LHCb-DP-2012-004}. 
A difference of $1.0\%$ between the two results is observed and is taken as the systematic uncertainty.

An uncertainty of $1.0\%$ is assigned to the determination of the fraction of candidates from $\bquark$-hadron decays, which accounts for the uncertainty of the prompt $\jpsi$ and $\bbbar$ production cross-sections.
The uncertainty introduced by the limited statistics of the simulated samples used to determine the efficiencies is estimated to be negligible. 
The $1.1\%$ uncertainty on $\BF(\jpsi \to \mup \mun)$ is propagated to the cross-section. 
The systematic uncertainty due to the luminosity calibration is $3.9\%$.
The total systematic uncertainty\linebreak
is $6.1\%$.

\begin{table}[h]
\caption{Summary of the systematic uncertainties on the measurement of the $\jpsi$ pair production cross-section.}\label{tab:sys}
\begin{center}
\begin{tabular}{lc}
Source & Uncertainty[\%] \\
\hline
Signal shape & 1.6 \\
Data/simulation difference & 1.0 \\
Tracking efficiency & $0.8\times4$ \\
Muon PID efficiency & 2.3 \\
Trigger efficiency & 1.0 \\
Fraction of $\jpsi$ from $\bquark$-hadron candidates & 1.0 \\
$\BF(\jpsi \to \mup \mun)$ & 1.1 \\
Luminosity & 3.9 \\
\hline
Total & 6.1 \\
\end{tabular}
\end{center}
\end{table}

\section{Results and comparison to theory}
The $\jpsi$ pair production cross-section where both $\jpsi$ mesons are in 
the region \mbox{$2.0<y<4.5$} and \mbox{$\pt < 10 \gevc$} is measured to be
\begin{equation*}
   \sigma(\jpsi\jpsi) = 15.2 \pm 1.0\stat \pm 0.9\syst \nb,
\end{equation*}
assuming negligible polarisation of the $\jpsi$ mesons.
The detection efficiency of $\jpsi$ mesons can be affected by the polarisation,
especially by the polarisation parameter $\lambda_{\theta}$
in the helicity frame~\cite{LHCb-PAPER-2015-037,LHCb-PAPER-2013-008}.
If a value of $\lambda_{\theta} = \pm 20\%$ is assumed for both of the $\jpsi$ mesons,
the $\jpsi$ pair production cross-section changes by $\pm 7\%$.
The ratio of the production cross-section of the $\jpsi$ pair to that of the inclusive
prompt $\jpsi$ is calculated to be
\begin{equation}
\frac{\sigma(\jpsi\jpsi)}{\sigma(\jpsi)} = (10.2 \pm 0.7\stat \pm 0.9\syst) \times 10^{-4},
\end{equation}
where the production cross-section of prompt $\jpsi$ mesons in the range \mbox{$2.0<y<4.5$} and
\mbox{$\pt<10\gevc$} is \mbox{$\sigma(\jpsi) = 14.94 \pm 0.02\stat \pm 0.91\syst \mub$}~\cite{LHCb-PAPER-2015-037},
and the systematic uncertainties of $\sigma(\jpsi\jpsi)$ and $\sigma(\jpsi)$ are treated as uncorrelated.
According to Eq.~\eqref{eq:DPS}, the ratio 
\begin{equation}
   \frac{1}{2} \frac{\sigma(\jpsi)^2}{\sigma(\jpsi\jpsi)} = 7.3 \pm 0.5\stat \pm 1.0\syst \mbarn.
\end{equation}
can be interpreted as \mbox{$\sigma_{\mathrm{eff}}$} if all $\jpsi$ pairs are produced through the~DPS~process.

The~results on $\jpsi$~pair production are compared with a data\nobreakdash-driven prediction for 
the DPS~mechanism and several calculations performed within the SPS~mechanism. 
The~DPS prediction is calculated via Eq.~\eqref{eq:DPS} using 
the~measured \jpsi~production cross\nobreakdash-section
at \mbox{$\sqs=13\tev$}~\cite{LHCb-PAPER-2015-037} and the~effective cross-section 
\mbox{$\sigma_{\mathrm{eff}}=14.5\pm1.7^{+1.7}_{-2.3}\mbarn$} from 
Refs.~\cite{Abe:1997xk,Gamma3jets1}.

Theoretical predictions of the production cross-section of $\jpsi$ pairs are summarized 
in Table~\ref{tab:theory}.
The contribution from the SPS mechanism is calculated using several approaches:
the~state\nobreakdash-of\nobreakdash-art complete NLO colour\nobreakdash-singlet\,(NLO\,CS) computations~\cite{Sun:2014gca};
the~incomplete\,(no-loops) 
next\nobreakdash-to\nobreakdash-leading\nobreakdash-order 
colour\nobreakdash-singlet\,(NLO$^{\ast}$\,CS)~calculations~\mbox{\cite{Likhoded:2016zmk, 
Lansberg:2013qka, Lansberg:2014swa,Lansberg:2015lva,Shao:2012iz,Shao:2015vga}};
leading-order colour-singlet\,(LO\,CS)~\cite{Likhoded:2016zmk} 
and colour-octet\,(LO\,CO)~\cite{Shao:2012iz,Shao:2015vga} calculations
and 
the~approach based on the $k_{\mathrm{T}}$-factorisation 
method~\cite{Gribov:1984tu,Levin:1990gg,Andersson:2002cf,Andersen:2003xj,Andersen:2006pg},
with the~leading-order colour-singlet matrix 
element\,(LO\,$k_{\mathrm{T}}$)~\cite{Baranov:2011zz,Baranov:1993qv}.
Even~with the~leading\nobreakdash-order matrix element,
the~LO\,$k_{\mathrm{T}}$~approach includes 
a~large fraction of higher\nobreakdash-order contributions
via the~evolution of 
parton densities~\cite{Baranov:2011zz}.
Since~NLO$^{\ast}$\,CS~calculations are divergent at small transverse momentum of the \jpsi~pair, 
two approaches are used:
a~simple cut-off for $\pt(\jpsi\jpsi)$~\cite{Likhoded:2016zmk}\,(denoted as NLO$^{\ast}$\,CS$^{\prime}$), 
and a cut on the mass of any light parton pair\,(NLO$^{\ast}$\,CS$^{\prime\prime}$)~\cite{Lansberg:2013qka,
  Lansberg:2014swa,Lansberg:2015lva,Shao:2012iz,Shao:2015vga}.

Gluon densities from Refs.~\cite{Jung:2004gs,Jung:2010si,Jung:2012hy,Hautmann:2013tba,Jung:2014vaa} 
are used for the LO\,$k_{\mathrm{T}}$~approach, while {\sc{CT\,14}} parton distribution 
functions\,(PDF)~\cite{Dulat:2015mca} are used for 
LO\,CS and NLO$^{\ast}$\,CS$^{\prime}$~calculations,
{\sc{NNPDF\,3.0}}~NLO\,PDFs with $\alpha_s(M_Z)=0.118$~\cite{Ball:2014uwa} 
are used for LO\,CO and NLO$^{\ast}$\,CS$^{\prime\prime}$~predictions,
and {\sc{CTEQ6L1}} and {\sc{CTEQ6M}} PDFs~\cite{Lai:1999wy,Pumplin:2002vw} 
are used for NLO\,CS computations.
For LO\,CO predictions the long-distance matrix elements\,(LDMEs) are taken from 
Refs.~\cite{Butenschoen:2011yh,Sharma:2012dy,NRQCDpol3,
  Sun:2012vc,Bodwin:2014gia, Shao:2014yta,Kramer:2001hh,Braaten:1999qk}
and a smearing of transverse momenta of initial gluons, 
similar to that used in NLO$^{\ast}$\,CS$^{\prime\prime}$, is applied. 
The~production cross-section of \jpsi~pairs is sensitive to 
the~choice of parameters; for example, it varies by a factor between 0.8 and 3 
when varying the~factorisation and renormalisation 
scales by a~factor of two, or increases 
if the {\sc{CTEQ\,6L}}~PDF set~\cite{Nadolsky:2008zw} is used instead of 
the nominal PDFs.
The~contribution of LO\,CO is very sensitive to the choice of the LDME;
the~absolute cross-section varies from the~minimum of~0.11\nb, based on 
LDME set from Ref.~\cite{Butenschoen:2011yh} to the~maximum of 0.70\nb, 
calculated using LDME set from Ref.~\cite{Bodwin:2014gia}, 
while most of the~predictions cluster around 0.5\nb.
The feed-down from $\psitwos\to\jpsi X$ decays is included in the LO\,$k_{\mathrm{T}}$, 
LO\,CO and NLO$^{\ast}$\,CS$^{\prime\prime}$ calculations 
and not in the LO\,CS and NLO$^{\ast}$\,CS$^{\prime}$~calculations.
Likewise, a~tiny contribution from $\jpsi\chi_{\cquark}$~production with subsequent
decay $\chi_{\cquark}\to\jpsi\g$~\cite{Likhoded:2016zmk} 
is included in the NLO$^{\ast}$\,CS$^{\prime}$~and LO\,CO~results 
but neglected in the NLO$^{\ast}$\,CS$^{\prime\prime}$~calculations.

\begin{table}[tb]
  \centering
  \caption{ \small
    Summary of the theoretical predictions and the measurement of $\sigma(\jpsi\jpsi)$
    for different regions of transverse momentum of the $\jpsi$ pair.
    For~SPS predictions, the~first uncertainty 
    accounts for the~variation of PDFs and gluon densities, 
    while the~second one corresponds to the~variation of 
    the~factorisation and renormalisation scales. 
    For the LO\,CO~predictions the~third uncertainty corresponds to the~choice of 
    LDMEs from Refs.\mbox{~\cite{Butenschoen:2011yh,Sharma:2012dy,NRQCDpol3,
        Sun:2012vc,Bodwin:2014gia, Shao:2014yta,Kramer:2001hh,Braaten:1999qk}}.
    For NLO\,CS~predictions~\cite{Sun:2014gca} the~uncertainty corresponds
    to the~variation of the~factorisation and renormalization scales.
    For the DPS prediction the first uncertainty 
    is due to the~measured prompt \jpsi~production cross-section~\cite{LHCb-PAPER-2015-037} 
    and the~second is due to the uncertainty in $\sigma_{\mathrm{eff}}$~\cite{Abe:1997xk,Gamma3jets1}.
  } \label{tab:theory}
  \vspace*{3mm}
  \begin{tabular*}{1.00\textwidth}{@{\hspace{0.1mm}}l@{\extracolsep{\fill}}ccc@{\hspace{0.1mm}}}
    \multirow{2}{*}{}  
    & \multicolumn{3}{c}{ $\sigma(\jpsi\jpsi)~\left[\!\nb\right]$ }\\ 
    \cmidrule{2-4}
    &  no $\pt$ cut
    &  $\pt>1\gevc$
    &  $\pt>3\gevc$
    \\
    \hline
    \\[-1em]
    LO\,CS~\cite{Likhoded:2016zmk} 
    & $\phantom{0}1.3\pm0.1^{+3.2}_{-0.1}$ 
    & --- 
    & --- 
    \\
    \\[-1em]
    LO\,CO~\cite{Shao:2012iz,Shao:2015vga}
    &  $0.45\pm0.09^{+1.42+0.25}_{-0.36-0.34}$
    &  --- 
    &  ---
    \\
    \\[-1em]
    LO\,$k_{\mathrm{T}}$~\cite{Baranov:2011zz} 
    & $\phantom{0}6.3^{+3.8+3.8}_{-1.6-2.6}$  
    & $\phantom{0}5.7^{+3.4+3.2}_{-1.5-2.1}$
    & $\phantom{0}2.7^{+1.6+1.6}_{-0.7-1.0}$
    \\
    \\[-1em]
    NLO$^{\ast}$\,CS$^{\prime}$~\cite{Likhoded:2016zmk} 
    & ---  
    & $\phantom{0}4.3\pm0.1^{+9.9}_{-0.9}$
    & $\phantom{0}1.6\pm0.1^{+3.3}_{-0.3}$
    \\
    \\[-1em]
    NLO$^{\ast}$\,CS$^{\prime\prime}$~\cite{Lansberg:2013qka,Lansberg:2014swa,Lansberg:2015lva,Shao:2012iz,Shao:2015vga} 
    &  $15.4\pm2.2^{+51\phantom{.}}_{-12}$
    &  $14.8\pm1.7^{+53\phantom{.}}_{-12}$
    &  $\phantom{0}6.8\pm0.6^{+22\phantom{.}}_{-5}$
    \\ 
    \\[-1em] 
    NLO\,CS~\cite{Sun:2014gca} 
    &  $11.9^{+4.6}_{-3.2}$
    &  --- 
    &  --- 
    \\ 
    \\[-1em] 
    DPS~\cite{Abe:1997xk,Gamma3jets1,LHCb-PAPER-2015-037} 
    & $\phantom{0}8.1\pm0.9^{+1.6}_{-1.3}$
    & $\phantom{0}7.5\pm0.8^{+1.5}_{-1.2}$
    & $\phantom{0}4.9\pm0.5^{+1.0}_{-0.8}$
    \\
    \hline
    \\[-1em]
    Data
    & $          15.2\pm1.0\pm0.9$
    & $          13.5\pm0.9\pm0.9$
    & $\phantom{0}8.3\pm0.6\pm0.5$
  \end{tabular*}
\end{table}

While the~predictions for the production 
cross\nobreakdash-section of \jpsi~pairs are significantly 
affected by the~theory uncertainties, the~shapes of 
the~differential cross\nobreakdash-sections are very stable and 
practically invariant with respect to the~choice of 
PDFs, scales and LDMEs.  
In contrast, the~smearing of gluon transverse momenta 
for NLO$^{\ast}$\,CS$^{\prime\prime}$ and LO\,CO~models
does not affect the~production cross-section, 
but significantly affects some differential 
distributions.

\begin{figure}[tb]
\begin{center}
\includegraphics[width=0.495\linewidth]{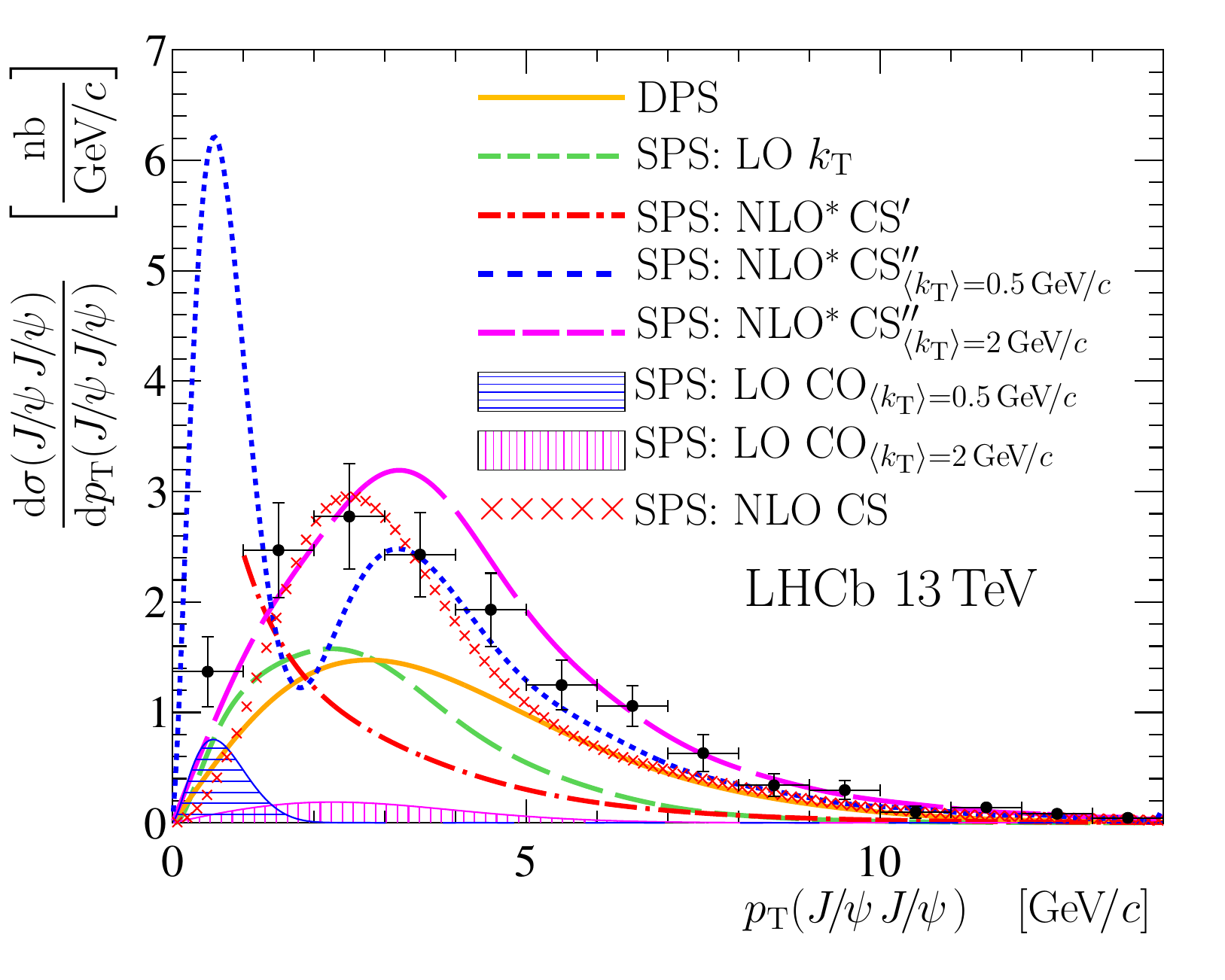}
\end{center}
  \caption { \small
     Comparisons between measurements and theoretical predictions for the differential cross-sections as a function of $p_{\mathrm{T}}(\jpsi\jpsi)$.
    The (black) points with error bars represent the measurements.
  }
  \label{fig:cmppt}
\end{figure}

\begin{figure}[tb]
\begin{center}
\includegraphics[width=0.495\linewidth]{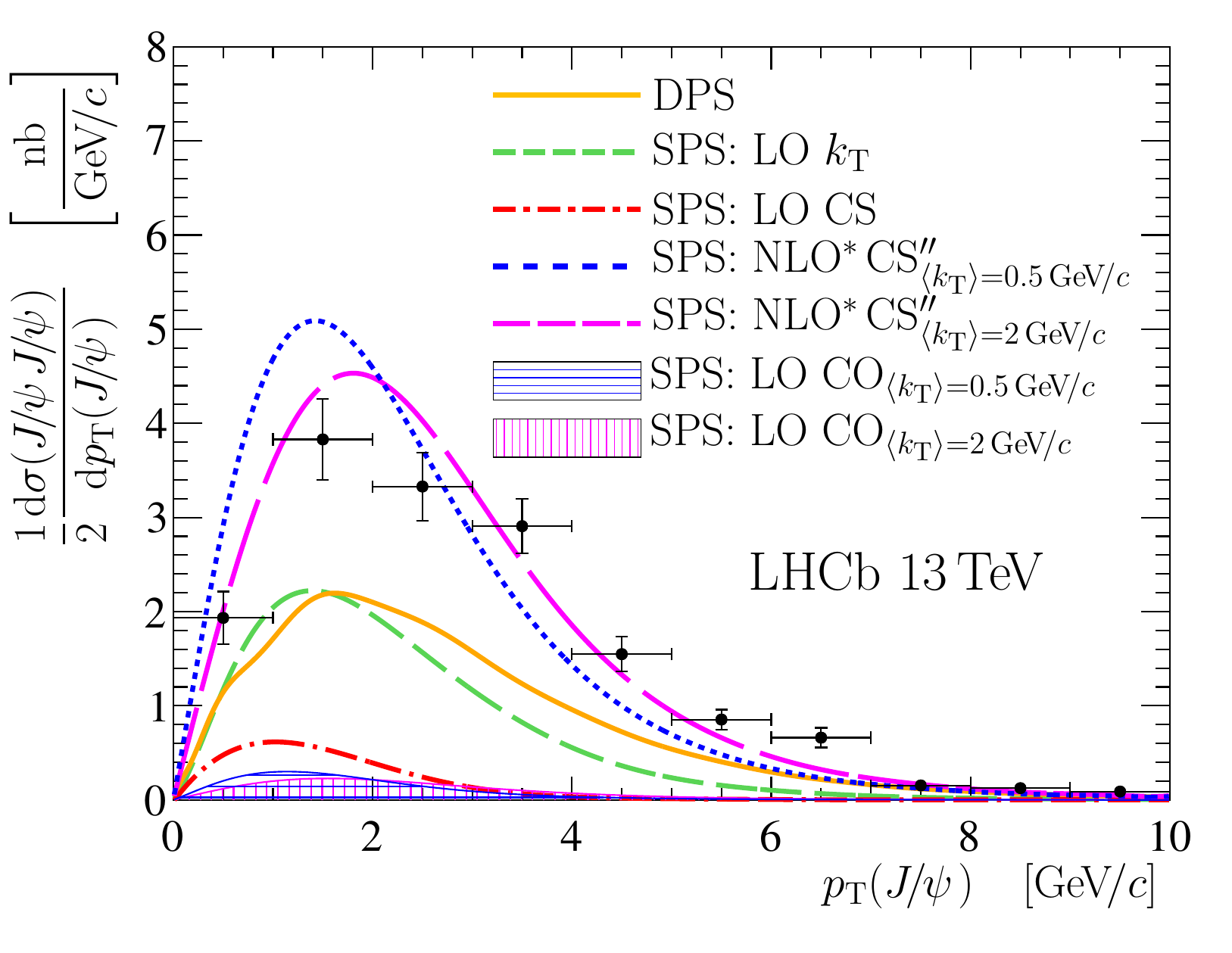}
\includegraphics[width=0.495\linewidth]{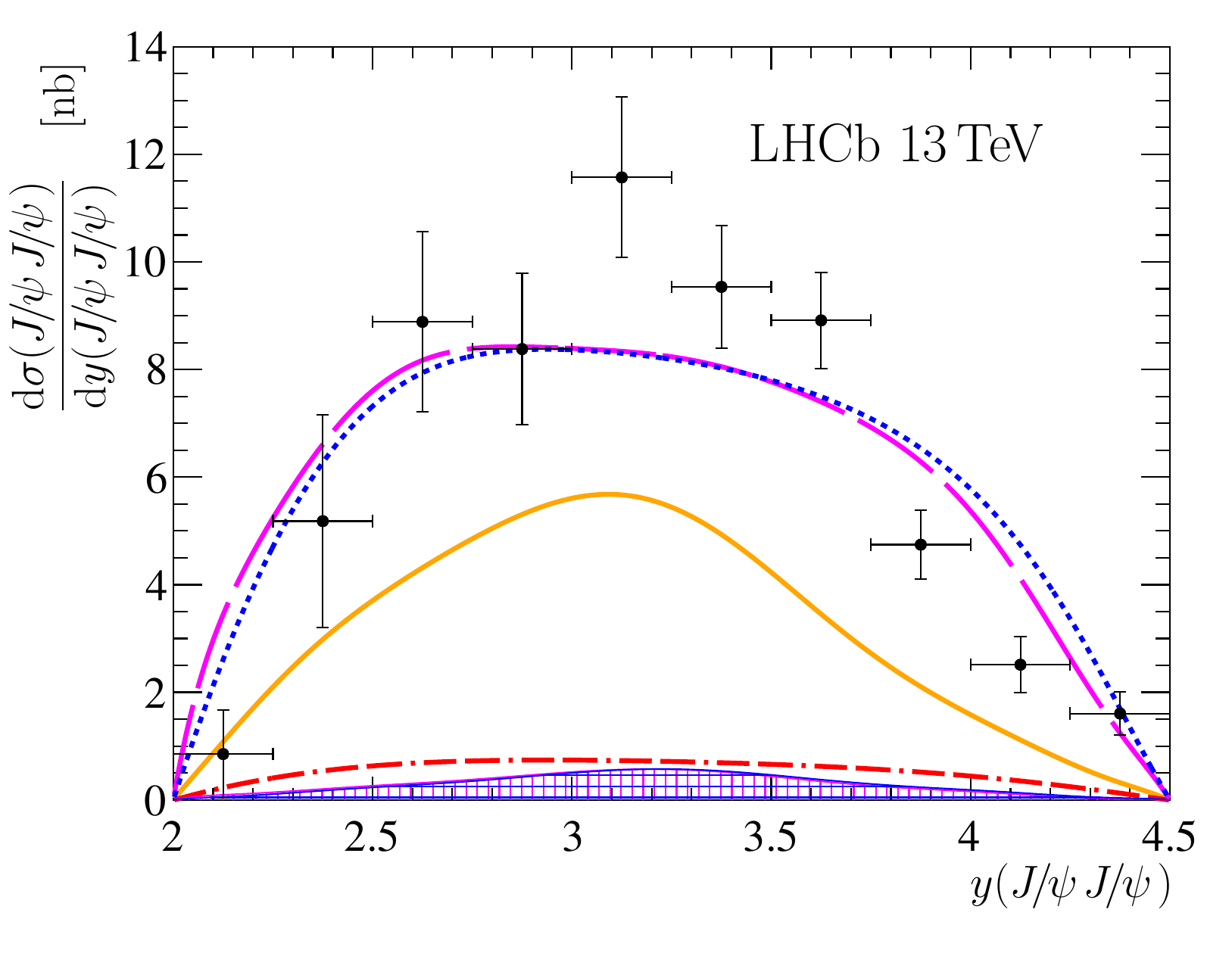}
\includegraphics[width=0.495\linewidth]{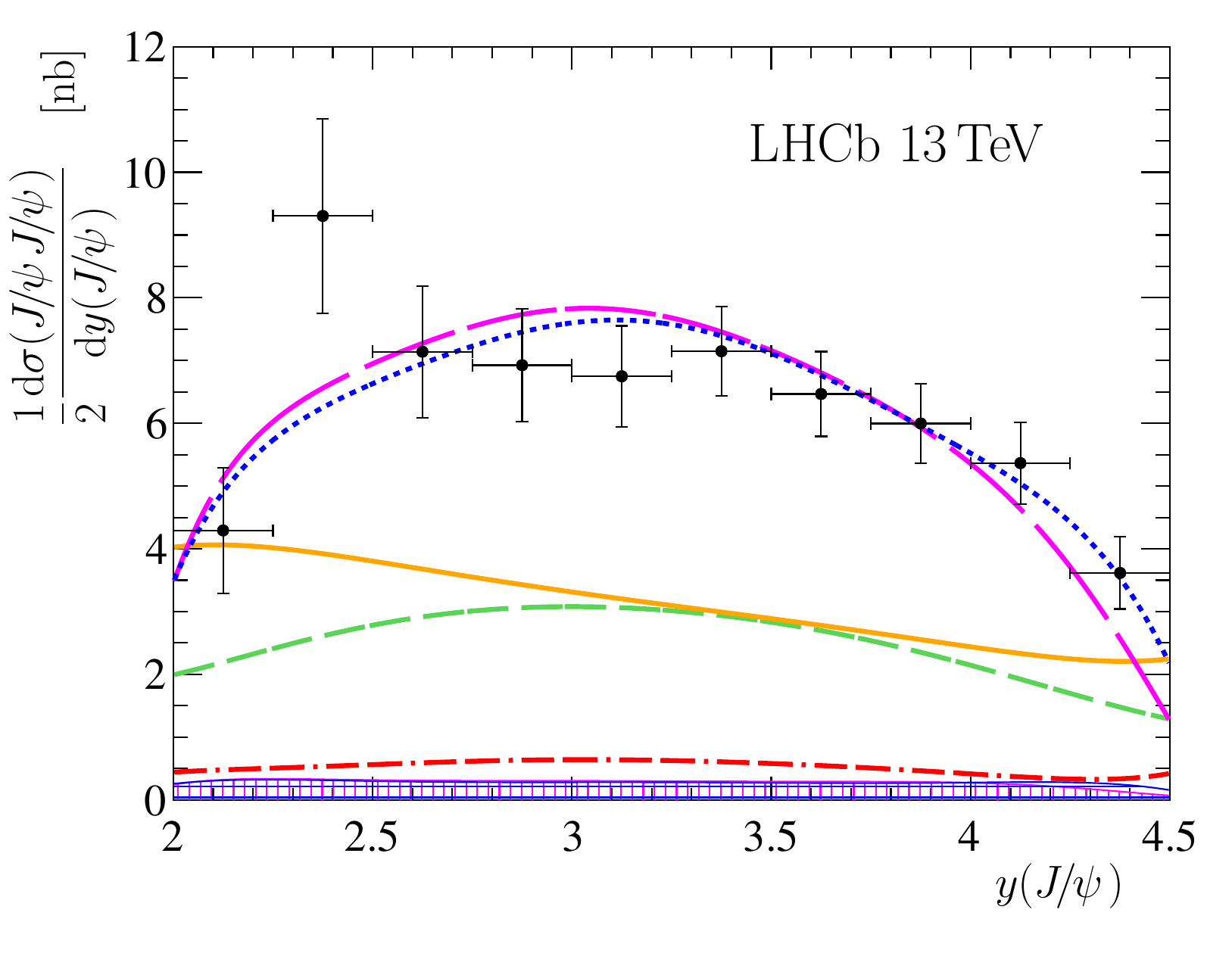}
\end{center}
  \caption { \small
     Comparisons between measurements and theoretical predictions for the differential cross-sections as functions of (top left)~$p_{\mathrm{T}}(\jpsi)$, (top right)~$y(\jpsi\jpsi)$ and (bottom)~$y(\jpsi)$.
    The (black) points with error bars represent the measurements.
  }
  \label{fig:cmp1}
\end{figure}

\begin{figure}[tb]
\begin{center}
\includegraphics[width=0.495\linewidth]{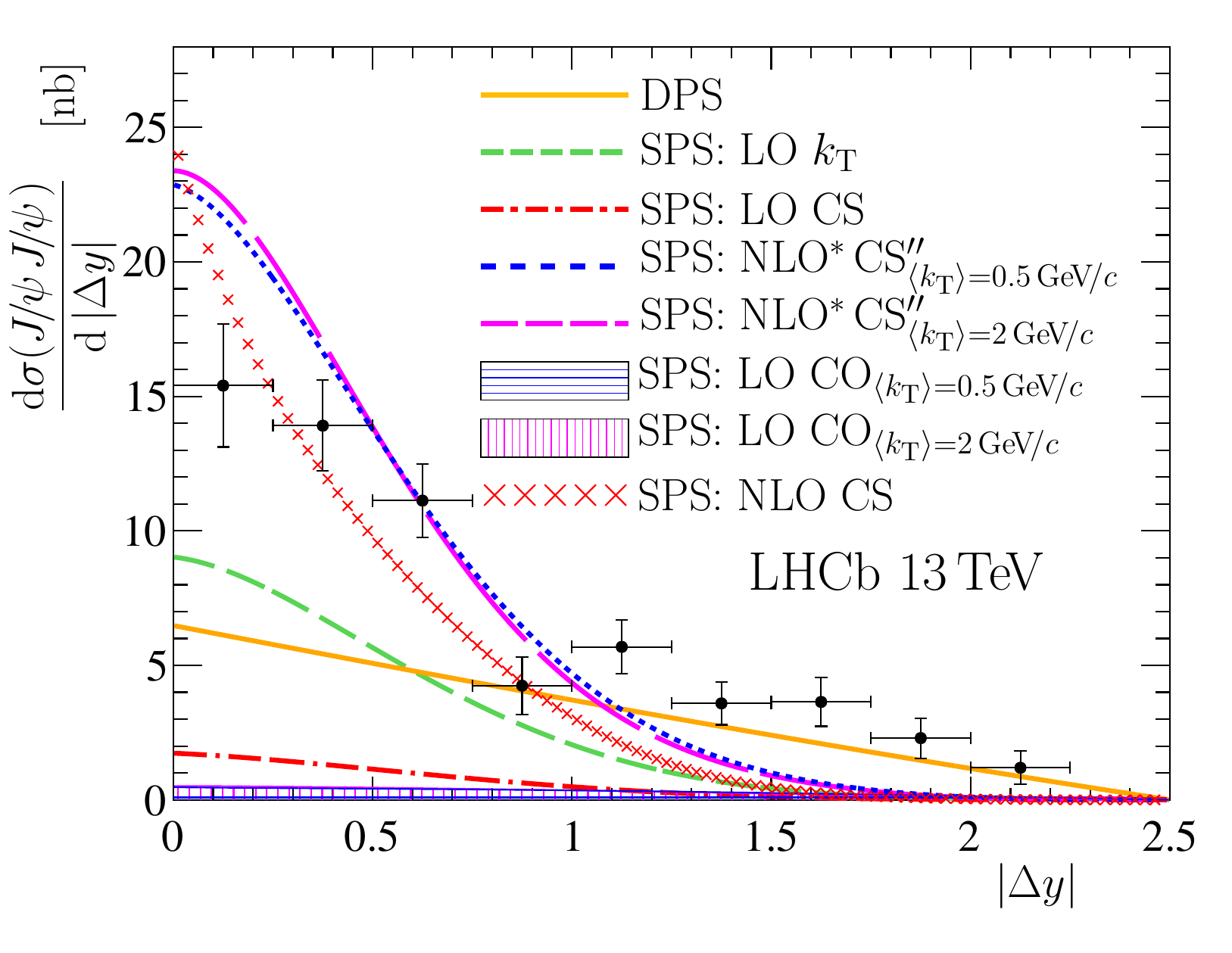}
\includegraphics[width=0.495\linewidth]{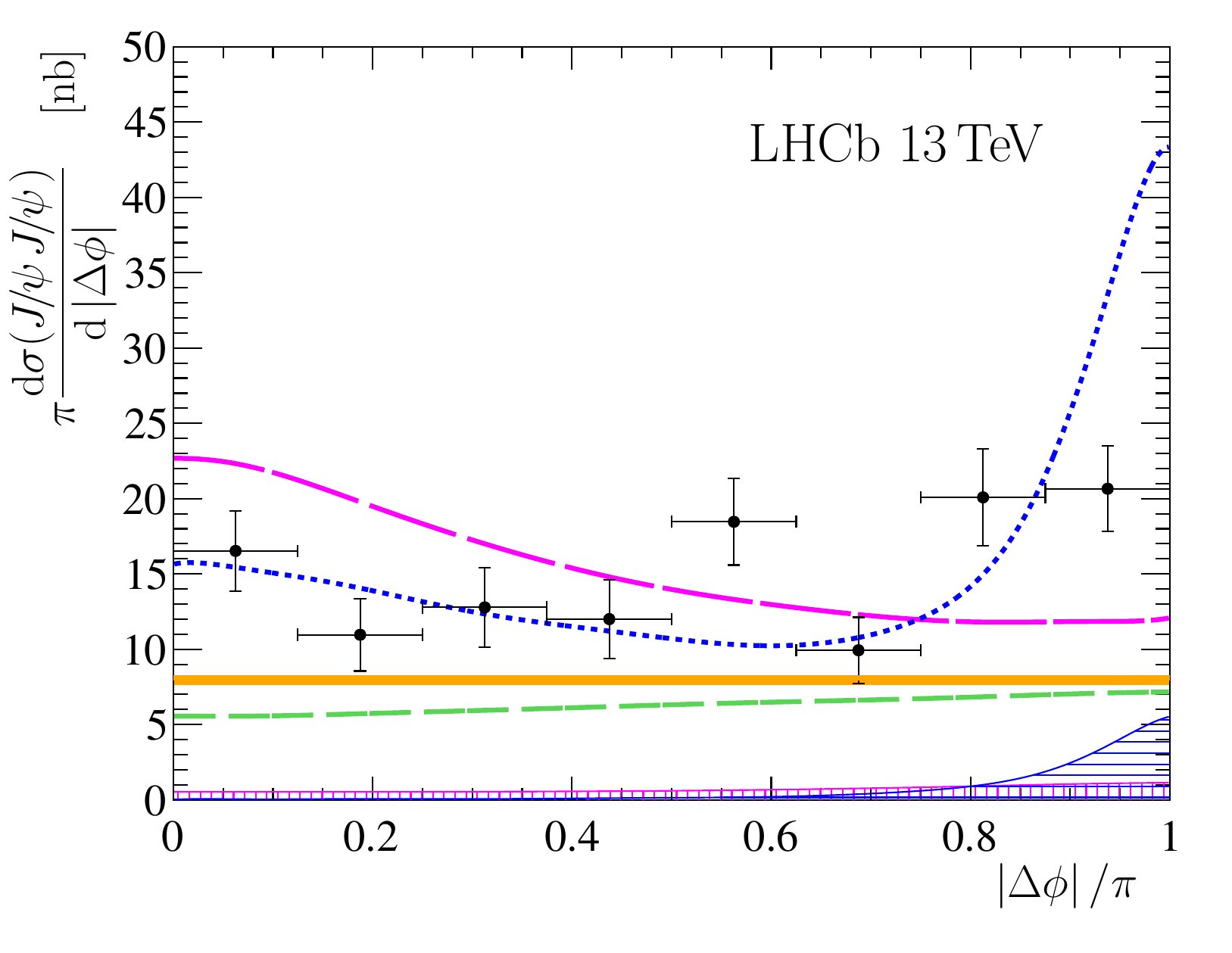}
\includegraphics[width=0.495\linewidth]{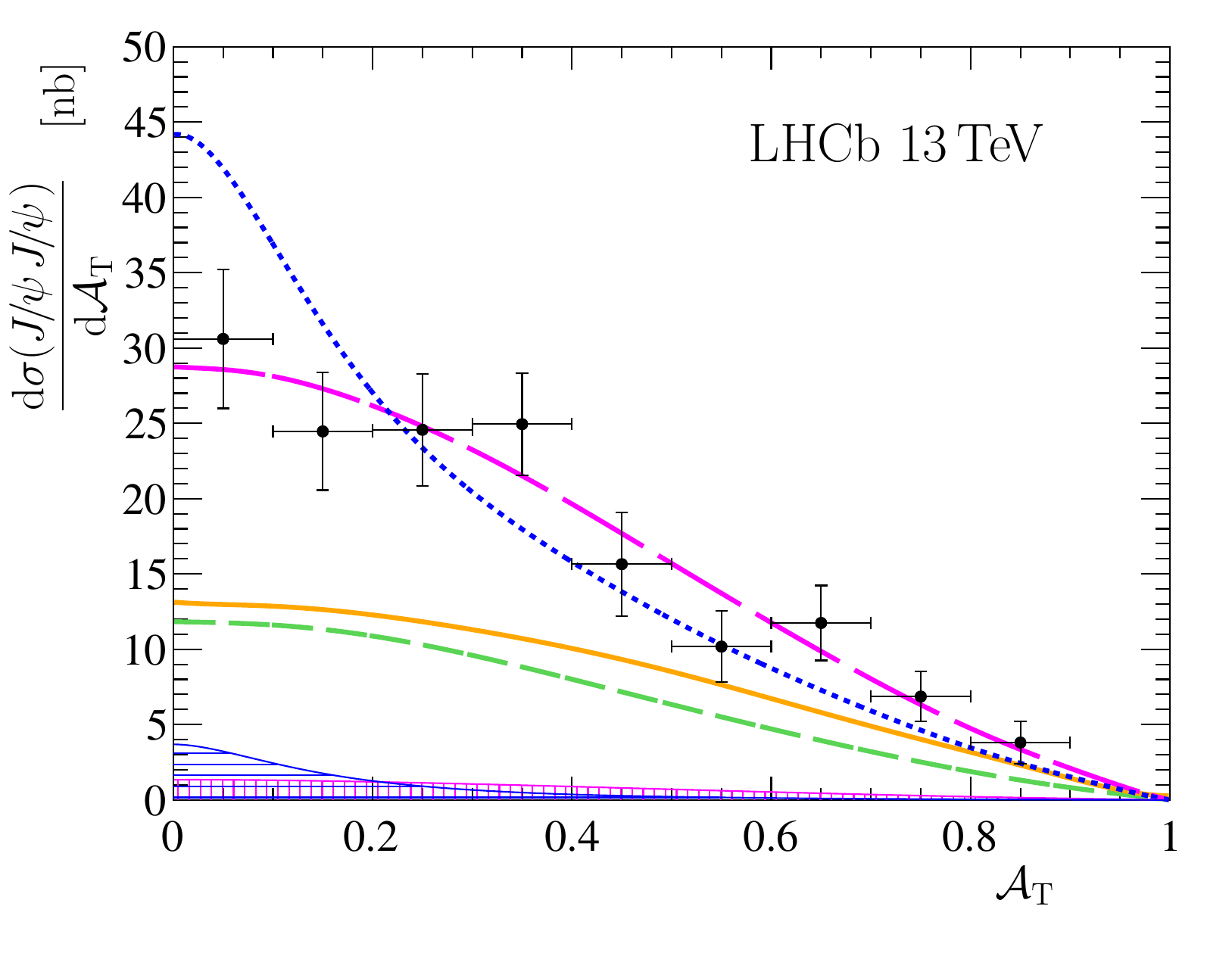}
\includegraphics[width=0.495\linewidth]{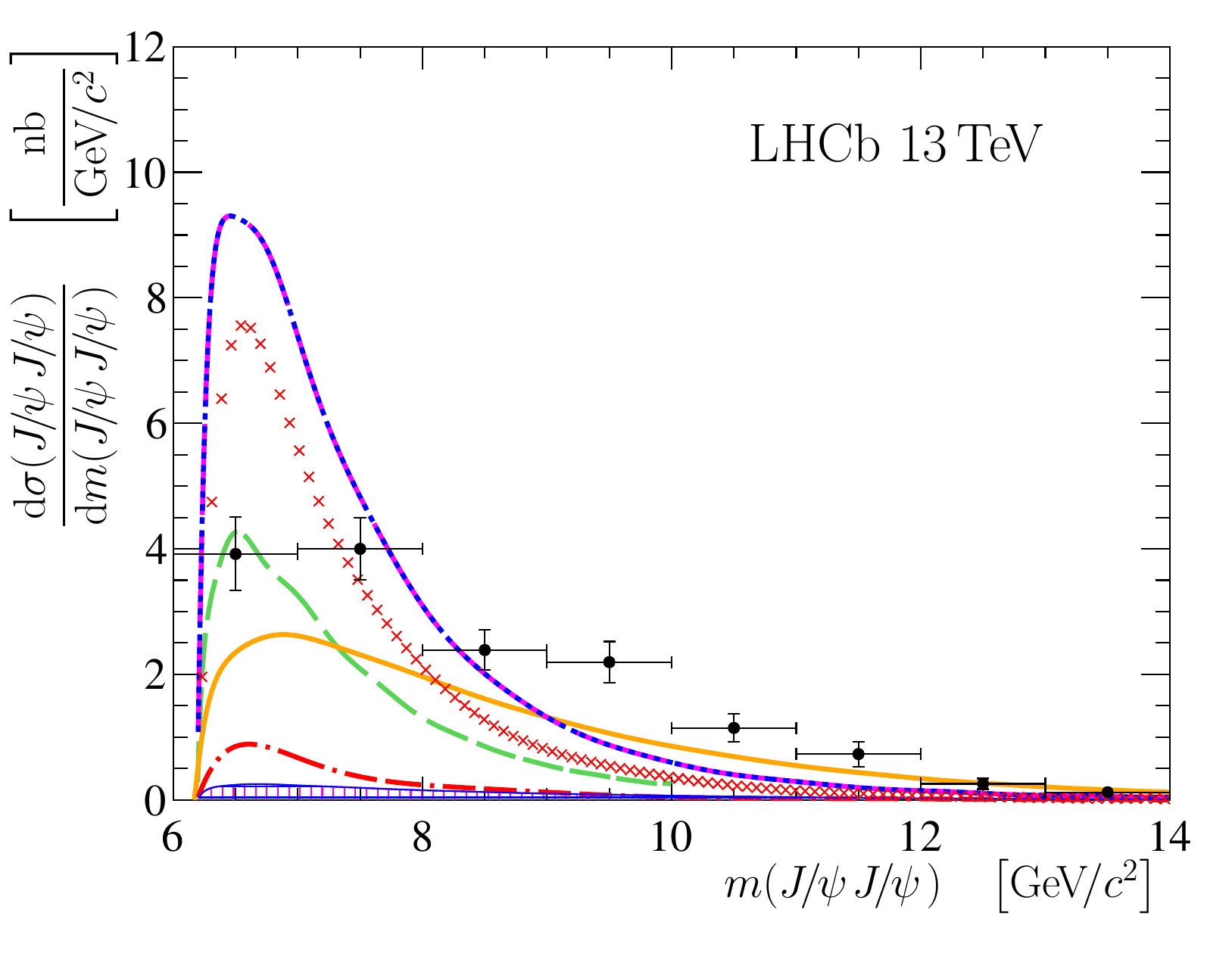}
\end{center}
  \caption { \small
     Comparisons between measurements and theoretical predictions for the differential cross-sections as functions of (top left)~$\left| \Delta y \right|$, (top right)~$\left| \Delta \phi \right|$, (bottom left)~$\mathcal{A}_{\mathrm{T}}$ and (bottom right)~$m(\jpsi\jpsi)$.
    The (black) points with error bars represent the measurements.
  }
  \label{fig:cmp2}
\end{figure}

\begin{figure}[tb]
\begin{center}
\includegraphics[width=0.495\linewidth]{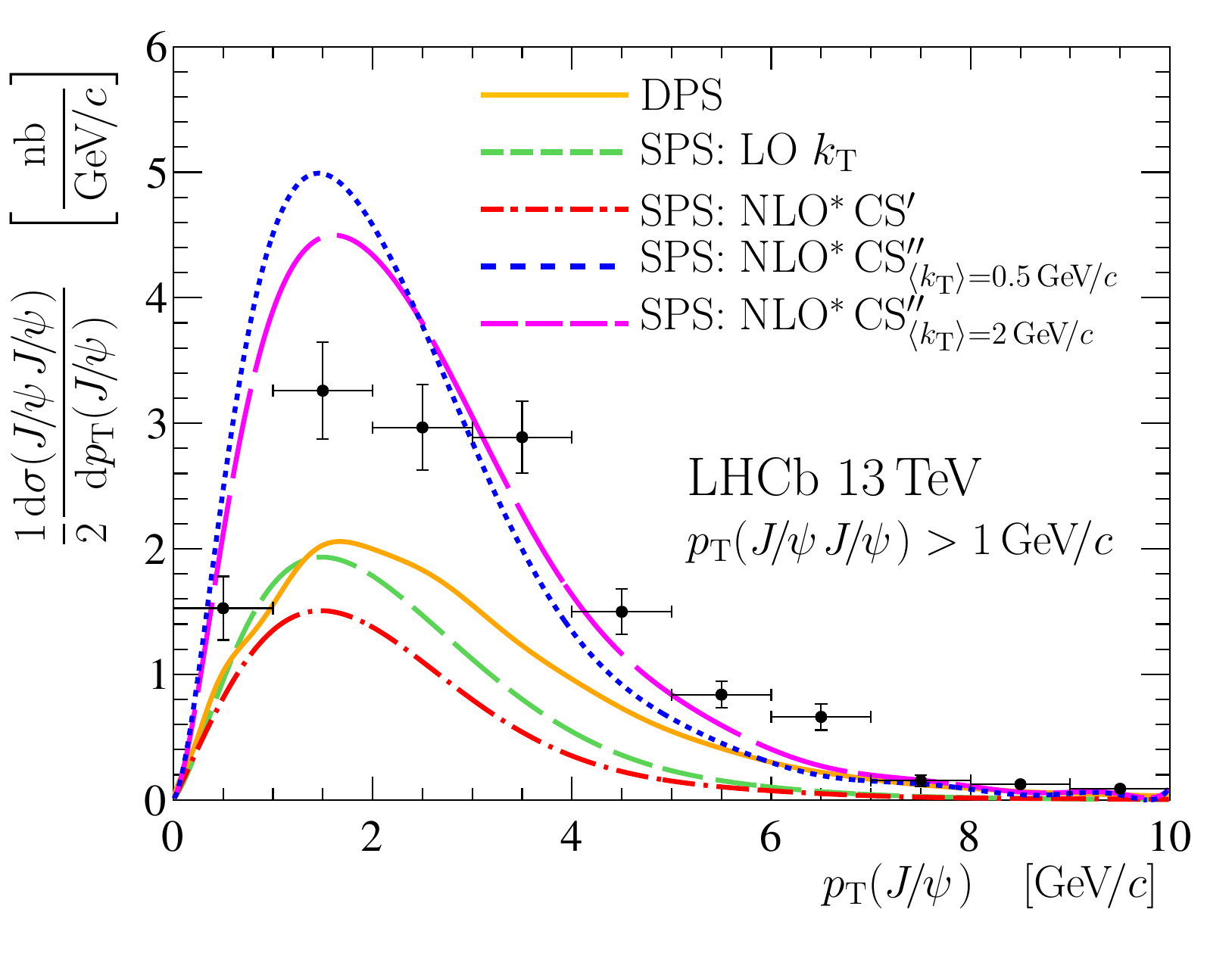}
\includegraphics[width=0.495\linewidth]{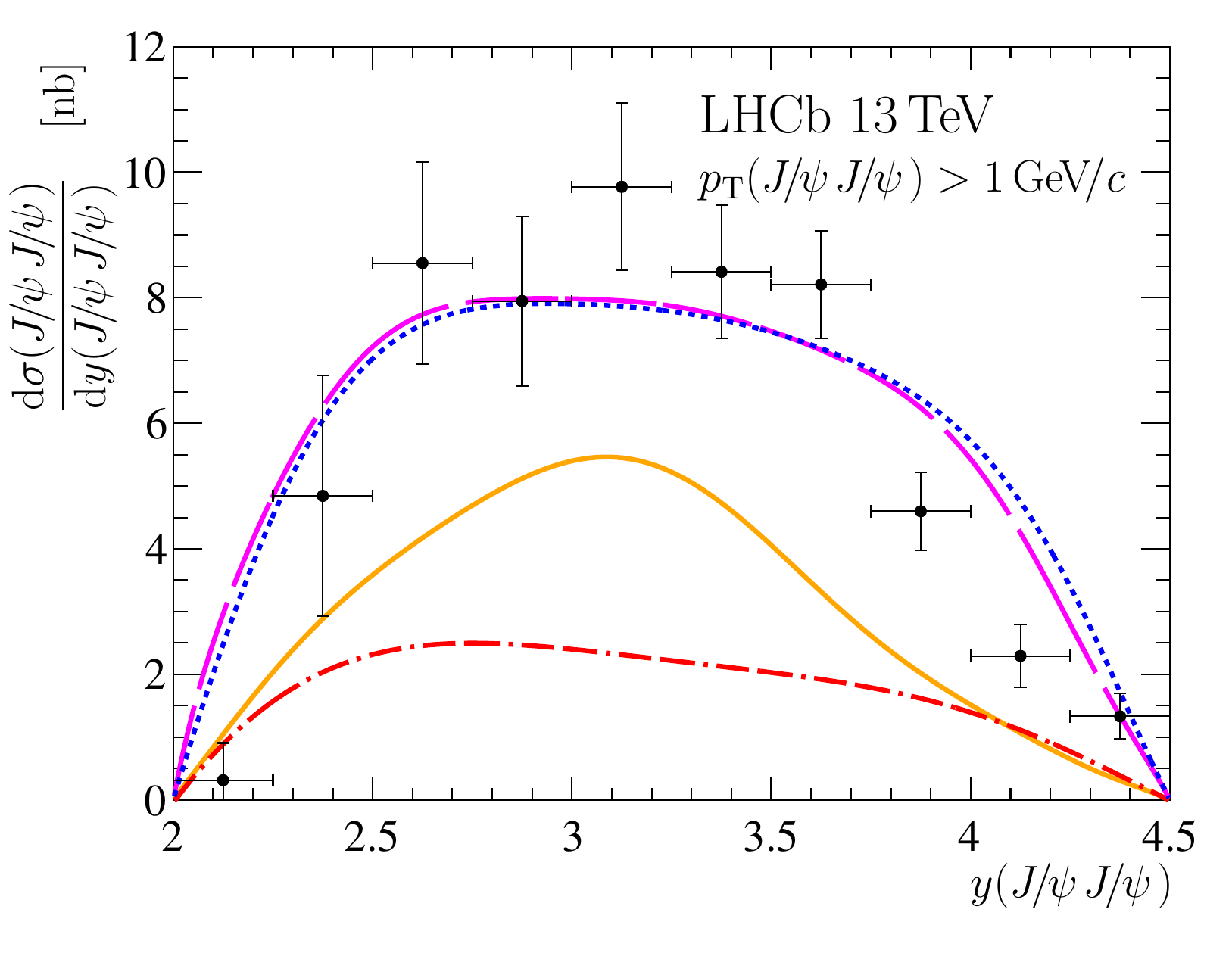}
\includegraphics[width=0.495\linewidth]{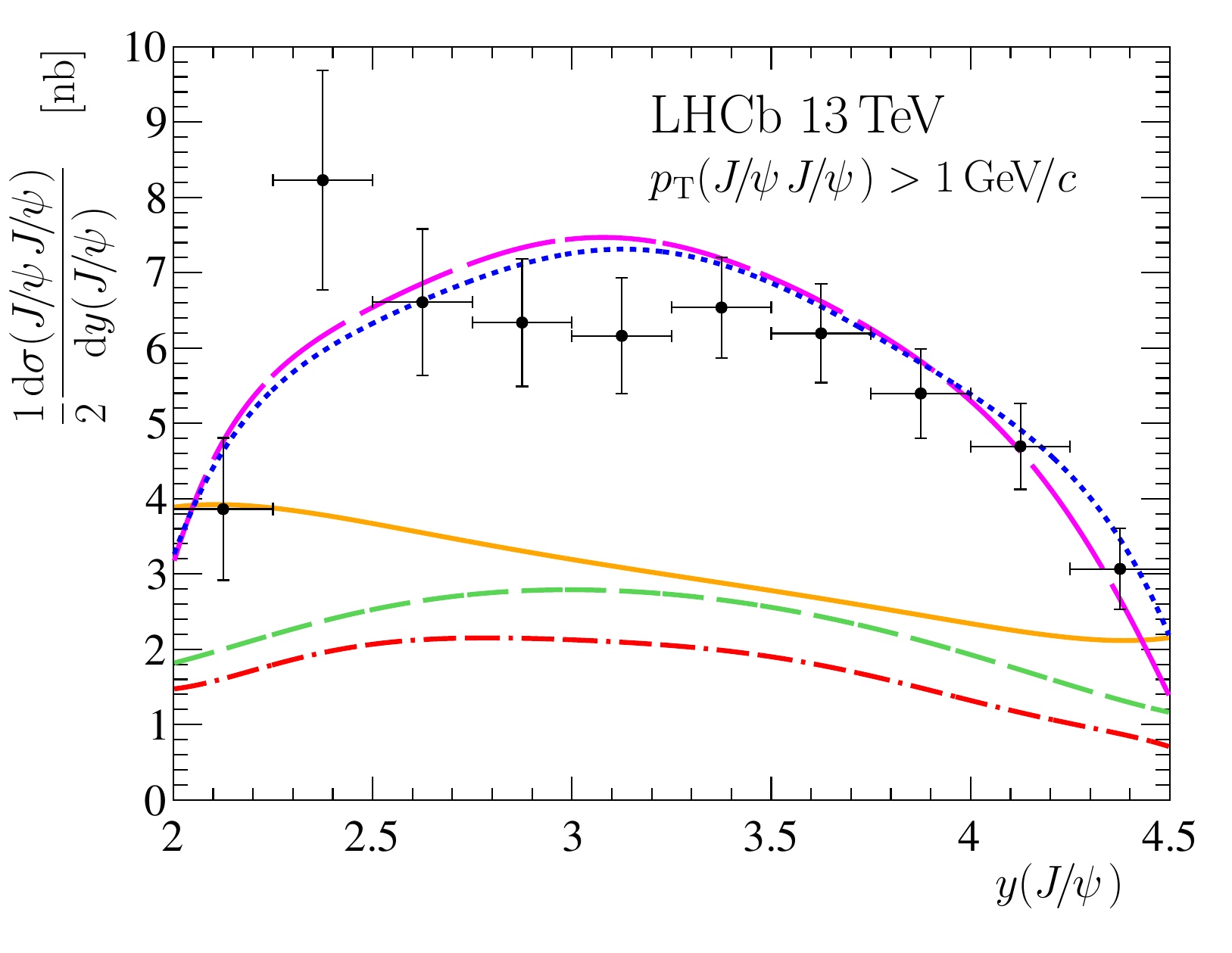}
\end{center}
  \caption { \small
     Comparisons between measurements and theoretical predictions with \mbox{$p_{\mathrm{T}}(\jpsi\jpsi)>1\gevc$} for the differential cross-sections as functions of (top left)~$p_{\mathrm{T}}(\jpsi)$, (top right)~$y(\jpsi\jpsi)$ and (bottom)~$y(\jpsi)$.
    The (black) points with error bars represent the measurements.
  }
  \label{fig:cmp1GeV1}
\end{figure}

\begin{figure}[tb]
\begin{center}
\includegraphics[width=0.495\linewidth]{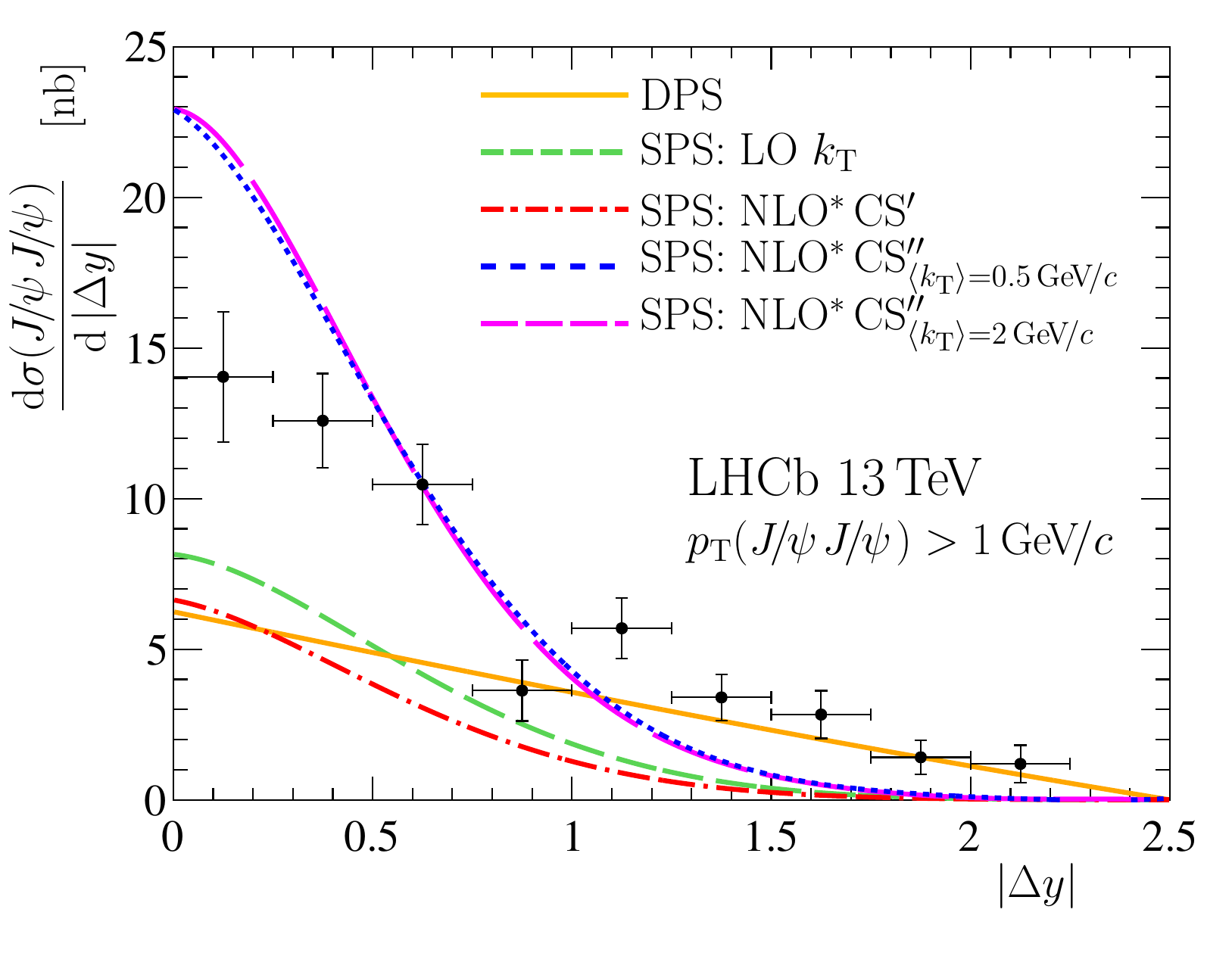}
\includegraphics[width=0.495\linewidth]{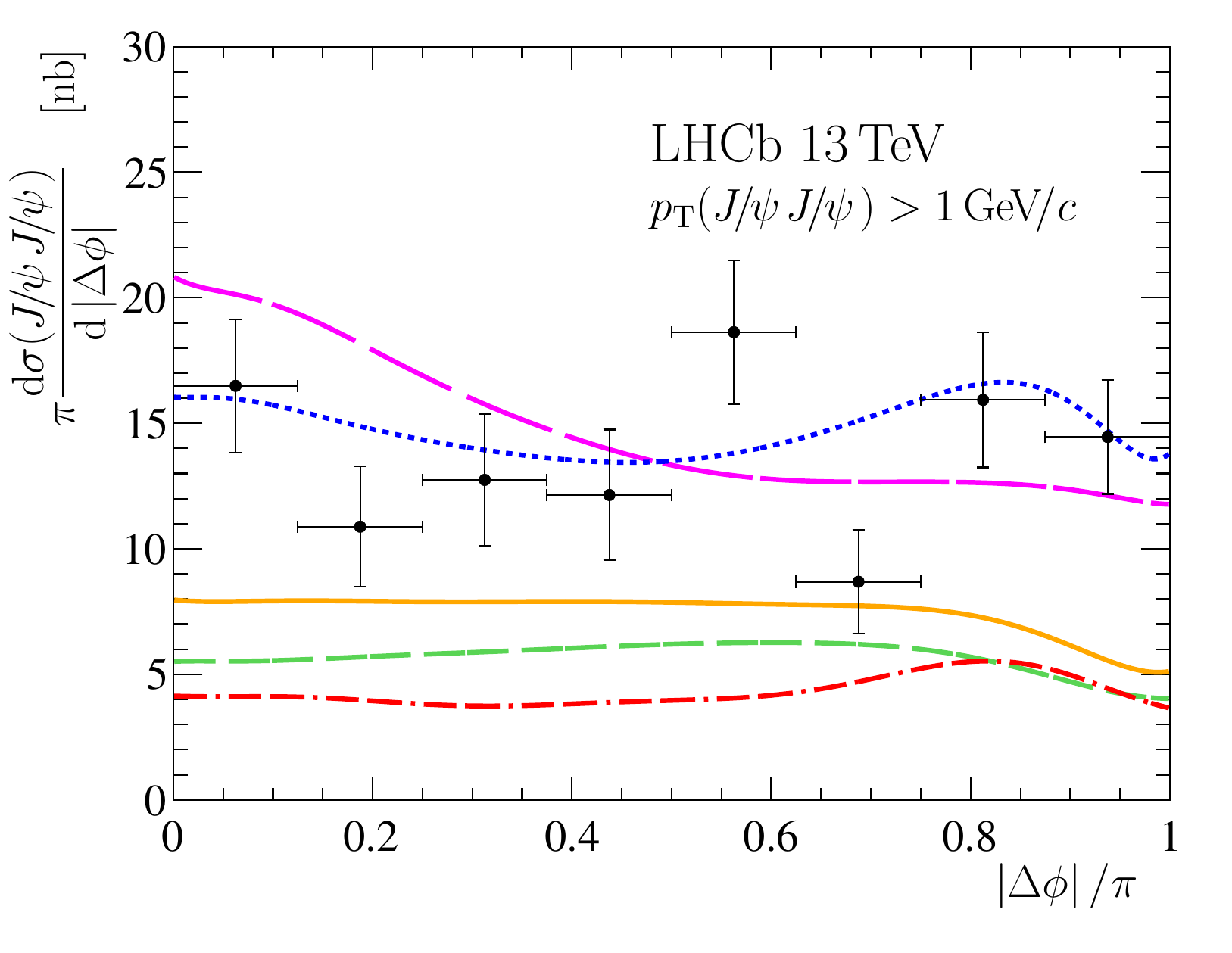}
\includegraphics[width=0.495\linewidth]{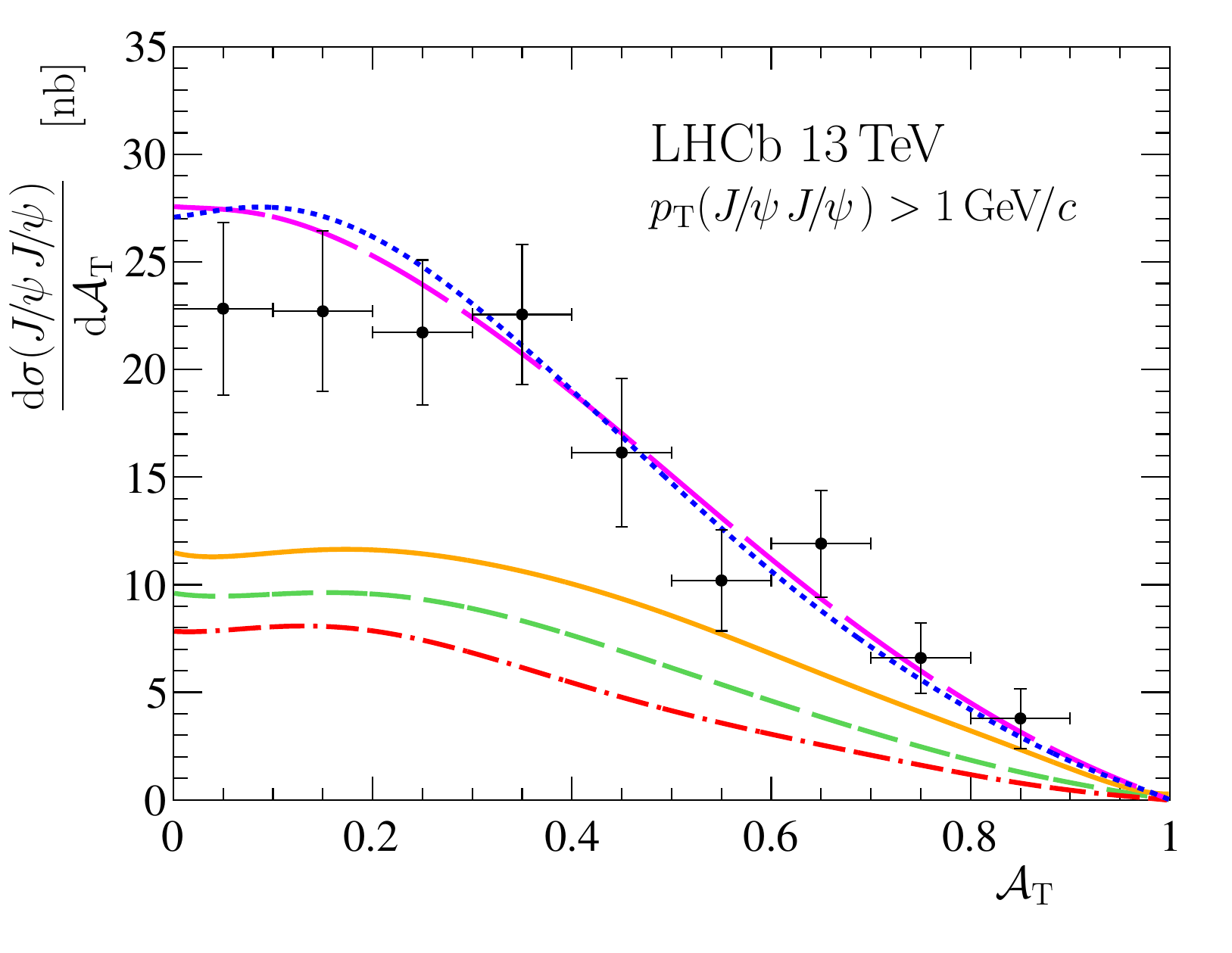}
\includegraphics[width=0.495\linewidth]{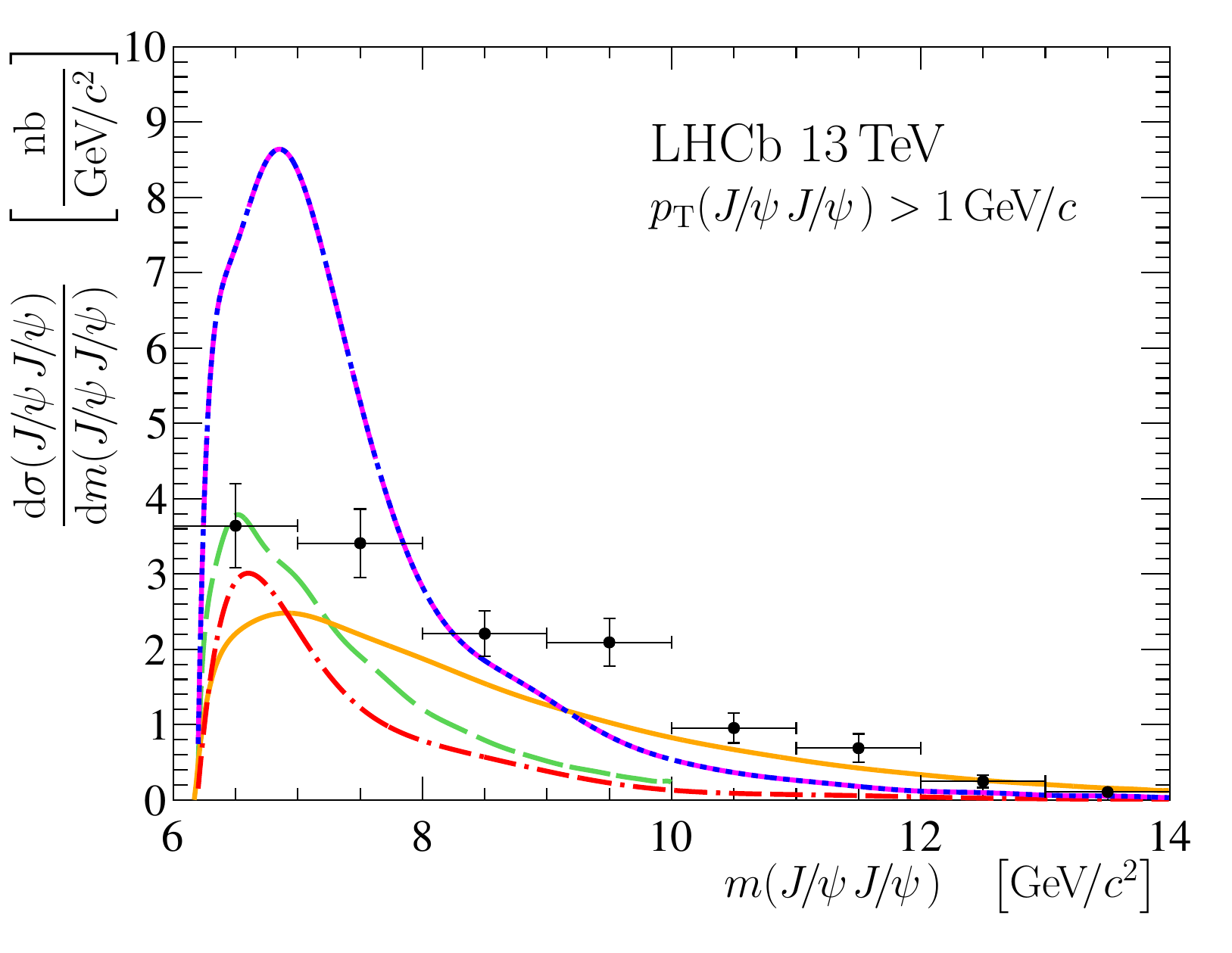}
\end{center}
  \caption { \small
     Comparisons between measurements and theoretical predictions with \mbox{$p_{\mathrm{T}}(\jpsi\jpsi)>1\gevc$} for the differential cross-sections as functions of (top left)~$\left| \Delta y \right|$, (top right)~$\left| \Delta \phi \right|$, (bottom left)~$\mathcal{A}_{\mathrm{T}}$ and (bottom right)~$m(\jpsi\jpsi)$.
    The (black) points with error bars represent the measurements.
  }
  \label{fig:cmp1GeV2}
\end{figure}

\begin{figure}[tb]
\begin{center}
\includegraphics[width=0.495\linewidth]{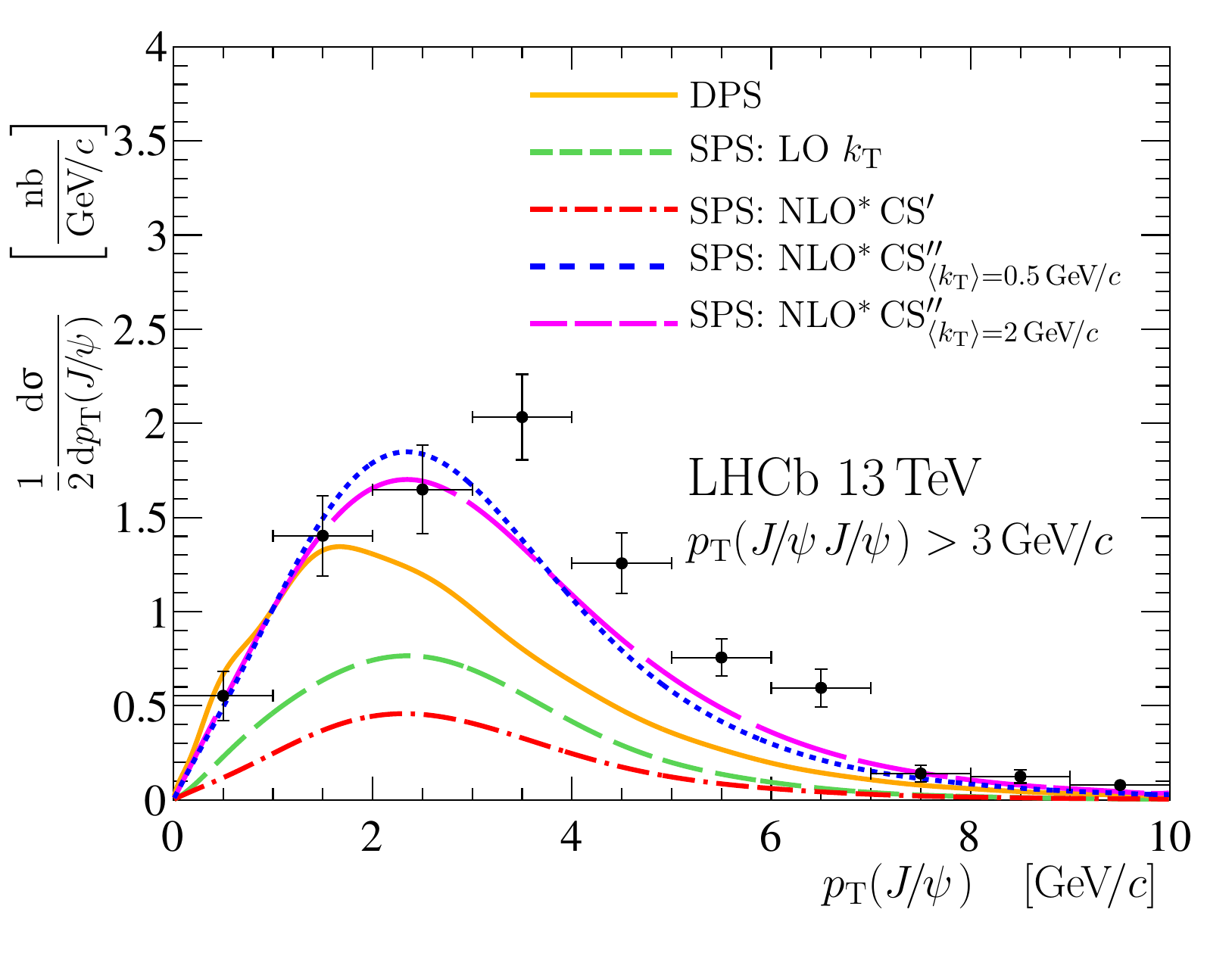}
\includegraphics[width=0.495\linewidth]{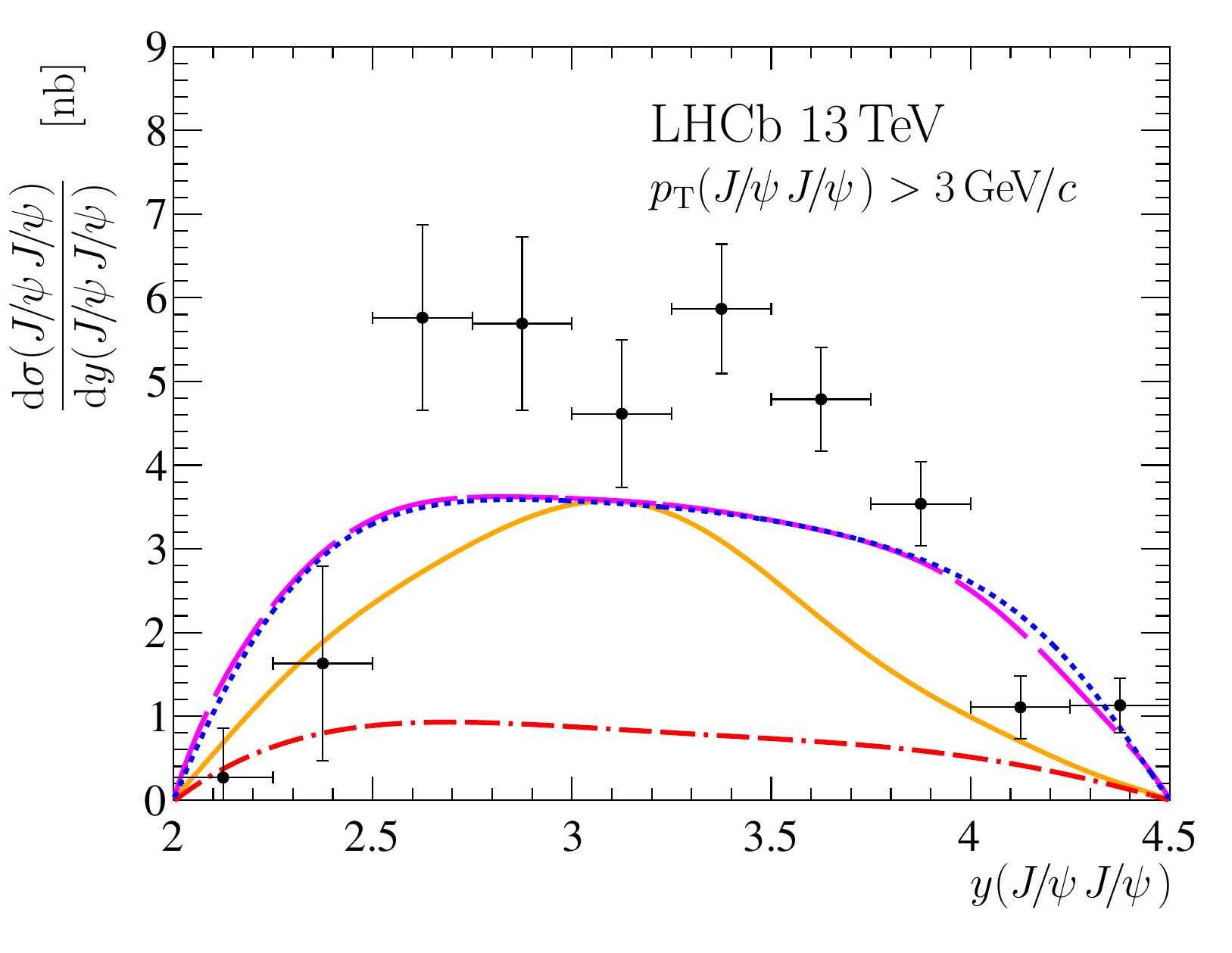}
\includegraphics[width=0.495\linewidth]{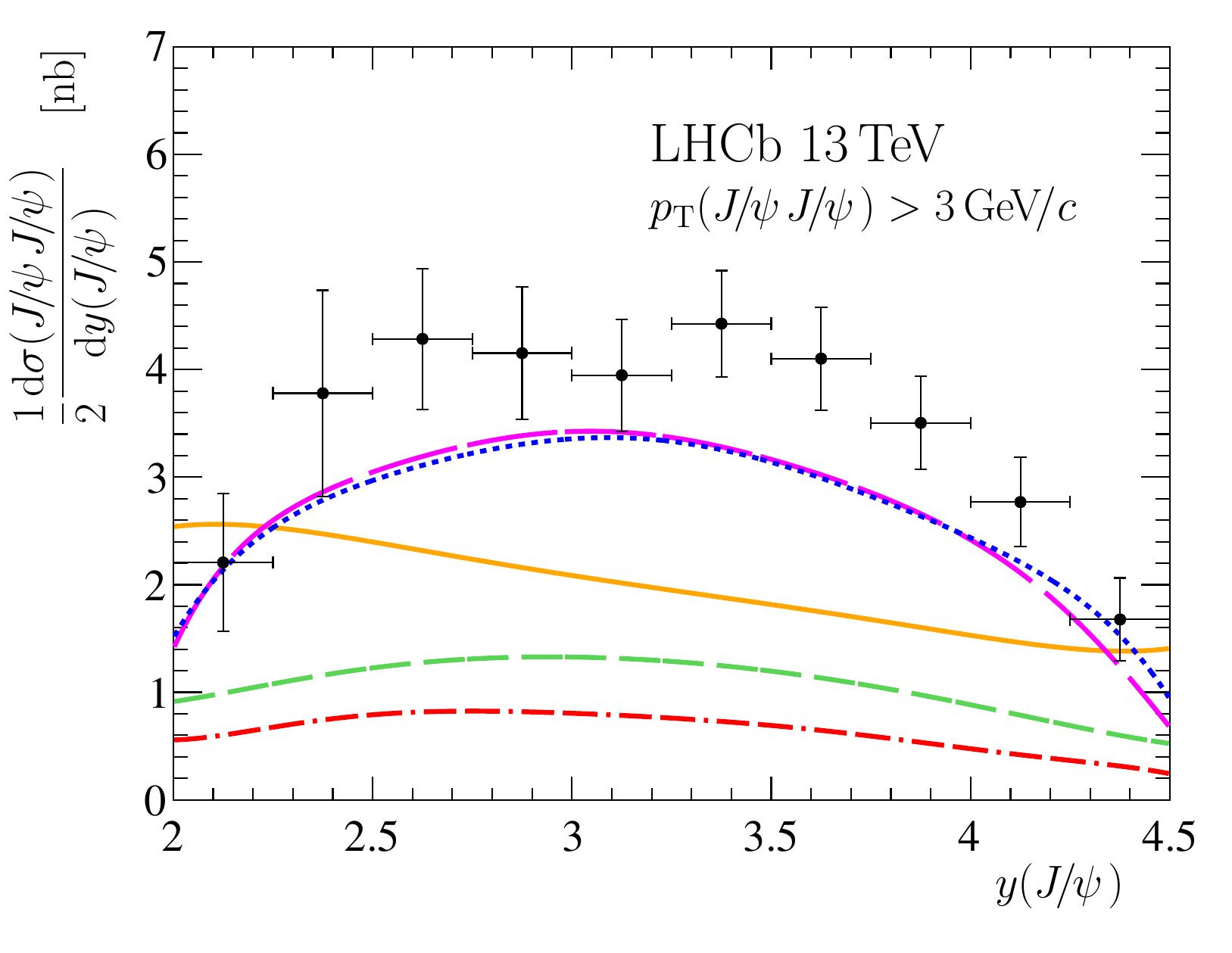}
\end{center}
  \caption { \small
     Comparisons between measurements and theoretical predictions with \mbox{$p_{\mathrm{T}}(\jpsi\jpsi)>3\gevc$} for the differential cross-sections as functions of (top left)~$p_{\mathrm{T}}(\jpsi)$, (top right)~$y(\jpsi\jpsi)$ and (bottom)~$y(\jpsi)$.
    The (black) points with error bars represent the measurements.
  }
  \label{fig:cmp3GeV1}
\end{figure}

\begin{figure}[tb]
\begin{center}
\includegraphics[width=0.495\linewidth]{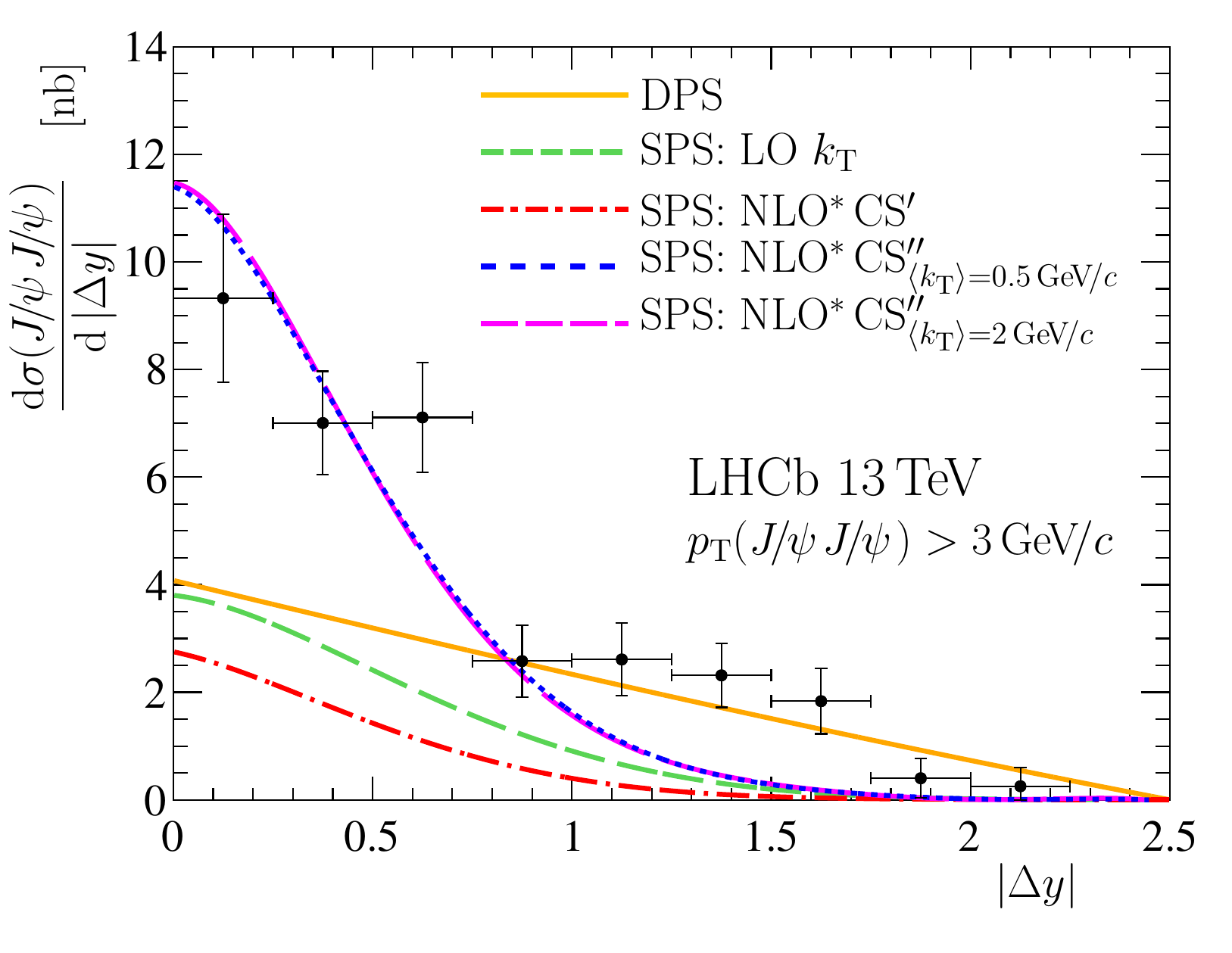}
\includegraphics[width=0.495\linewidth]{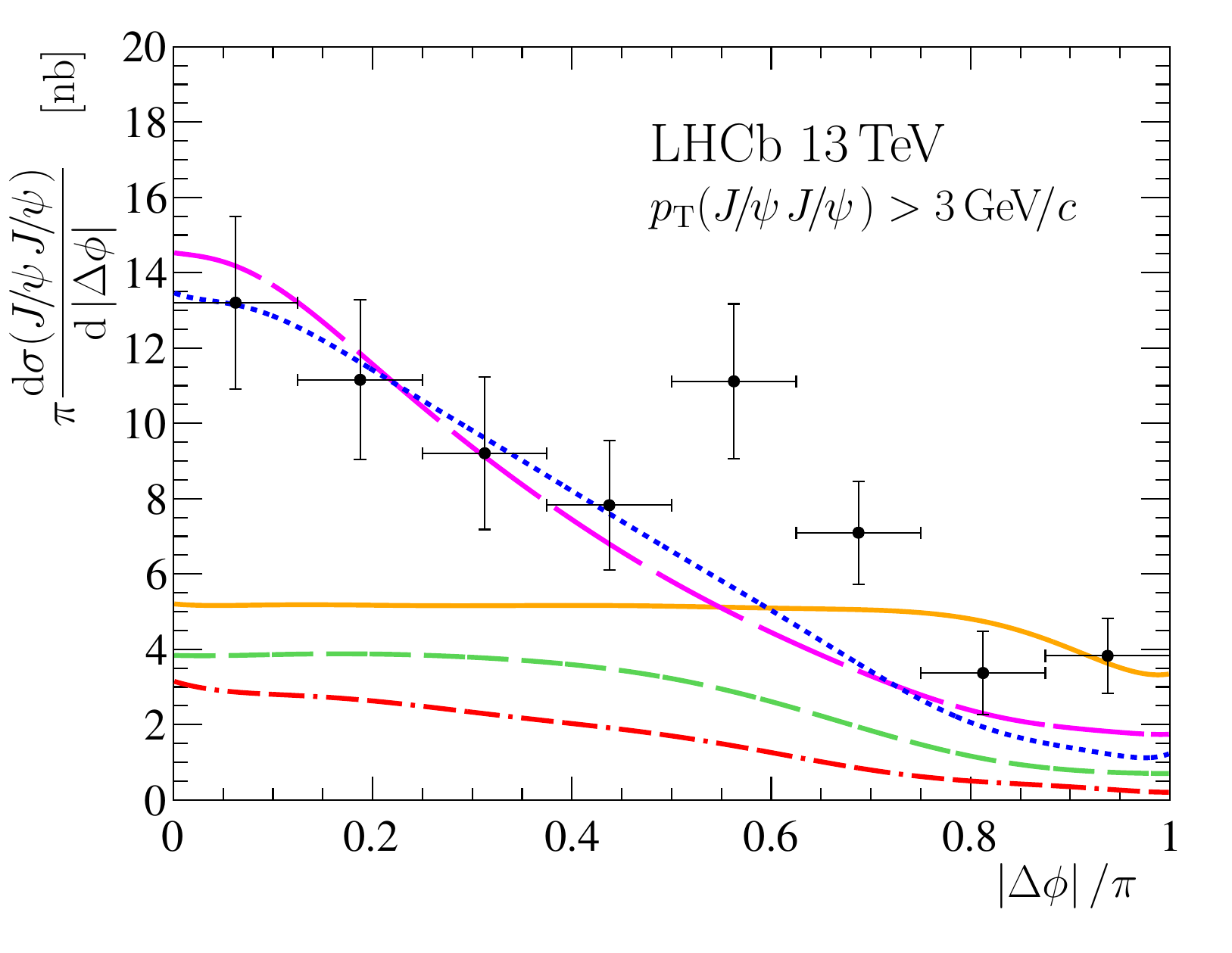}
\includegraphics[width=0.495\linewidth]{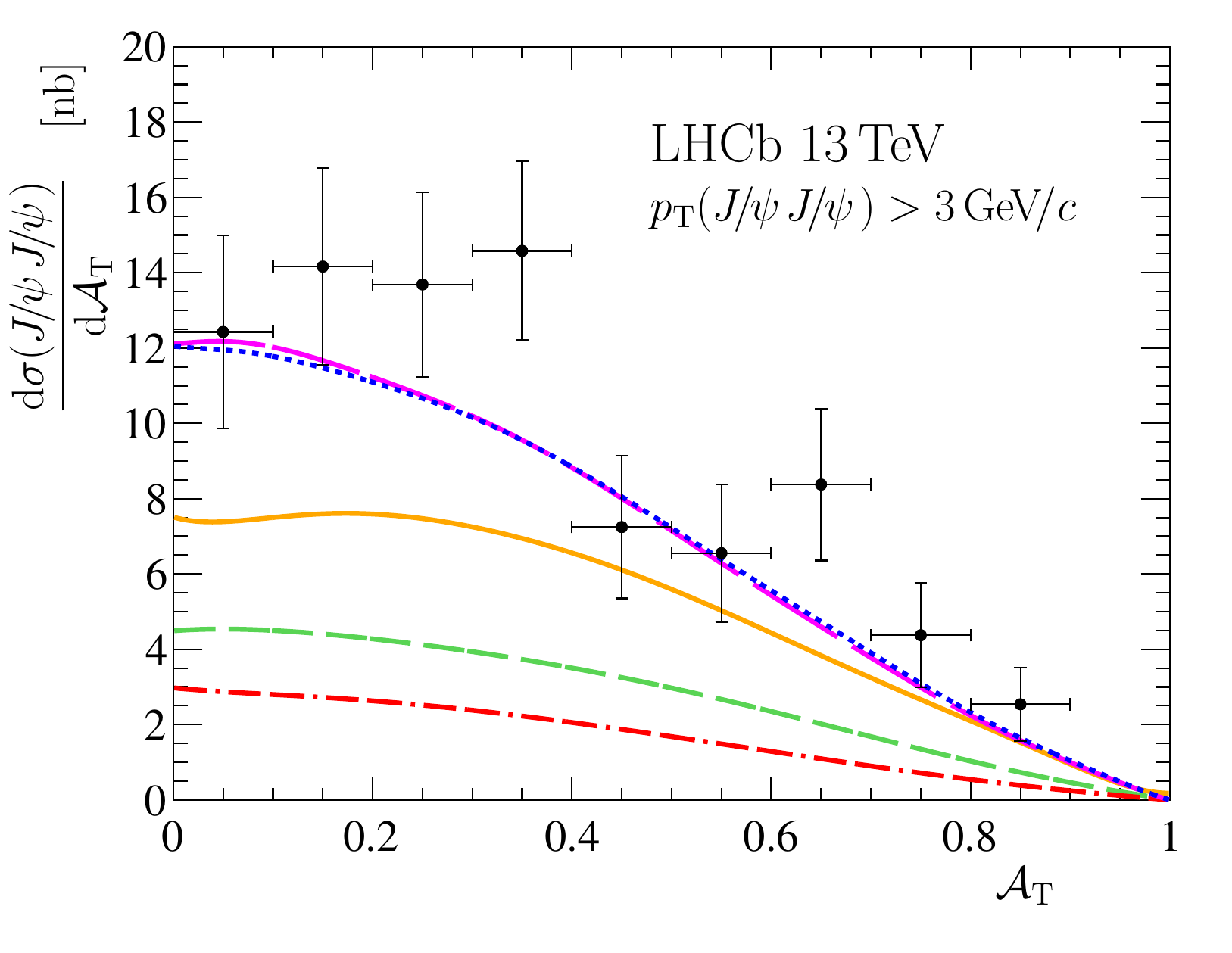}
\includegraphics[width=0.495\linewidth]{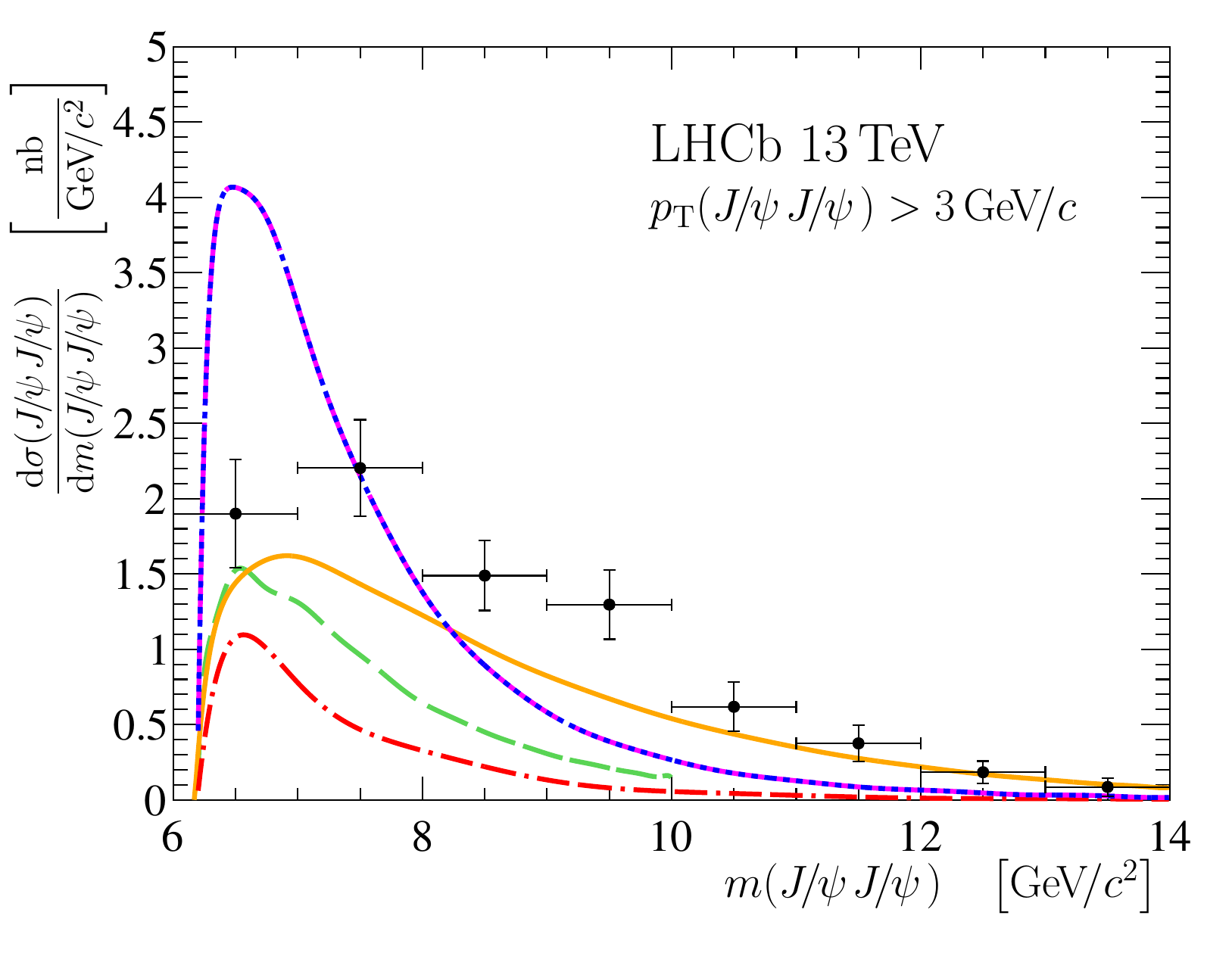}
\end{center}
  \caption { \small
     Comparisons between measurements and theoretical predictions with \mbox{$p_{\mathrm{T}}(\jpsi\jpsi)>3\gevc$} for the differential cross-sections as functions of (top left)~$\left| \Delta y \right|$, (top right)~$\left| \Delta \phi \right|$, (bottom left)~$\mathcal{A}_{\mathrm{T}}$ and (bottom right)~$m(\jpsi\jpsi)$.
    The (black) points with error bars represent the measurements.
  }
  \label{fig:cmp3GeV2}
\end{figure}

The~measured differential production cross-sections of $\jpsi$ pairs 
as a function of several kinematic variables are compared to 
the~theoretical predictions.
For each variable $v$, the differential production cross-section of $\jpsi$ pairs is calculated as
\begin{equation*}
  \dfrac{{\rm d}\sigma(\jpsi\jpsi)}{{\rm d}v} = 
  \dfrac{1}{{\cal{L}} \times \BF(\jpsi \to \mup \mun)^{2}} \times \dfrac{\Delta N^{{\rm cor}}_i}{\Delta v_i},
\end{equation*}
where $\Delta N^{{\rm cor}}_i$ is the number of efficiency-corrected signal candidates 
in bin $i$, and $\Delta v_i$ is the corresponding bin width. 
The luminosity uncertainty and the uncertainty introduced by $\BF(\jpsi \to \mup \mun)$ are common to all bins and are fully correlated.
The~tracking efficiency and muon PID efficiency uncertainties 
are strongly correlated.
In Figs.~\ref{fig:cmppt}$-$\ref{fig:cmp3GeV2} of the differential cross-sections,
only the statistical uncertainties are shown 
as the systematic ones are negligibly small and almost 100\% correlated.

The comparison between measurements and theoretical predictions 
is performed for the following kinematical variables: 
transverse momentum and rapidity of the \jpsi~pair, 
transverse momentum and rapidity of each \jpsi~meson,
differences in the~azimuthal angle and rapidity between 
the two \jpsi~mesons\,($\left|\Delta\phi\right|$ and $\left|\Delta y\right|$), 
the~mass of the \jpsi~pair   
and the~transverse momentum asymmetry, defined as 
\begin{equation*}
\mathcal{A}_{\mathrm{T}}\equiv\left| \dfrac{\pt(\jpsione) - \pt(\jpsitwo)}{\pt(\jpsione) + \pt(\jpsitwo)}\right|.
\end{equation*}
The~distributions for the~whole $p_{\mathrm{T}}(\jpsi\jpsi)$ range 
are presented in  Figs.~\ref{fig:cmppt}, \ref{fig:cmp1} and~\ref{fig:cmp2},
for $p_{\mathrm{T}}(\jpsi\jpsi)>1\gevc$ in~Figs.~\ref{fig:cmp1GeV1} and~\ref{fig:cmp1GeV2},
and for $p_{\mathrm{T}}(\jpsi\jpsi)>3\gevc$ in~Figs.~\ref{fig:cmp3GeV1} and~\ref{fig:cmp3GeV2}.

The~DPS predictions are obtained using a~large number of pseudoexperiments, 
where two uncorrelated $\jpsi$ mesons are produced according to 
the~measured differential distributions 
\mbox{${\rm d^2}\sigma\left(\jpsi\right)/{\mathrm{d}p_{\mathrm{T}}\mathrm{d}y}$}~\cite{LHCb-PAPER-2015-037}
for single prompt $\jpsi$ production, uniformly distributed over the~azimuthal angle~$\phi$.
For~LO\,CO and NLO$^{\ast}$\,CS$^{\prime\prime}$~models two values 
of Gaussian smearing of the~initial transverse momentum of gluon
$k_{\mathrm{T}}$ are used, 
namely $\left\langle k_{\mathrm{T}}\right\rangle=0.5$ and $2\gevc$.
The~$\pt(\jpsi\jpsi)$~distribution, shown in Fig.~\ref{fig:cmppt}, 
demonstrates the~large dependence of the~shape on 
the~choice of the $\left\langle k_{\mathrm{T}}\right\rangle$~parameter.
For the NLO$^{\ast}$\,CS$^{\prime\prime}$ approach~\cite{Lansberg:2013qka,Lansberg:2014swa,Lansberg:2015lva,Shao:2012iz,Shao:2015vga},
relatively large smearing of the initial gluon transverse momenta $\left\langle k_{\mathrm{T}}\right\rangle=2\gevc$
is required to eliminate peaking structures in the distribution.
The distributions of the variables $\pt(\jpsi\jpsi)$, $\left|\Delta \phi\right|$ and $\mathcal{A}_{\mathrm{T}}$,
predicted by the LO\,CS model, are trivial,
$\pt(\jpsi\jpsi)\sim0$, 
$\left|\Delta \phi\right|\sim\pi$ and  
$\mathcal{A}_{\mathrm{T}}\sim0$, and omitted from the~plots.
A similar trivial pattern is expected for the LO\,CO~model, 
but due to the $k_{\mathrm{T}}$\nobreakdash-smearing, 
the~actual shape of the distributions strongly depends on 
the~choice of the $\left\langle k_{\mathrm{T}}\right\rangle$~parameter.
The~NLO$^{\ast}$\,CS$^{\prime\prime}$~model also demonstrates 
a~large dependence on the $\left\langle k_{\mathrm{T}}\right\rangle$~parameter
for $\left| \Delta \phi \right|/\pi$~distribution.

Neither the DPS~model with the~given value of the $\sigma_{\mathrm{eff}}$ parameter, 
nor any of the~SPS~models can describe simultaneously 
the~measured cross\nobreakdash-section and the~differential shapes.
However, the~sum of the DPS and SPS contributions 
can adequately describe both the~measured production cross-sections and 
the~differential distributions.
To~discriminate between the SPS and DPS contributions,
the~differential distribution for each variable $v$
is fitted with the~simple two\nobreakdash-component model
\begin{equation}
  \dfrac{\mathrm{d}\sigma}{\mathrm{d}v} =  
  \sigma_{\mathrm{DPS}} F_{\mathrm{DPS}}(v) + 
  \sigma_{\mathrm{SPS}} F_{\mathrm{SPS}}(v), 
  \label{eq:dps_fit}
\end{equation}
where  $F_{\mathrm{DPS}}$ and $F_{\mathrm{SPS}}$
are templates for the DPS and SPS~models and $\sigma_{\mathrm{DPS}}$ 
and $\sigma_{\mathrm{SPS}}$ are floating 
fit parameters representing the~DPS and SPS contributions.
The theory normalisation is not used in the fits.
The~DPS~fraction $f_{\mathrm{DPS}}$ is defined as 
\begin{equation}
  f_{\mathrm{DPS}} \equiv \dfrac{ \sigma_{\mathrm{DPS}} }
  { \sigma_{\mathrm{SPS}} + \sigma_{\mathrm{DPS}}}.
\end{equation}
Some~distributions give little discrimination between SPS and DPS. 
The percentages of the DPS~component obtained from 
the~fits for the~most discriminating variables 
are presented in Table~\ref{tab:dps_fit1}. 
The~fit results are presented in the Appendix.
All the fits indicate a large DPS~contribution 
to the~\jpsi~pair production process. 
The~inclusion of the CO~component in the~fit 
does not have a large effect on the~determination 
of the DPS~fraction $f_{\mathrm{DPS}}$, and 
the~fraction of the CO~component determined 
in such a fit procedure is significantly 
smaller than the CS~contribution.
The value of $\sigma_{\mathrm{SPS}}$, calculated as $(1-f_{\mathrm{DPS}}) \times \sigma(\jpsi\jpsi)$,
is smaller than expectations from the  
NLO$^{\ast}$\,CS$^{\prime\prime}$~\cite{Lansberg:2013qka,Lansberg:2014swa,Lansberg:2015lva,Shao:2012iz,Shao:2015vga} and NLO\,CS~\cite{Sun:2014gca} approaches 
and roughly agrees with the NLO$^{\ast}$\,CS$^{\prime}$~\cite{Likhoded:2016zmk} 
and LO\,$k_{\mathrm{T}}$~\cite{Baranov:2011zz} predictions.

\begin{table}[t]
  \centering
  \caption{ \small
    Percentages of the DPS~component, $f_{\mathrm{DPS}}$,
    determined with the simple two\nobreakdash-component fit to different distributions 
    for different SPS models. 
  } \label{tab:dps_fit1}
  \vspace*{3mm}
  \begin{tabular*}{1.00\textwidth}{@{\hspace{0.1mm}}l@{\extracolsep{\fill}}cccccc@{\hspace{0.1mm}}}
    \multirow{2}{*}{Variable} 
    & \multirow{2}{*}{LO\,CS}
    & \multirow{2}{*}{LO\,$k_{\mathrm{T}}$} 
    & \multirow{2}{*}{NLO$^{\ast}$\,CS$^{\prime}$}
    & \multicolumn{2}{c}{NLO$^{\ast}$\,CS$^{\prime\prime}$}
    & \multirow{2}{*}{NLO\,CS}
    \\ 
    & 
    & 
    & 
    & $\left\langle k_{\mathrm{T}}\right\rangle=2\gevc$ 
    & $\left\langle k_{\mathrm{T}}\right\rangle=0.5\gevc$ 
    \\
    \hline 
    \\[-1em]
    \multicolumn{7}{c}{ no $\pt(\jpsi\jpsi)$ cut} 
    \\
    \hline 
    \\[-1em]
$\pt(\jpsi\jpsi)$
    &  ---
    &  $78\pm2\phantom{0}$ 
    &  --- 
    &  $86\pm55$
    &  $81\pm7\phantom{0}$
    &  --- 
    \\
$y(\jpsi\jpsi)$ 
    &  $83\pm39$
    &  ---
    &  ---  
    &  $75\pm37$
    &  $68\pm34$
    &  ---
    \\
$m(\jpsi\jpsi)$ 
    &  $76\pm 7\phantom{0}$
    &  $74\pm 7\phantom{0}$ 
    &  --- 
    & \multicolumn{2}{c}{$78\pm 7\phantom{0}$}
    &  $77\pm 7\phantom{0}$
    \\
$\left| \Delta y \right|$ 
    &  $59\pm21$
    &  $61\pm18$ 
    &  ---
    &  $63\pm18$
    &  $61\pm18$
    &  $69\pm16$ 
    \\
    \hline 
    \\[-1em]
    \multicolumn{7}{c}{ $\pt(\jpsi\jpsi)>1\gevc$} 
    \\
    \hline 
    \\[-1em]
$y(\jpsi\jpsi)$ 
    &  ---
    &  --- 
    &  $75\pm24$
    &  $71\pm38$
    &  $68\pm34$
    &  --- 
    \\
$m(\jpsi\jpsi)$ 
    &  --- 
    &  $73\pm8\phantom{0}$
    &  $76\pm7\phantom{0}$
    & \multicolumn{2}{c}{$88\pm1$}
    &  --- 
    \\
$\left| \Delta y \right|$ 
    &  ---
    &  $57\pm20$ 
    &  $59\pm19$ 
    &  $60\pm18$
    &  $60\pm19$
    &  ---
    \\
    \hline 
    \\[-1em]
    \multicolumn{7}{c}{ $\pt(\jpsi\jpsi)>3\gevc$} 
    \\
    \hline 
    \\[-1em]
$y(\jpsi\jpsi)$ 
    &  ---
    &  --- 
    &  $77\pm18$
    &  $64\pm38$
    &  $64\pm35$
    &  ---
    \\   
$m(\jpsi\jpsi)$ 
    &  --- 
    &  $76\pm10$
    &  $84\pm7\phantom{0}$
    & \multicolumn{2}{c}{$87\pm2\phantom{0}$}
    &  ---
    \\ 
$\left| \Delta y \right|$ 
    &  ---
    & $42\pm25$
    & $53\pm21$
    & $53\pm21$
    & $53\pm21$
    &  ---
  \end{tabular*}
\end{table}

The~value $\sigma_{\mathrm{DPS}}$ determined with Eq.~\eqref{eq:dps_fit}
is converted to $\sigma_{\mathrm{eff}}$,
\begin{equation}
  \sigma_{\mathrm{eff}} = 
  \dfrac{1}{2}\dfrac{ \sigma\left(\jpsi\right)^2}{\sigma_{\mathrm{DPS}}},
  \label{eq:sigmadps}
\end{equation}
where $\sigma(\jpsi)$ is the production cross-section of prompt \jpsi~mesons
from Ref.~\cite{LHCb-PAPER-2015-037}.
The values obtained for $\sigma_{\mathrm{eff}}$ are summarized in Table~\ref{tab:dps_fit2}.
Values between $8.8$ and $12.5\mbarn$ are found for the models considered in this analysis.
These~values are slightly larger than those measured from central \jpsi~pair production at LHC, 
$\sigma_{\mathrm{eff}}=8.2\pm2.2\mbarn$~\cite{Lansberg:2014swa} and 
\mbox{$\sigma_{\mathrm{eff}}=6.3\pm1.9\mbarn$}~\cite{ATLASDJ},  
and significantly exceed the~values obtained by the \dzero~collaboration from analysis of 
\jpsi~pair production, \mbox{$\sigma_{\mathrm{eff}}=4.8\pm2.5\mbarn$}~\cite{D0DJ}, 
and $\PUpsilon\jpsi$~production, 
\mbox{$\sigma_{\mathrm{eff}}=2.2\pm1.1\mbarn$}~\cite{Abazov:2015fbl}.
On the other hand, they are smaller than 
the values of $\sigma_{\mathrm{eff}}$ measured by 
the~LHCb collaboration in the processes of multiple associated 
heavy quark production~\cite{LHCb-PAPER-2012-003,LHCb-PAPER-2015-046},
in particular $\sigma_{\mathrm{eff}} \sim 15\mbarn$ measured for various $\jpsi + \cquark\cquarkbar$ production processes~\cite{LHCb-PAPER-2012-003}
and $\sigma_{\mathrm{eff}} = 18.0 \pm 1.8 \mbarn$ measured for the $\PUpsilon(1S) + D^{0,+}$ production processes~\cite{LHCb-PAPER-2015-046}.   

\begin{table}[t]
  \centering
  \caption{ \small
    Summary of the $\sigma_{\mathrm{eff}}$ values (in ${\mathrm{mb}}\xspace$) from DPS~fits for different SPS models. 
    The~uncertainty is statistical only, originating from 
    the~statistical uncertainty in 
    $\sigma_{\mathrm{DPS}}$\,(and \mbox{${\mathrm{d}\sigma\left(\jpsi\jpsi\right)}/{\mathrm{d}v}$}).
    The~common systematic uncertainty of 12\%, accounting 
    for the~systematic uncertainty of $\sigma\left(\jpsi\jpsi\right)$ and 
    the~total uncertainty for $\sigma(\jpsi)$, is not shown.
  } \label{tab:dps_fit2}
  \vspace*{3mm}
  \begin{tabular*}{0.90\textwidth}{@{\hspace{1mm}}l@{\extracolsep{\fill}}cccc@{\hspace{1mm}}}
    \multirow{2}{*}{Variable} 
    & \multirow{2}{*}{LO\,$k_{\mathrm{T}}$}
    & \multicolumn{2}{c}{NLO$^{\ast}$\,CS$^{\prime\prime}$} 
    & \multirow{2}{*}{NLO\,CS}
    \\
    & 
    & ${\left\langle k_{\mathrm{T}}\right\rangle=2\gevc}$    
    & ${\left\langle k_{\mathrm{T}}\right\rangle=0.5\gevc}$    
    \\
    \hline 
    \\[-1em]
$\pt(\jpsi\jpsi)$
    & $9.7 \pm 0.5$ 
    & $8.8 \pm 5.6$ 
    & $9.3 \pm 1.0$ 
    & --- 
    \\
$y(\jpsi\jpsi)$
    & --- 
    & $11.9 \pm 7.5$ 
    & $10.0 \pm 5.0$ 
    & --- 
    \\
$m(\jpsi\jpsi)$
    & $10.6 \pm 1.1$
    & \multicolumn{2}{c}{$10.2 \pm 1.0$}
    & $10.4 \pm 1.0$ 
    \\
$\left| \Delta y \right|$ 
    & $12.5 \pm 4.1$
    & $12.2 \pm 3.7$
    & $12.4 \pm 3.9$    
    & $11.2 \pm 2.9$
  \end{tabular*}
\end{table}

\clearpage

\section{Summary}
\label{sec:result}
The $\jpsi$ pair production cross-section with both $\jpsi$ mesons
in the region $2.0<y<4.5$ and $\pt < 10 \gevc$ is measured
to be $15.2 \pm 1.0\stat \pm 0.9\syst \nb$,
using $pp$ collision data collected by $\lhcb$ at $\sqs = 13 \tev$,
corresponding to an integrated luminosity of $279 \invpb$.
The differential production cross-sections as functions of
$\pt(\jpsi\jpsi)$, $\pt(\jpsi)$, $m(\jpsi\jpsi)$, $y(\jpsi\jpsi)$,
$y(\jpsi)$, $|\Delta \phi|$, $|\Delta y|$ and ${\cal A}_{\rm T}$ 
are compared to theoretical predictions.
A fit to the differential cross-sections using simple DPS plus SPS models indicates a significant DPS contribution.
The data can be reasonably well described with a sum of DPS and SPS colour-singlet contributions,
with no evidence for a large SPS colour-octet contribution. 
The obtained SPS contribution is overestimated in the 
NLO$^{\ast}$\,CS$^{\prime\prime}$~\cite{Lansberg:2013qka,Lansberg:2014swa,Lansberg:2015lva,Shao:2012iz,Shao:2015vga} and NLO\,CS~\cite{Sun:2014gca} approaches 
and roughly agrees with the NLO$^{\ast}$\,CS$^{\prime}$~\cite{Likhoded:2016zmk} 
and LO\,$k_{\mathrm{T}}$~\cite{Baranov:2011zz} predictions. 
Good agreement with the data for the differential cross-sections calculated within the LO\,$k_{\mathrm{T}}$~\cite{Baranov:2011zz} 
and NLO$^{\ast}$\,CS$^{\prime}$~\cite{Likhoded:2016zmk}~approaches 
indicates that a significant part of high-order contributions 
can be properly accounted via the evolution of parton densities~\cite{Baranov:2011zz}. 
Relatively large smearing of initial gluon transverse momenta 
$\langle k_{\mathrm{T}} \rangle=2\gevc$ is preferred over $\langle k_{\mathrm{T}} \rangle=0.5\gevc$ 
for the NLO$^{\ast}$\,CS$^{\prime\prime}$~approach~\cite{Lansberg:2013qka,Lansberg:2014swa,Lansberg:2015lva,Shao:2012iz,Shao:2015vga}. 
An improvement in the precision for SPS~predictions is needed 
for a better discrimination between the different theory approaches. 
A large DPS contribution results in values of $\sigma_{\mathrm{eff}}$ that are 
smaller than the values of $\sigma_{\mathrm{eff}}$ measured previously by the~LHCb collaboration in the processes of multiple associated heavy quark production~\cite{LHCb-PAPER-2012-003,LHCb-PAPER-2015-046},
and slightly larger than those measured from central \jpsi~pair production at the CMS~\cite{CMSDJ} and ATLAS~\cite{ATLASDJ} experiments.

\section*{Acknowledgements}

\noindent 
We would like to thank K.-T.~Chao, J.-P.~Lansberg, A.K.~Likhoded and A.V.~Luchinsky for interesting discussions
on quarkonia and quarkonium-pair production,
and S.P.~Baranov,\linebreak
S.V.~Poslavsky, H.\nobreakdash-S.~Shao and L.\nobreakdash-P.~Sun for providing the SPS calculations.
We express our gratitude to our colleagues in the CERN
accelerator departments for the excellent performance of the LHC. We
thank the technical and administrative staff at the LHCb
institutes. We acknowledge support from CERN and from the national
agencies: CAPES, CNPq, FAPERJ and FINEP (Brazil); NSFC (China);
CNRS/IN2P3 (France); BMBF, DFG and MPG (Germany); INFN (Italy); 
FOM and NWO (The Netherlands); MNiSW and NCN (Poland); MEN/IFA (Romania); 
MinES and FASO (Russia); MinECo (Spain); SNSF and SER (Switzerland); 
NASU (Ukraine); STFC (United Kingdom); NSF (USA).
We acknowledge the computing resources that are provided by CERN, IN2P3 (France), KIT and DESY (Germany), INFN (Italy), SURF (The Netherlands), PIC (Spain), GridPP (United Kingdom), RRCKI and Yandex LLC (Russia), CSCS (Switzerland), IFIN-HH (Romania), CBPF (Brazil), PL-GRID (Poland) and OSC (USA). We are indebted to the communities behind the multiple open 
source software packages on which we depend.
Individual groups or members have received support from AvH Foundation (Germany),
EPLANET, Marie Sk\l{}odowska-Curie Actions and ERC (European Union), 
Conseil G\'{e}n\'{e}ral de Haute-Savoie, Labex ENIGMASS and OCEVU, 
R\'{e}gion Auvergne (France), RFBR and Yandex LLC (Russia), GVA, XuntaGal and GENCAT (Spain), Herchel Smith Fund, The Royal Society, Royal Commission for the Exhibition of 1851 and the Leverhulme Trust (United Kingdom).

% $Id: appendix.tex 90313 2016-04-06 08:01:40Z lafferty $
% ===============================================================================
% Purpose: appendix to the standard template: standard symbol alises from Ulrik
% Author: Tomasz Skwarnicki
% Created on: 2009-09-24
% ===============================================================================

\clearpage

{\noindent\normalfont\bfseries\Large Appendix}

\appendix
\section*{Fits to the differential cross-sections with SPS and DPS components}
\label{app:dpsfits}
\begin{figure}[tb]
\begin{center}
\includegraphics[width=0.495\linewidth]{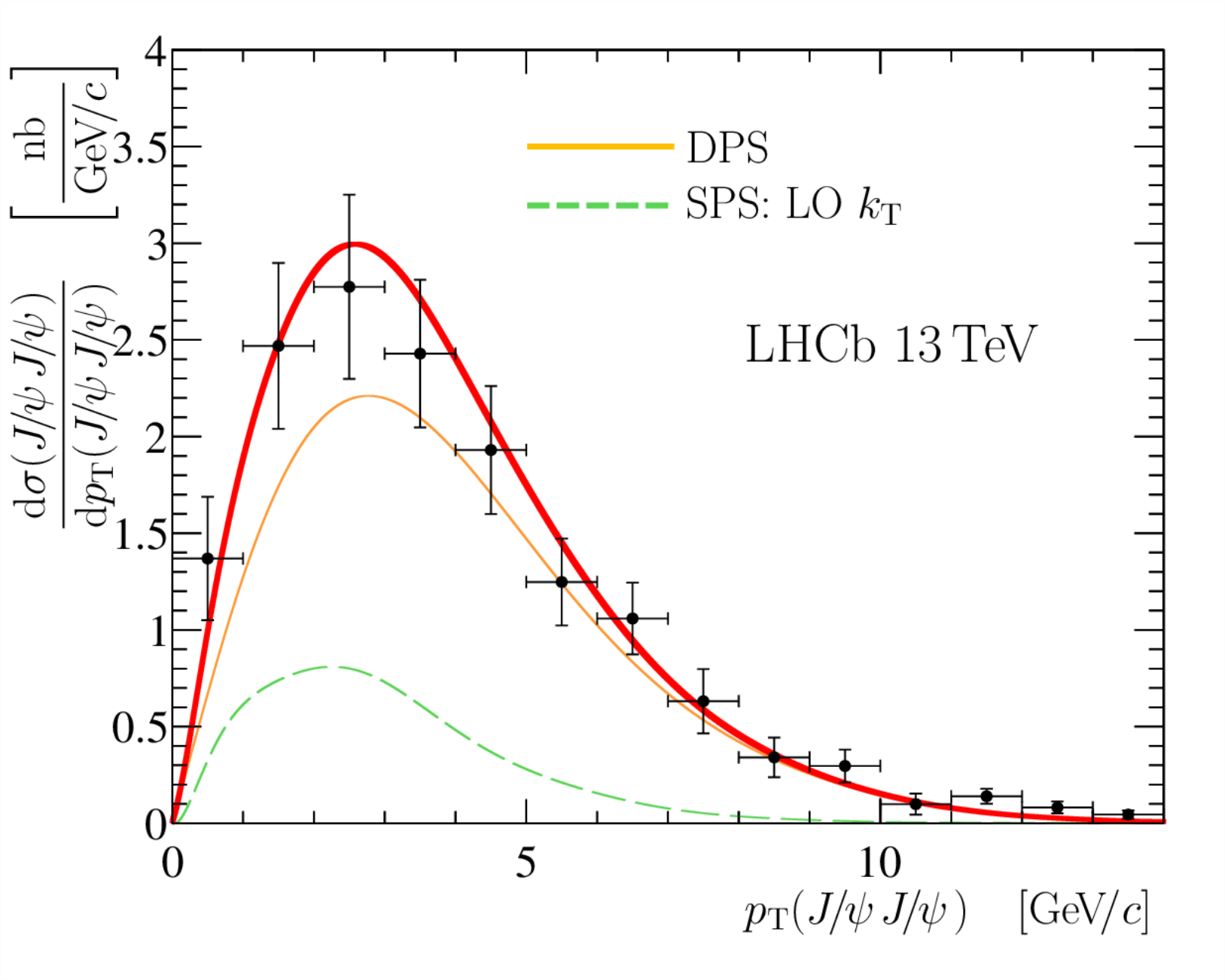}
\includegraphics[width=0.495\linewidth]{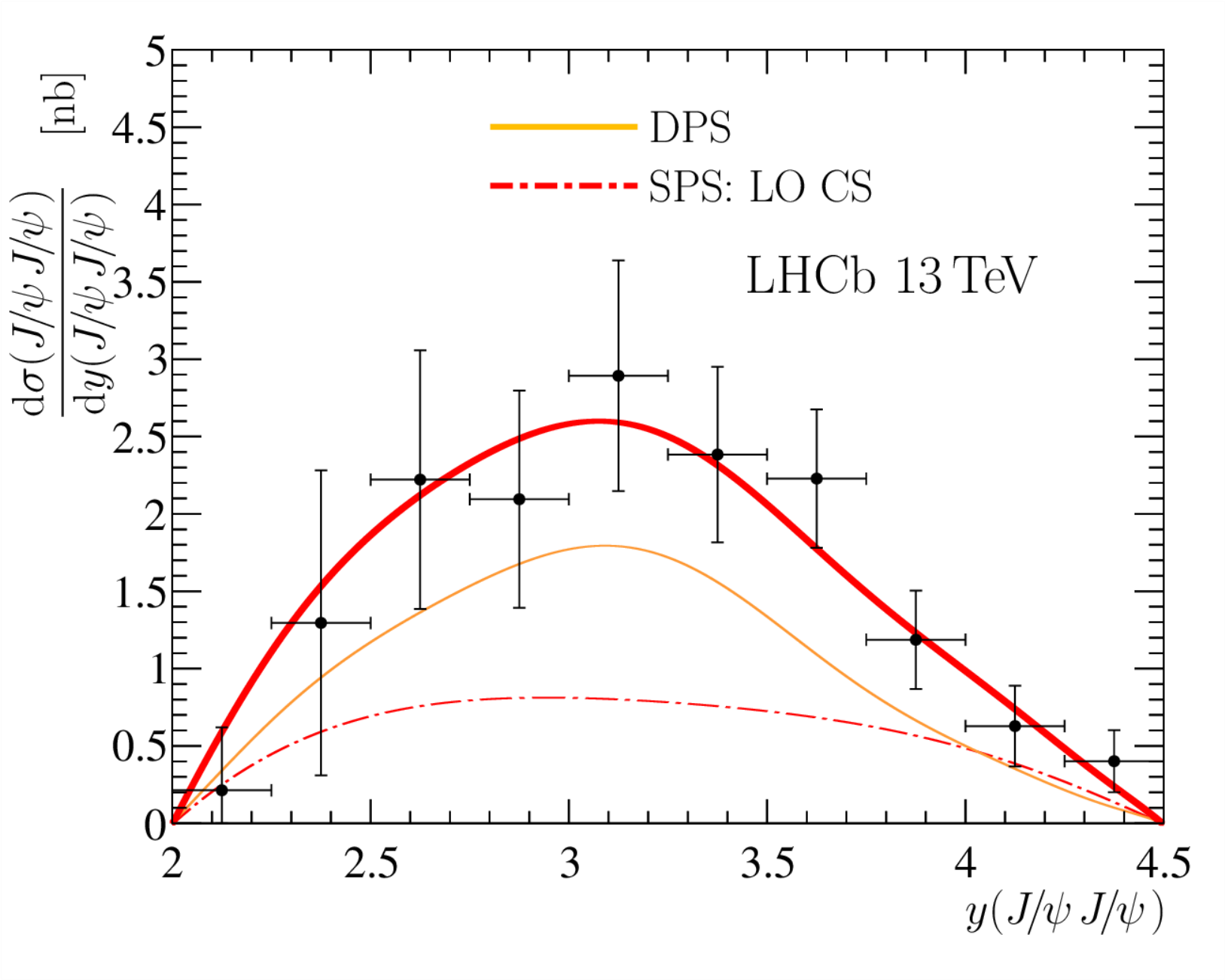}
\includegraphics[width=0.495\linewidth]{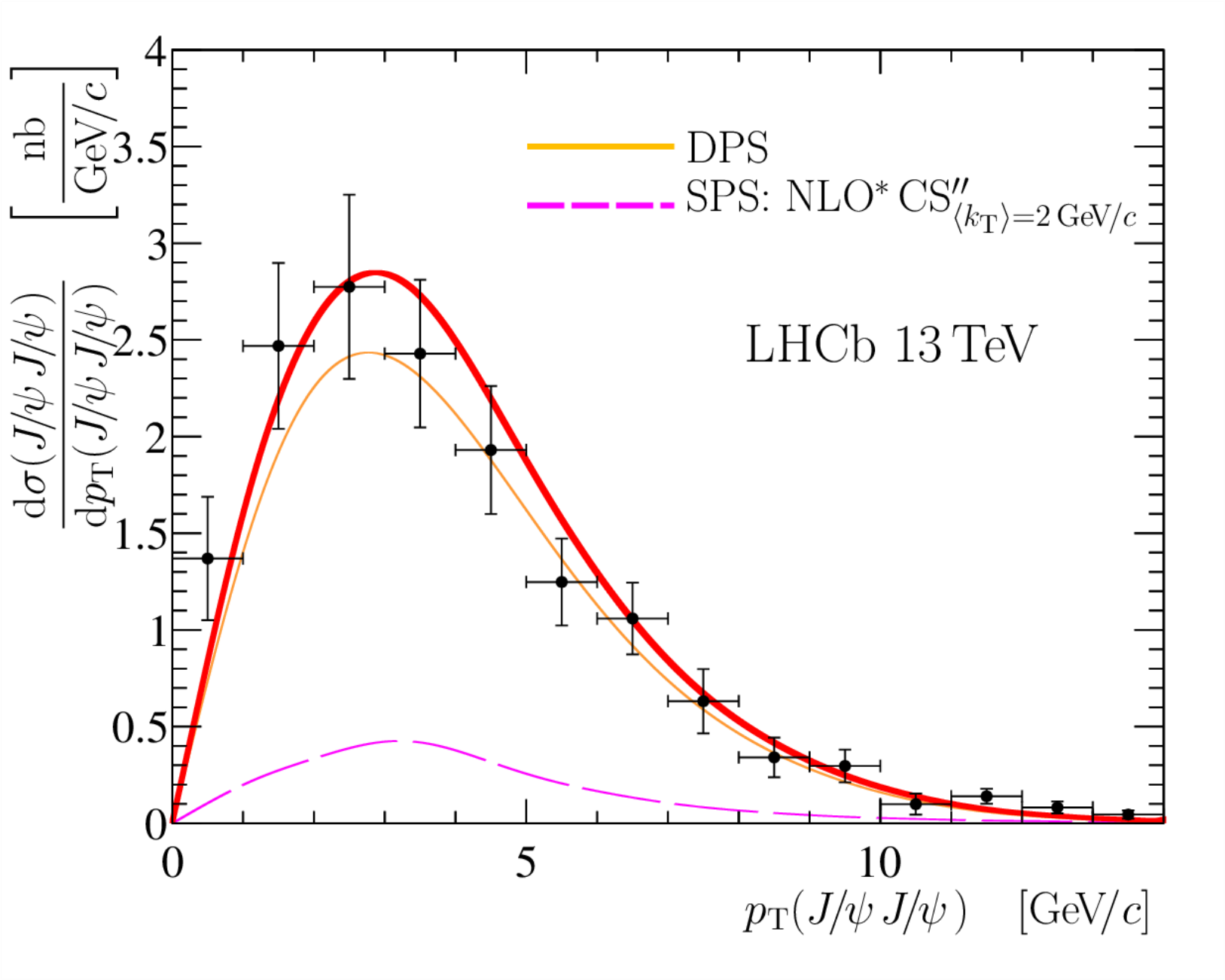}
\includegraphics[width=0.495\linewidth]{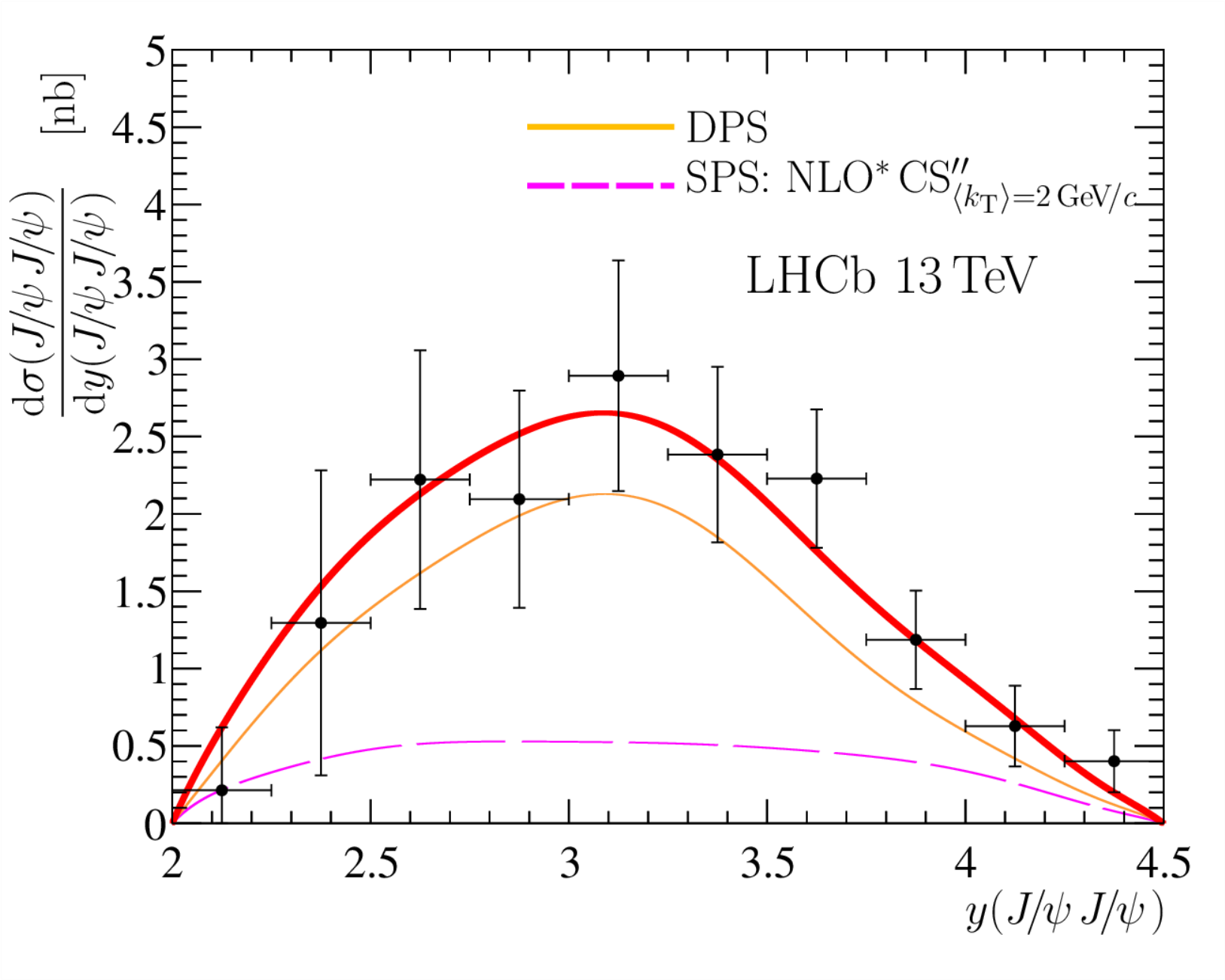}
\includegraphics[width=0.495\linewidth]{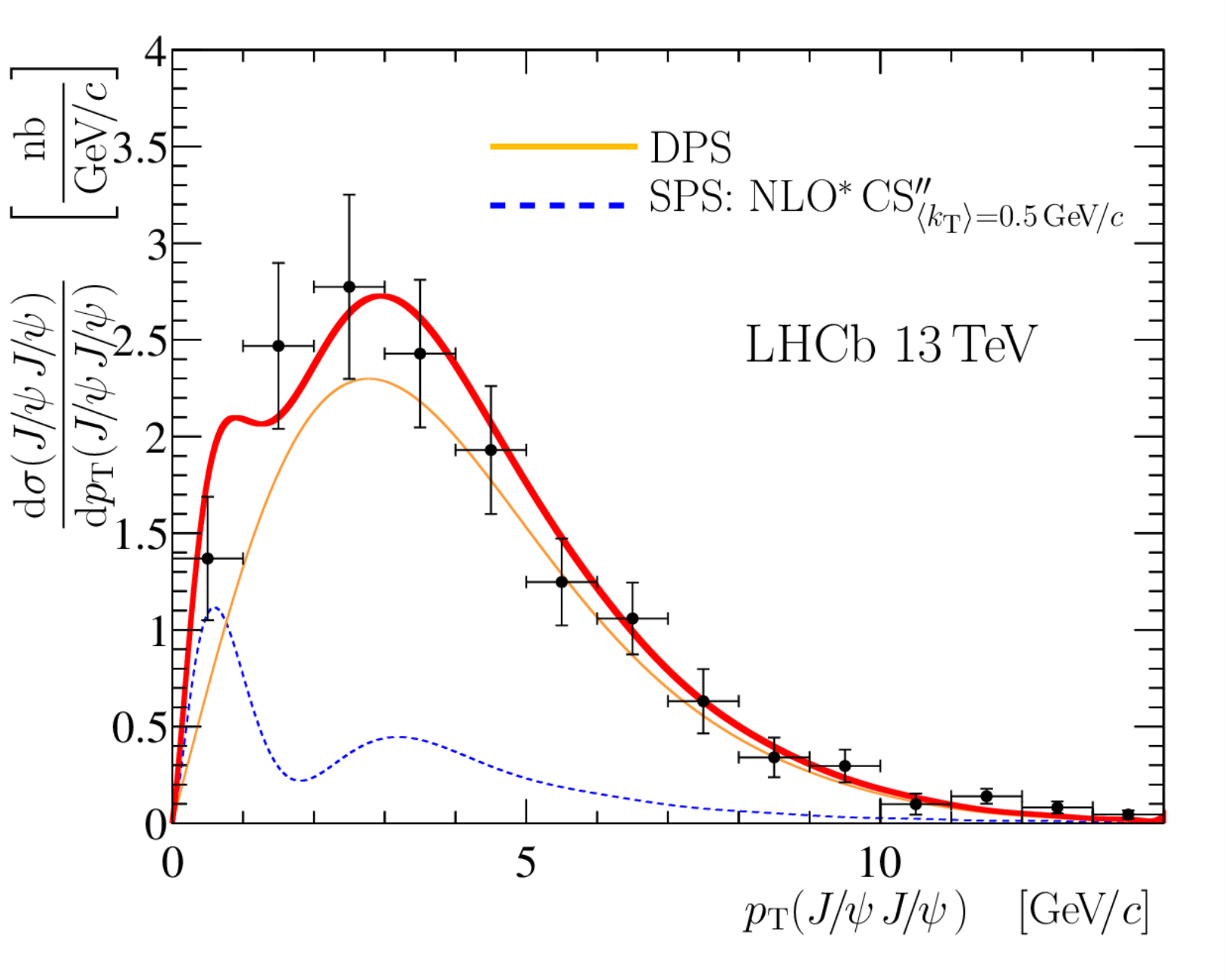}
\includegraphics[width=0.495\linewidth]{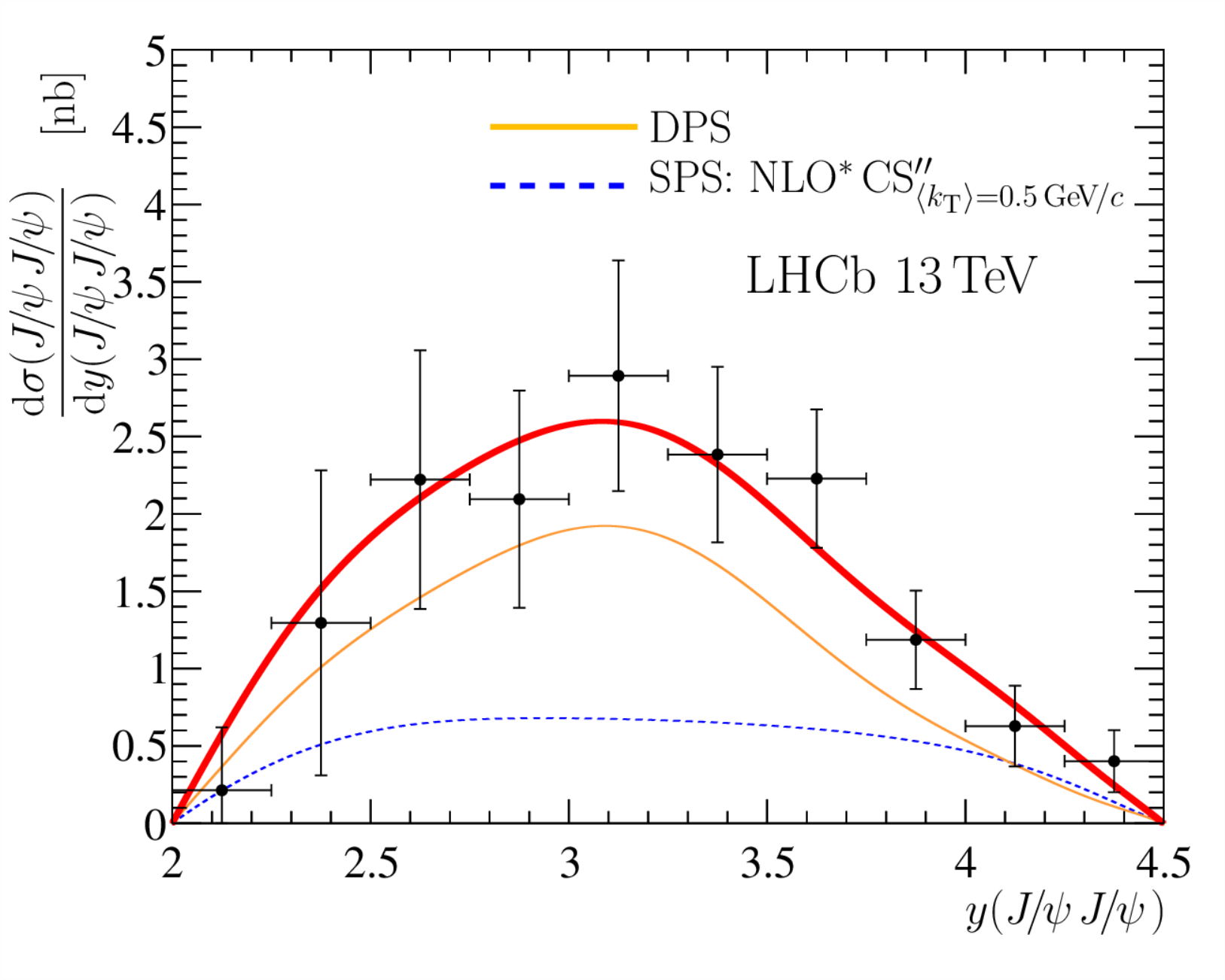}
\end{center}
  \caption { \small
    Result of templated DPS fit for 
    $\frac{ \mathrm{d}\sigma(\jpsi\jpsi)}{\mathrm{d} \pt(\jpsi\jpsi)}$ and 
    $\frac{ \mathrm{d}\sigma(\jpsi\jpsi)}{\mathrm{d} y(\jpsi\jpsi)}$. 
    The (black) points with error bars represent the data. 
    The total fit result is shown with the thick (red) solid line
    and the DPS component is shown with the thin (orange) solid line.
  }
  \label{fig:cmp:fits_pty_psipsi}
\end{figure}

\begin{figure}[tb]
\begin{center}
\includegraphics[width=0.495\linewidth]{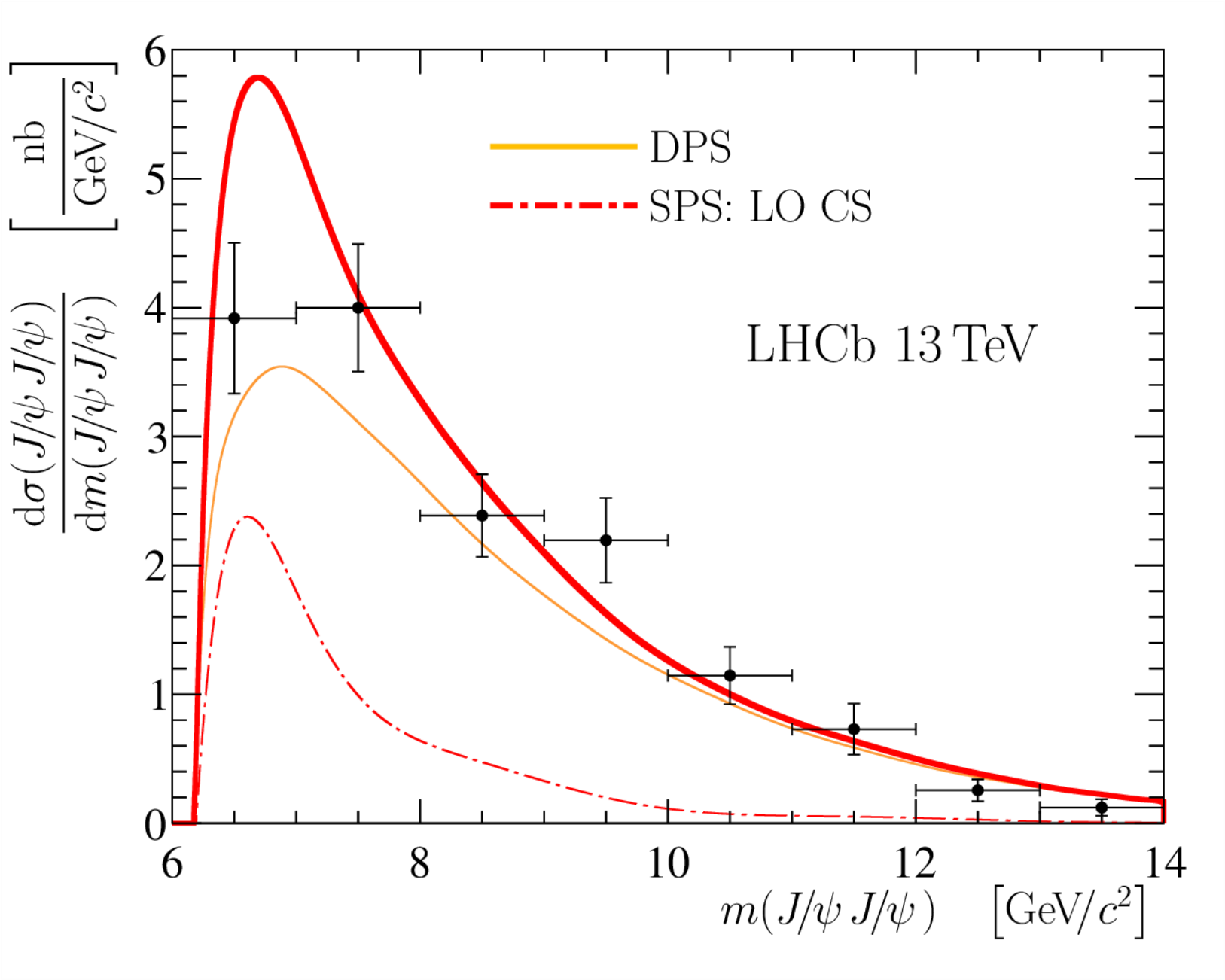}
\includegraphics[width=0.495\linewidth]{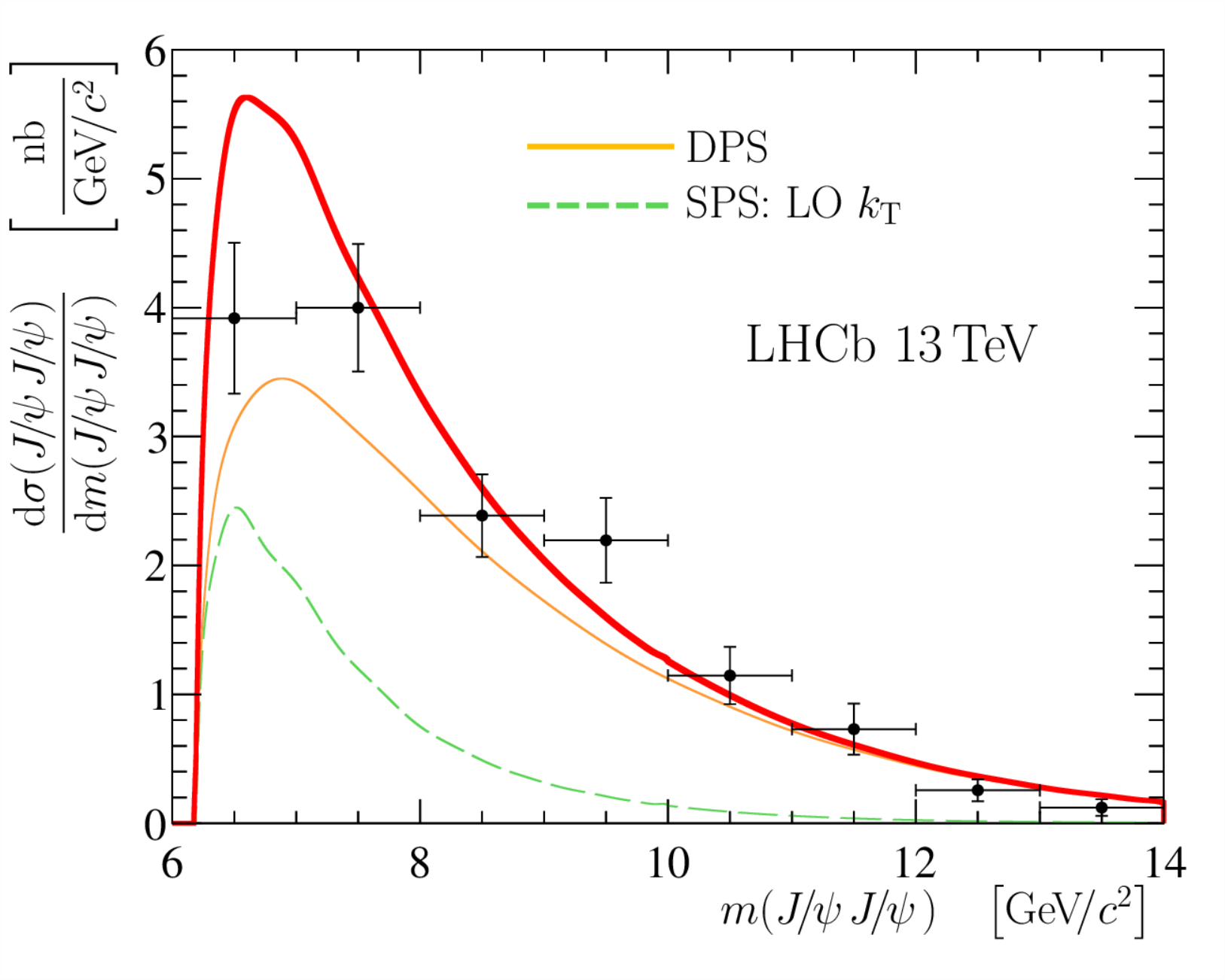}
\includegraphics[width=0.495\linewidth]{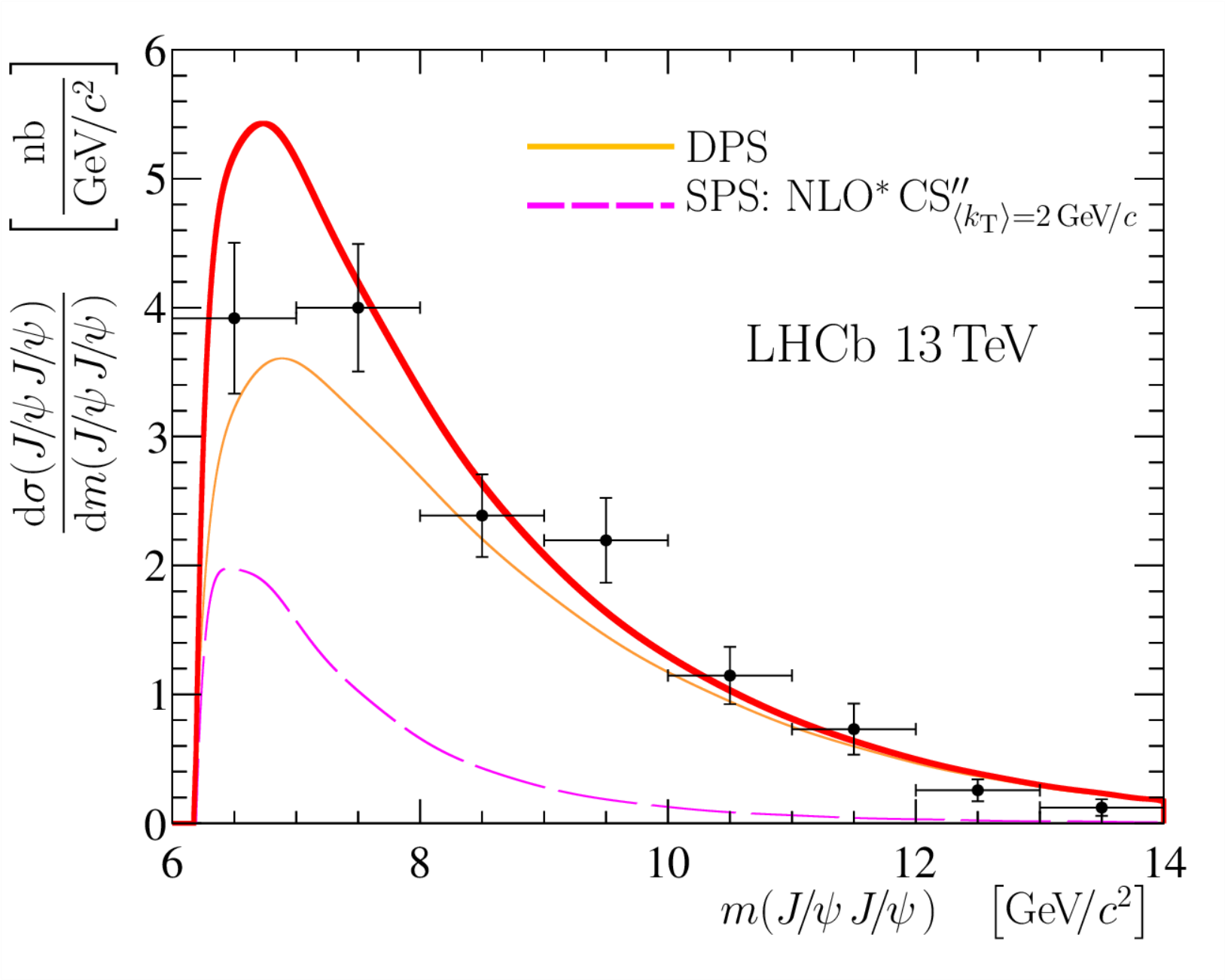}
\includegraphics[width=0.495\linewidth]{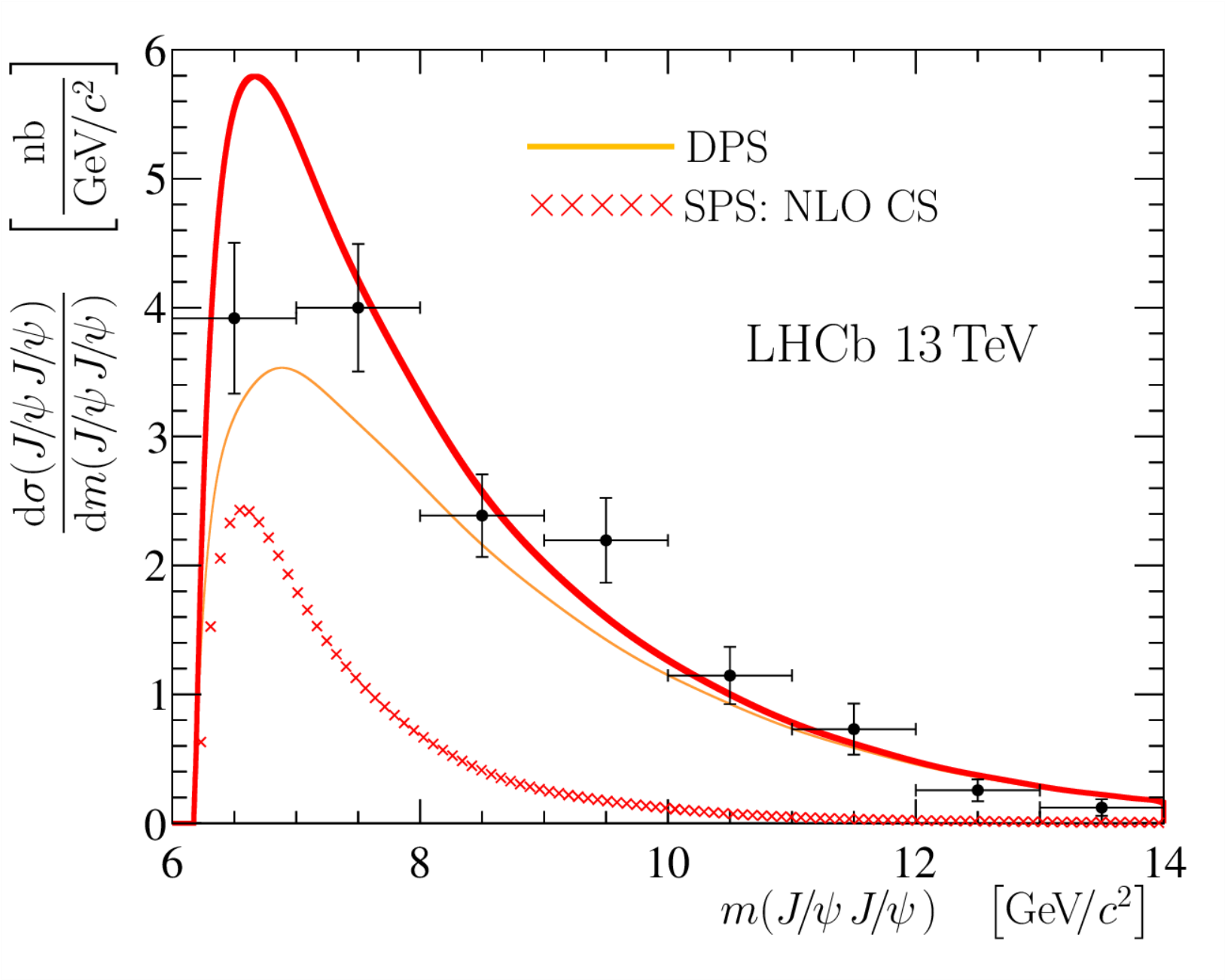}
\end{center}
  \caption { \small
    Result of templated DPS fit for 
    $\frac{ \mathrm{d}\sigma(\jpsi\jpsi)}{\mathrm{d} m(\jpsi\jpsi)}$.
    The (black) points with error bars represent the data. 
    The total fit result is shown with the thick (red) solid line
    and the DPS component is shown with the thin (orange) solid line.
  }
  \label{fig:cmp:fits_m_psipsi}
\end{figure}

\begin{figure}[tb]
\begin{center}
\includegraphics[width=0.495\linewidth]{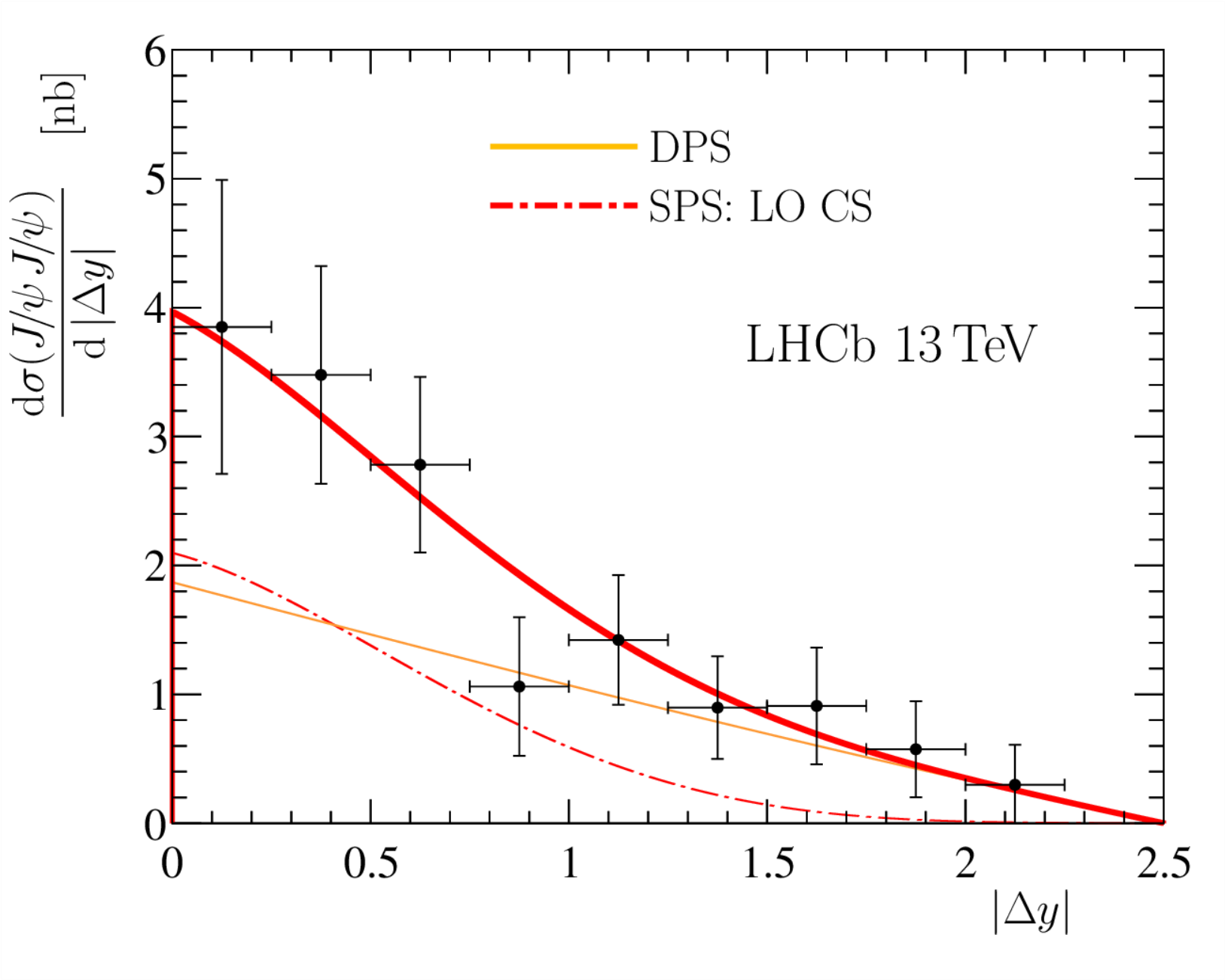}
\includegraphics[width=0.495\linewidth]{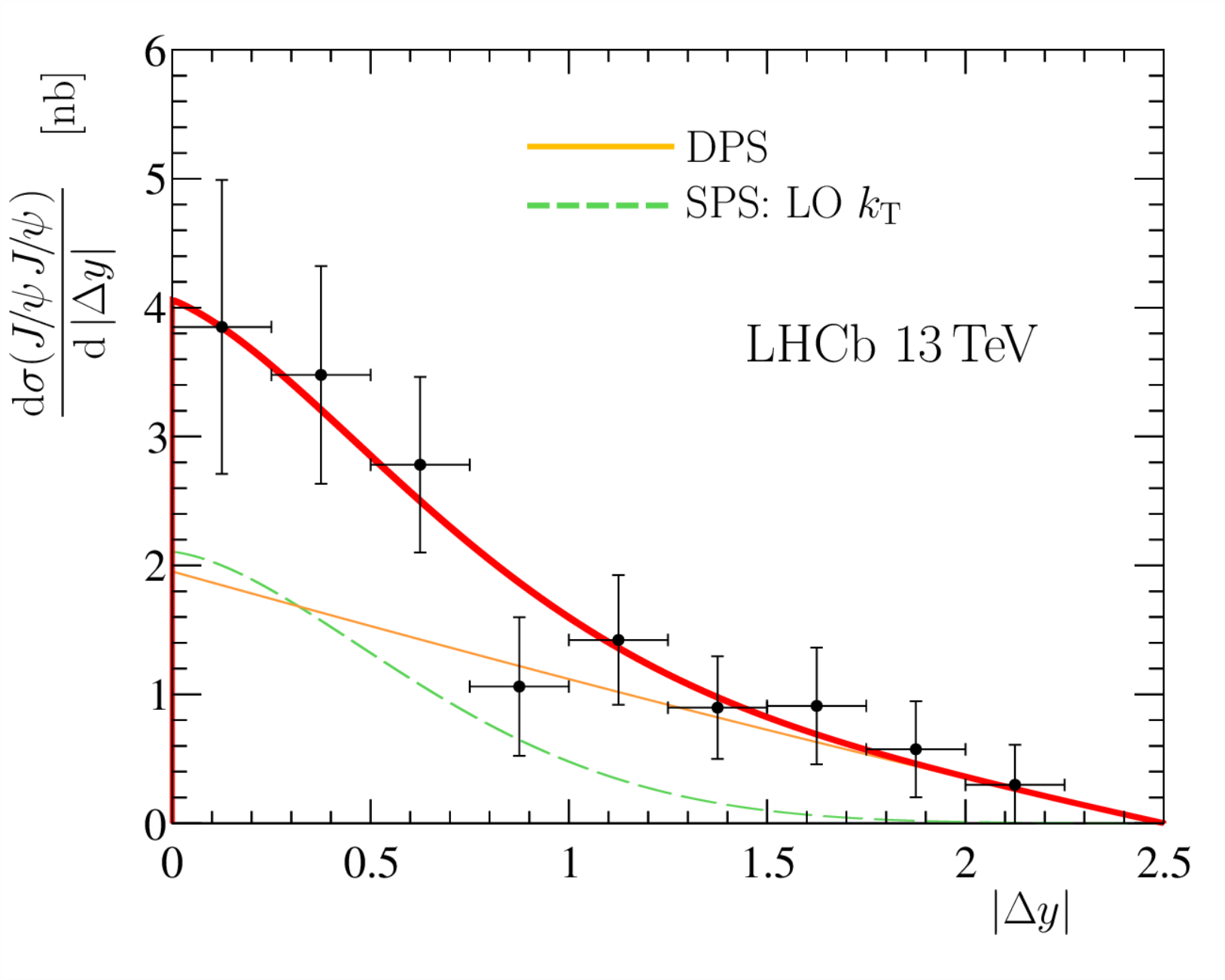}
\includegraphics[width=0.495\linewidth]{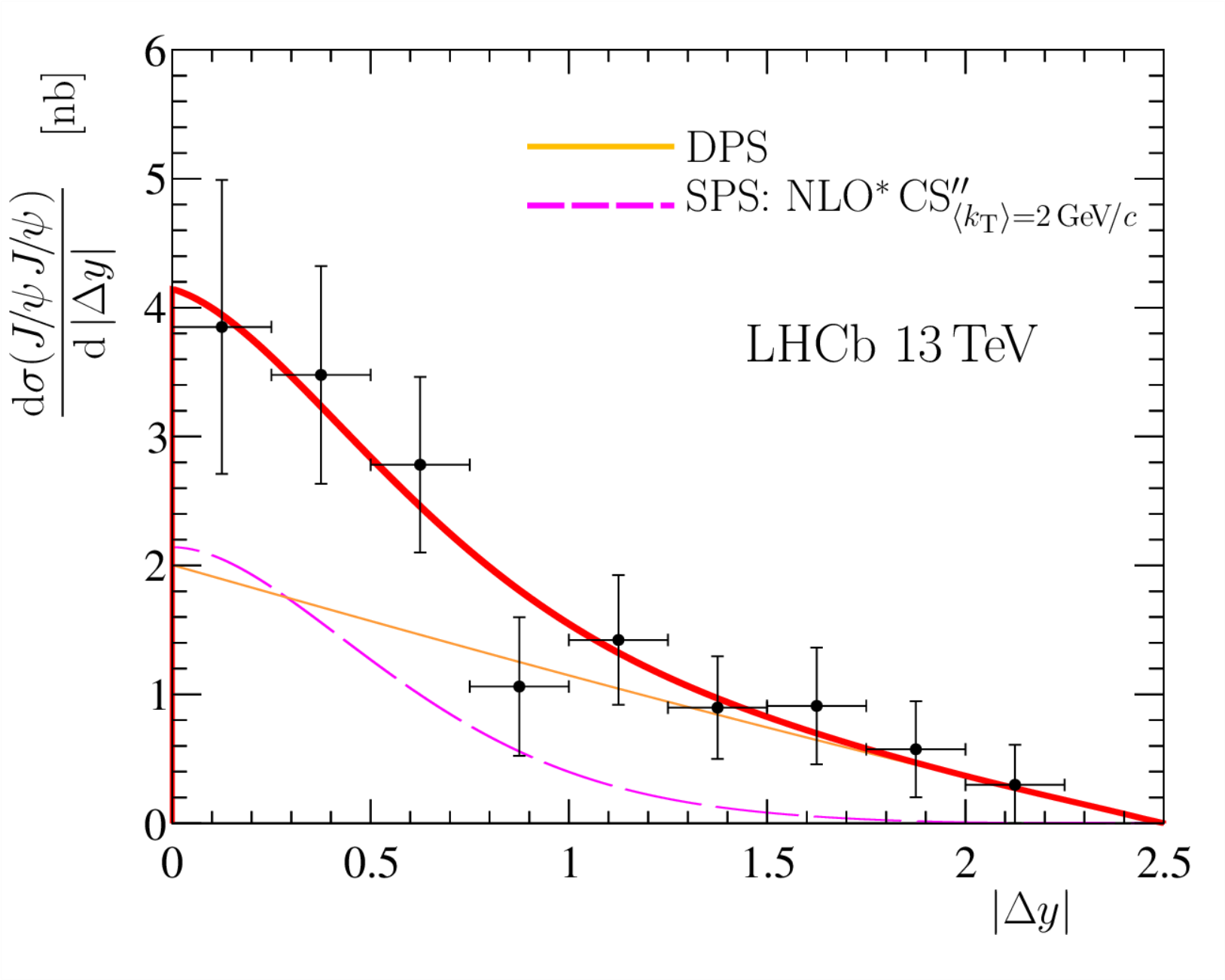}
\includegraphics[width=0.495\linewidth]{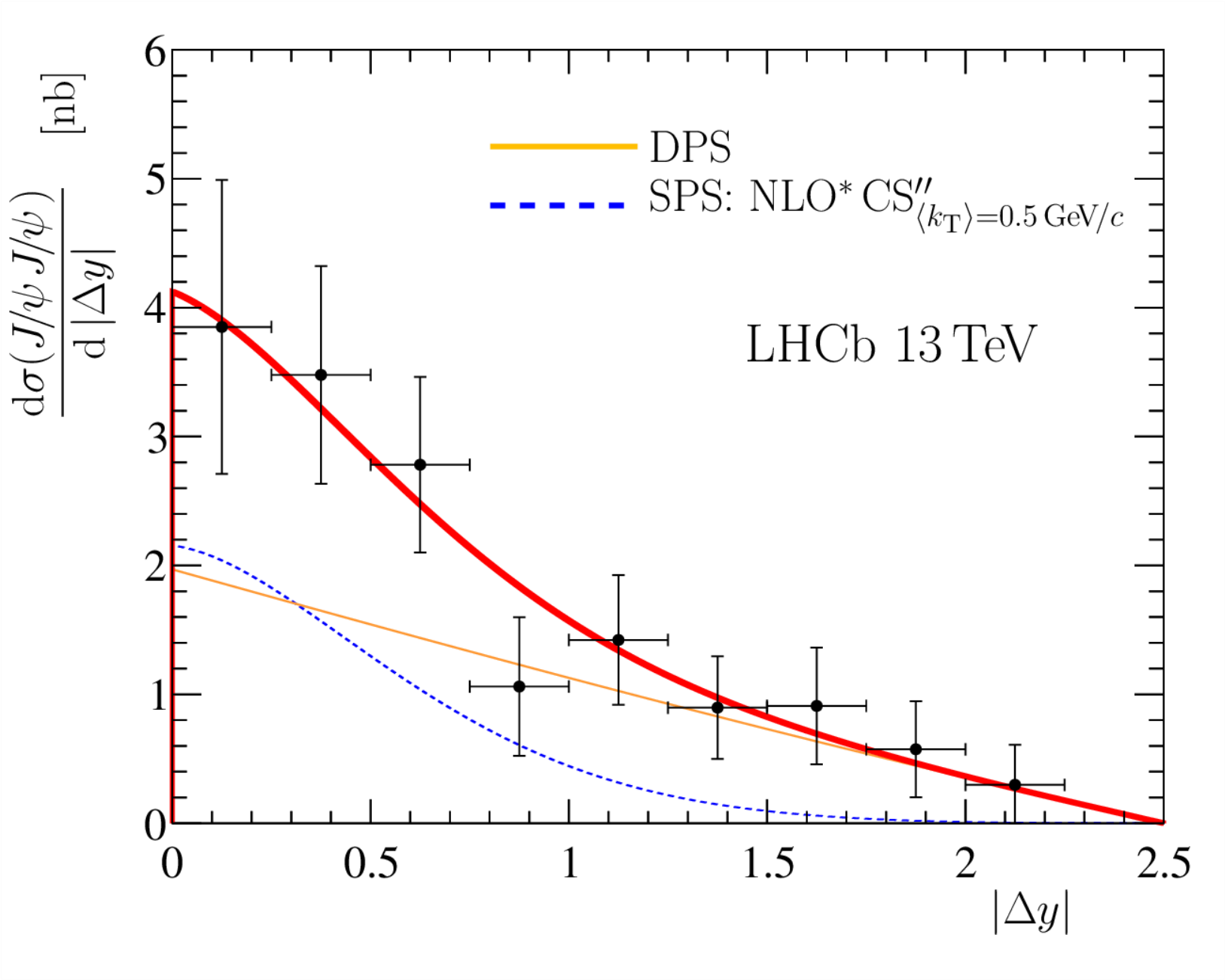}
\includegraphics[width=0.495\linewidth]{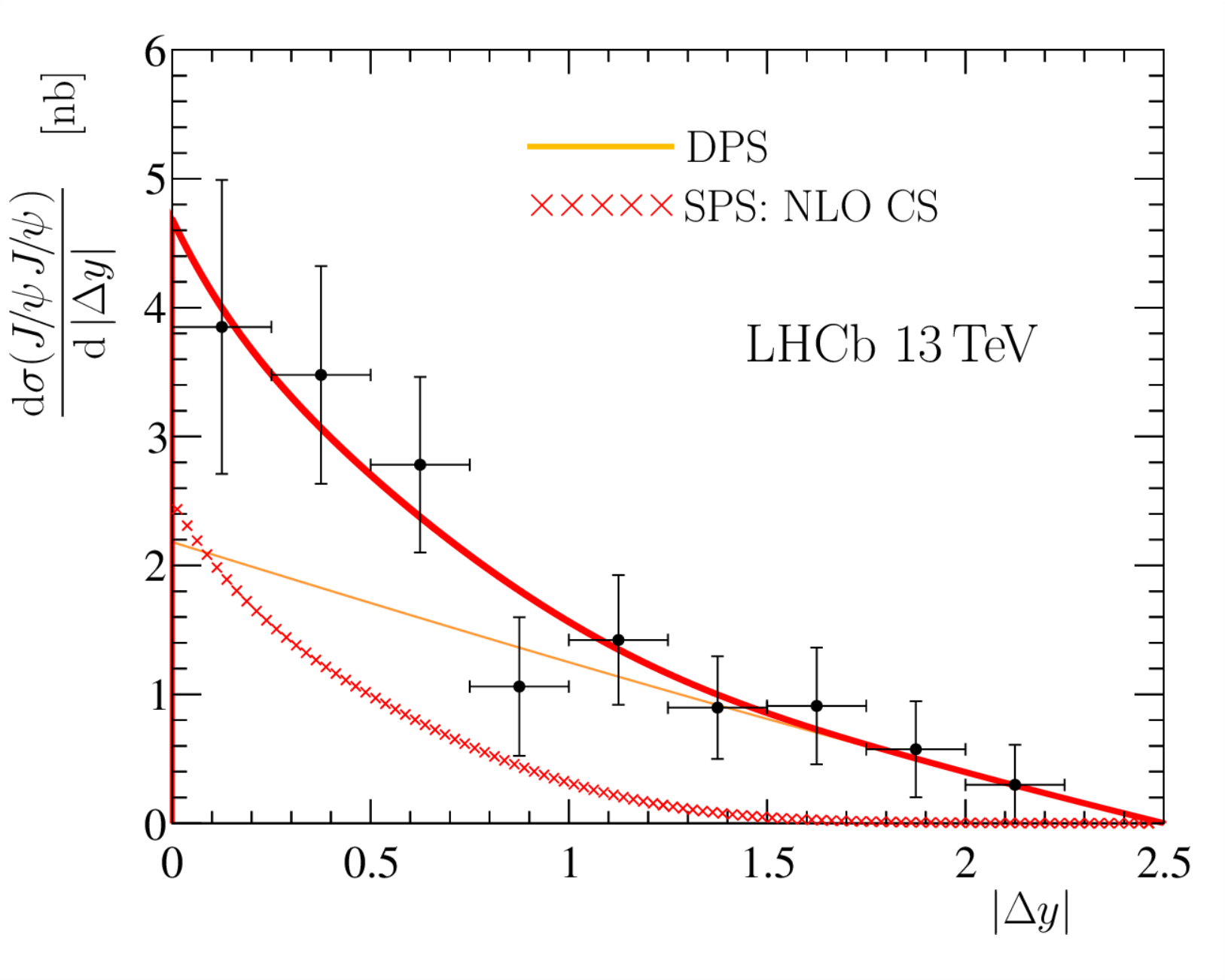}
\end{center}
  \caption { \small
    Result of templated DPS fit for 
    $\frac{ \mathrm{d}\sigma(\jpsi\jpsi)}{\mathrm{d} \left| \Delta y \right|}$.
    The (black) points with error bars represent the data. 
    The total fit result is shown with the thick (red) solid line
    and the DPS component is shown with the thin (orange) solid line.
  }
  \label{fig:cmp:fits_dy_psipsi}
\end{figure}

\begin{figure}[tb]
\begin{center}
\includegraphics[width=0.495\linewidth]{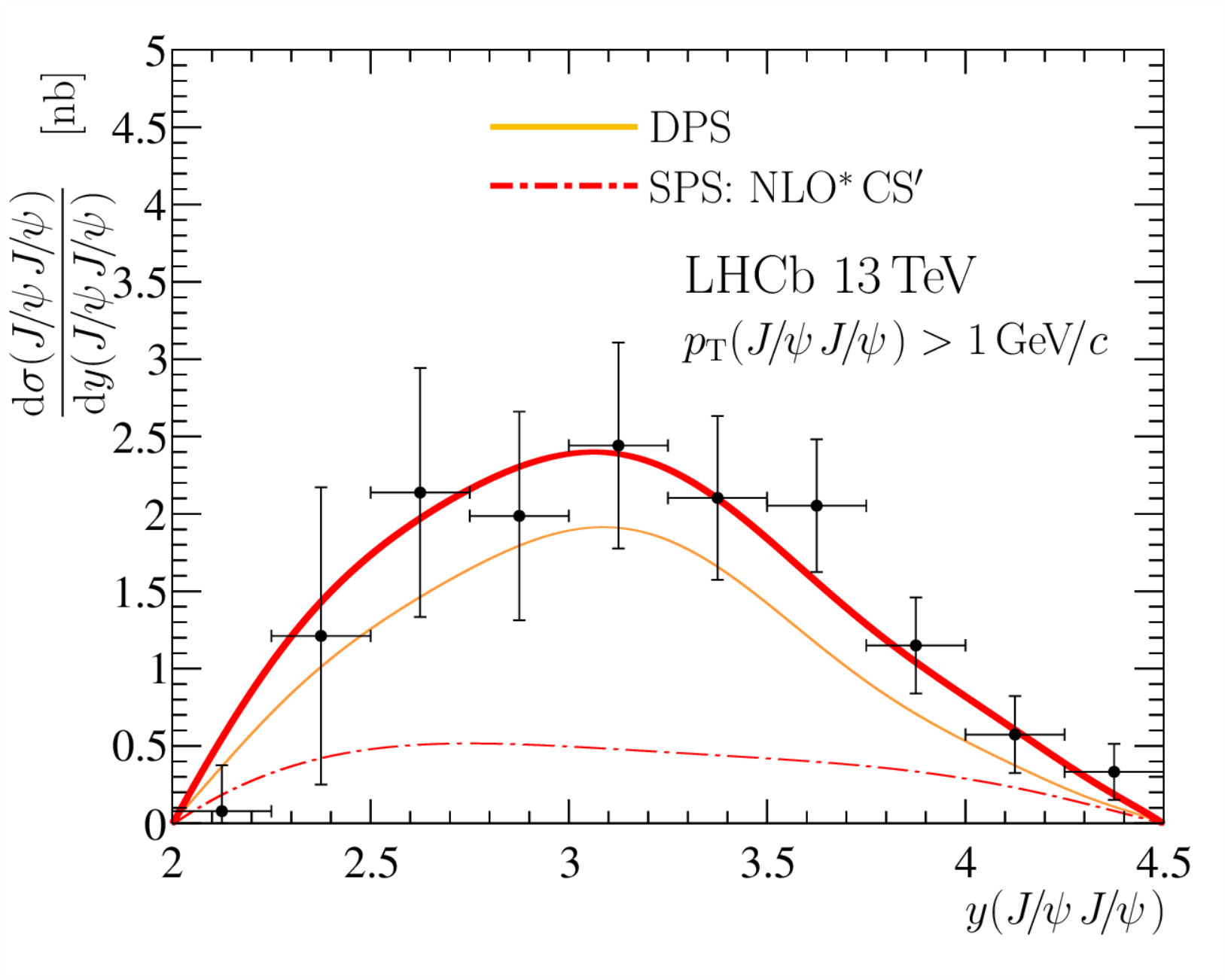}
\includegraphics[width=0.495\linewidth]{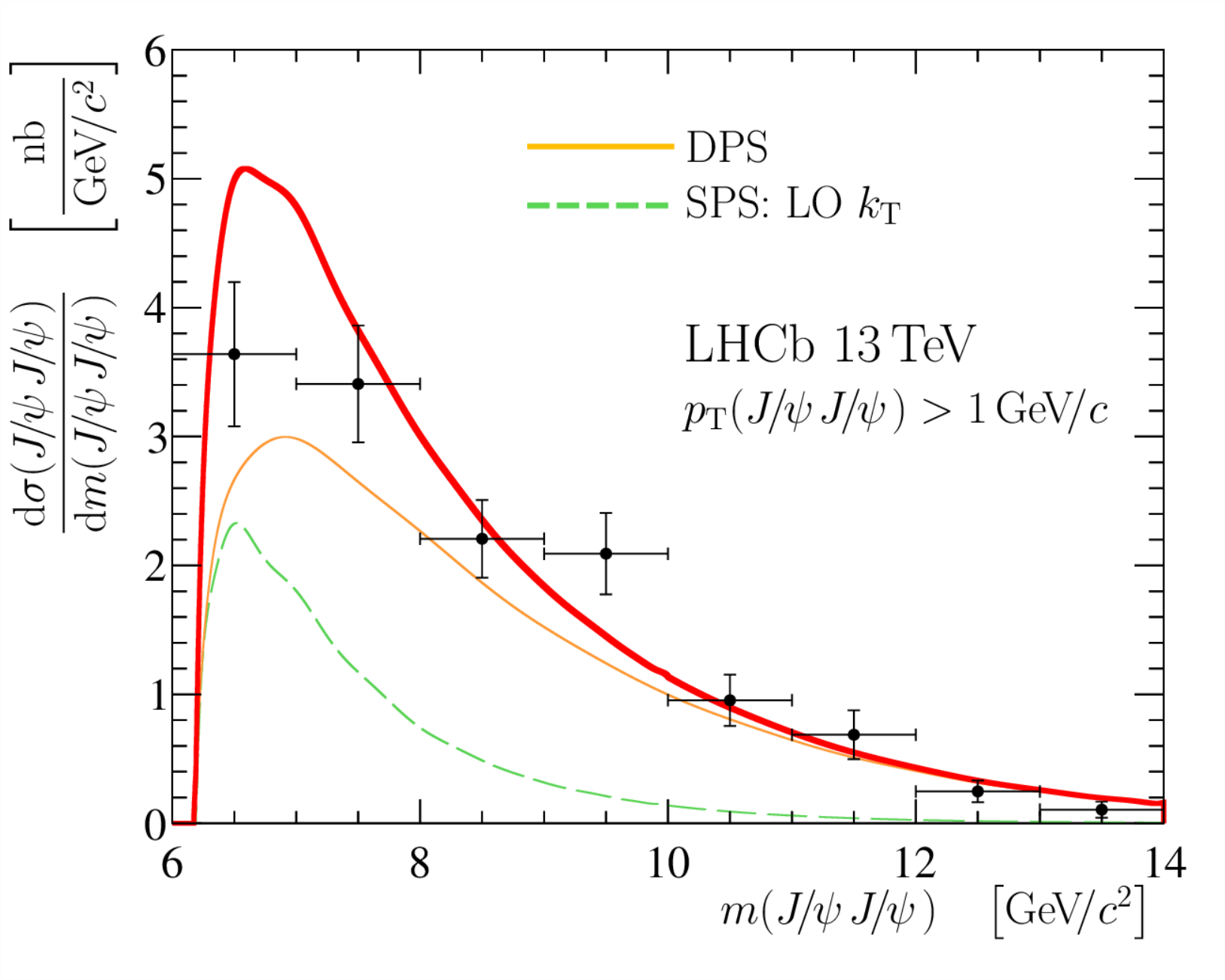}
\includegraphics[width=0.495\linewidth]{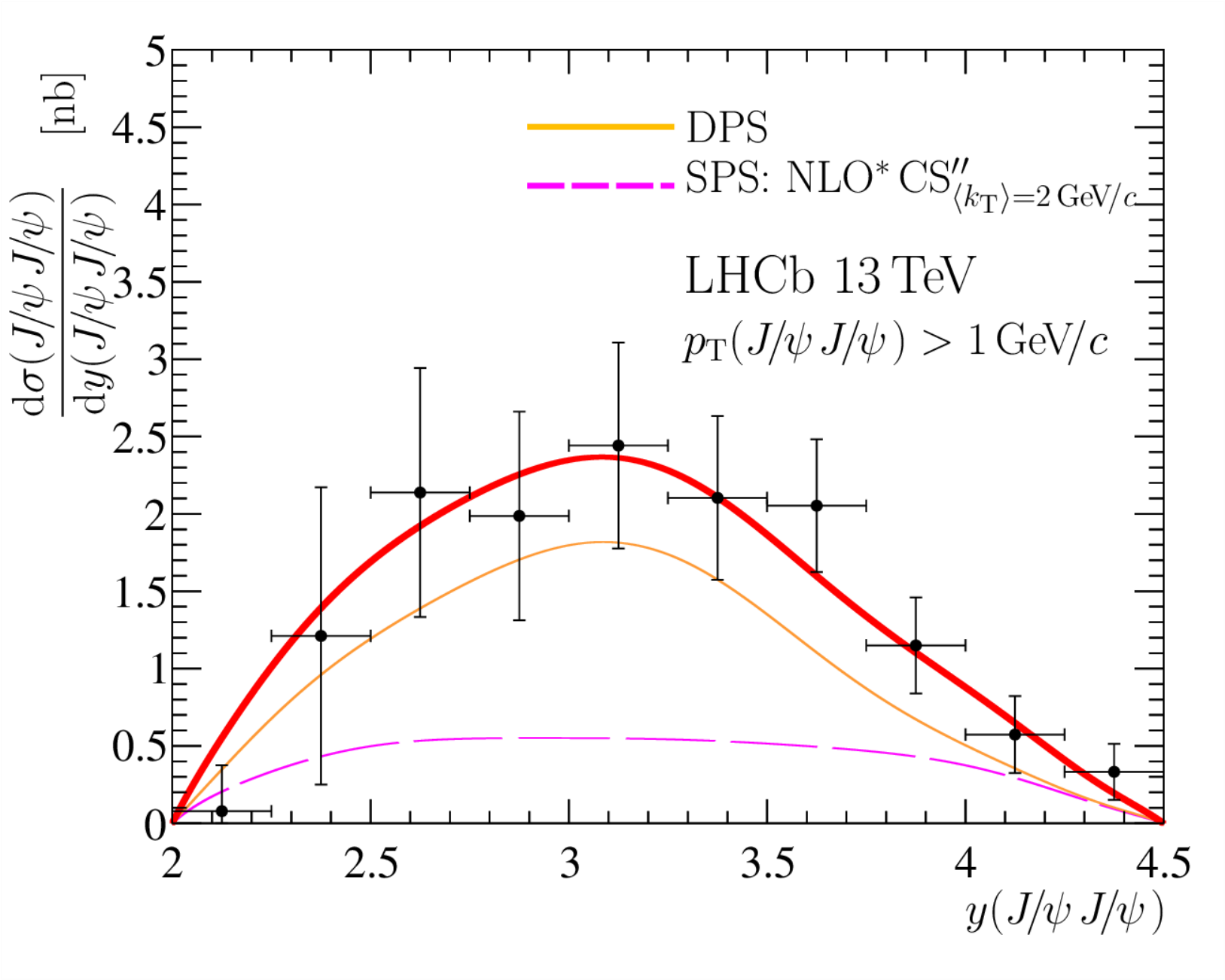}
\includegraphics[width=0.495\linewidth]{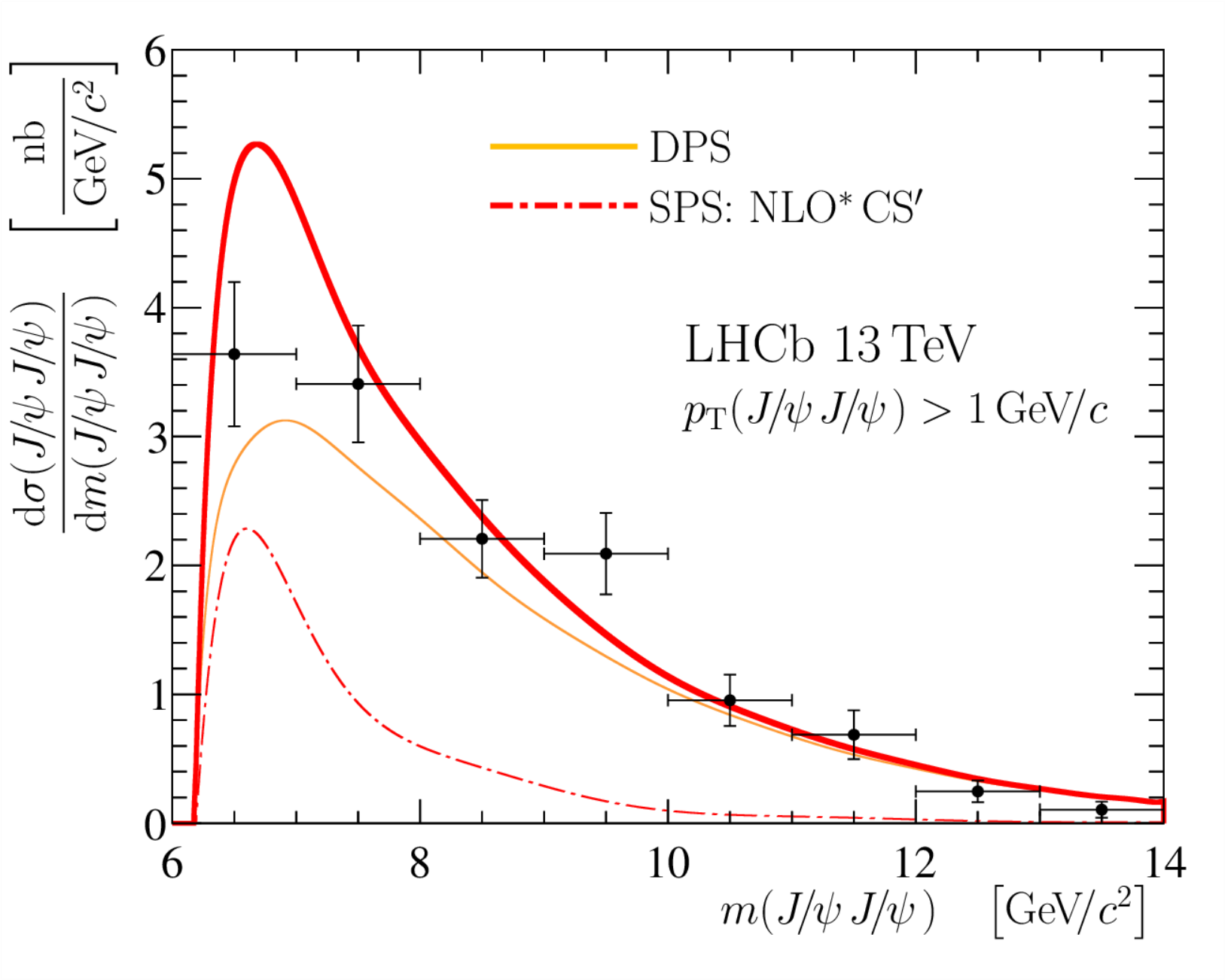}
\includegraphics[width=0.495\linewidth]{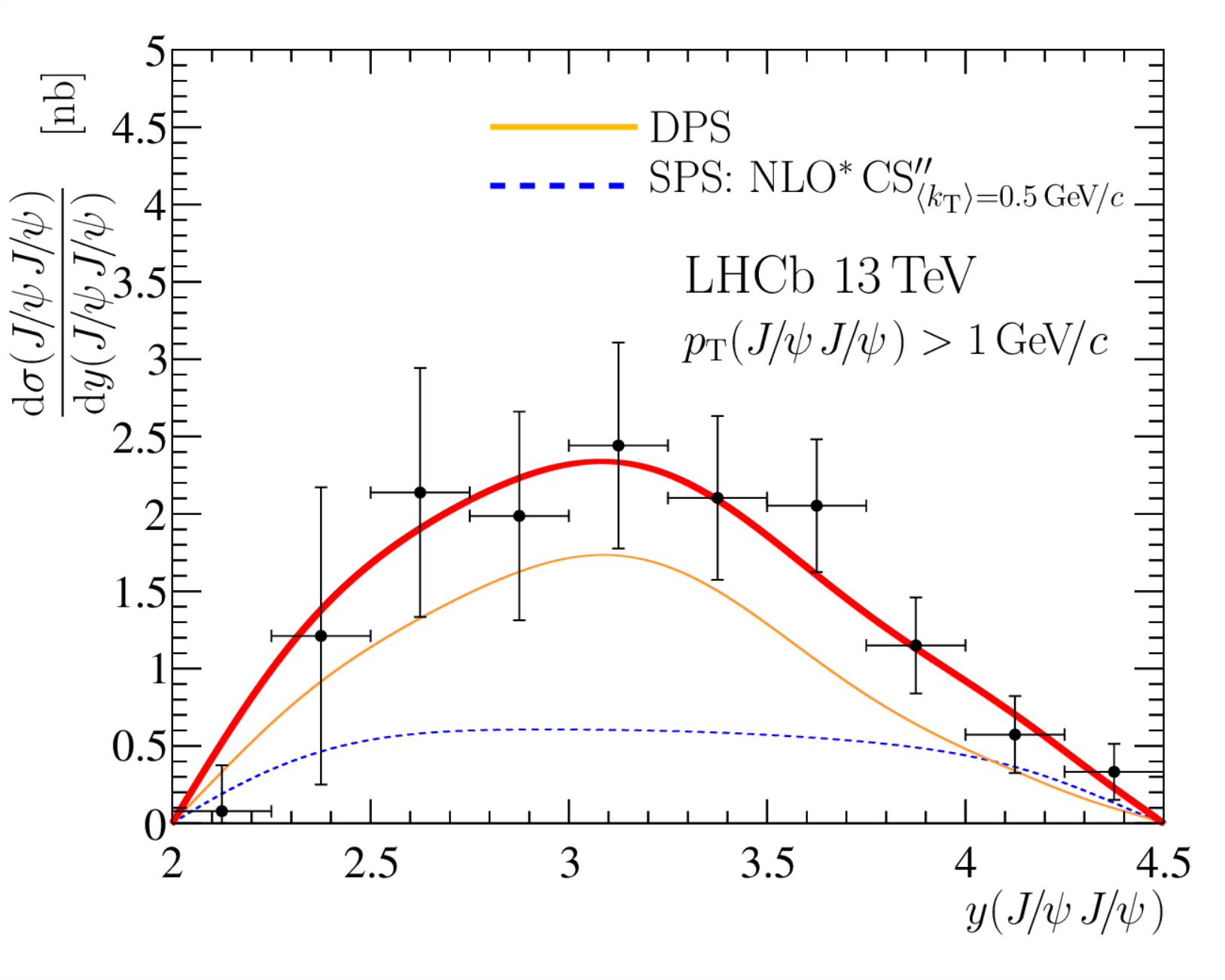}
\includegraphics[width=0.495\linewidth]{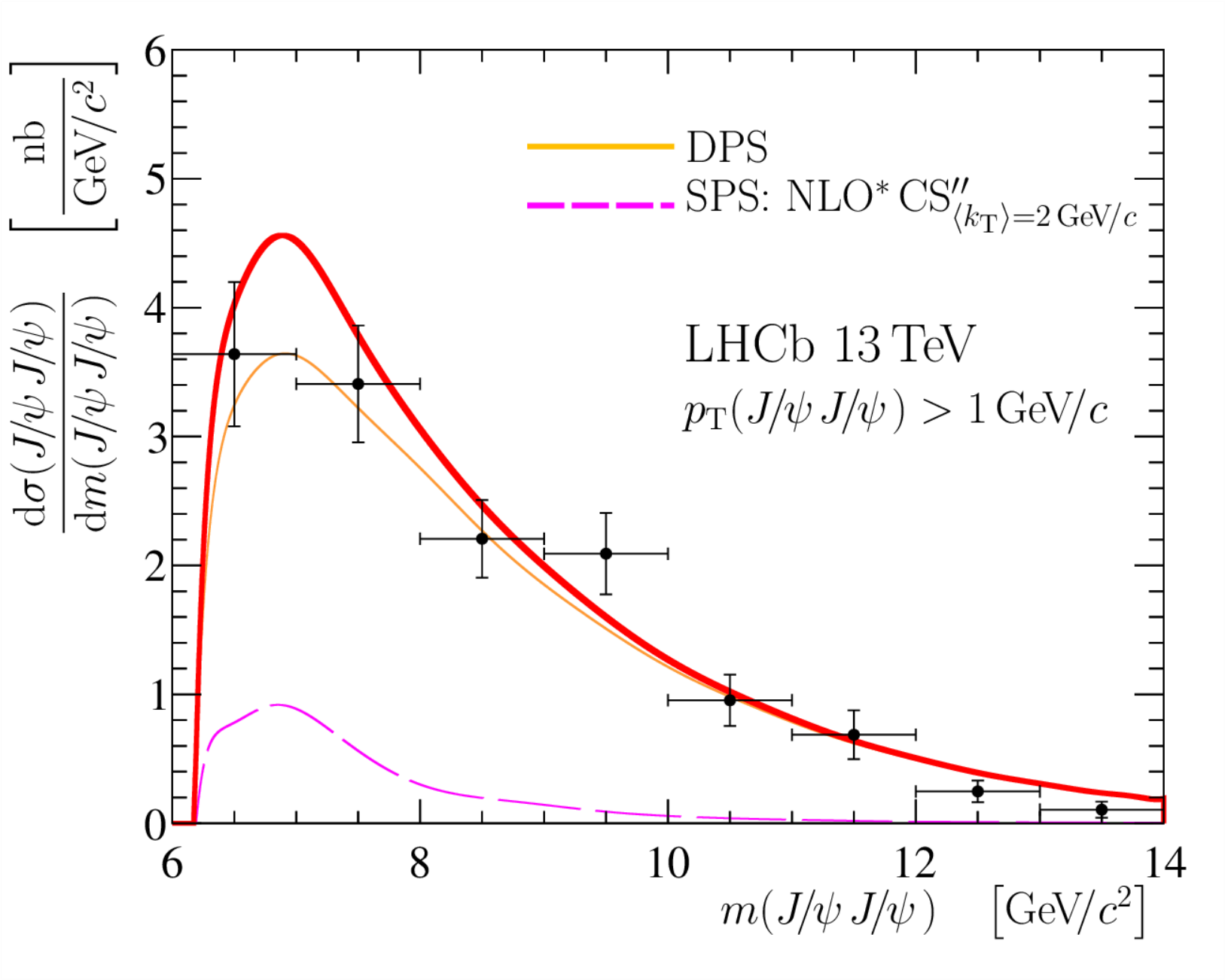}
\end{center}
  \caption { \small
    Result of templated DPS fit for 
    $\frac{ \mathrm{d}\sigma(\jpsi\jpsi)}{\mathrm{d} y(\jpsi\jpsi)}$ and
    $\frac{ \mathrm{d}\sigma(\jpsi)}{\mathrm{d} m(\jpsi\jpsi)}$
    for the $\pt(\jpsi\jpsi)>1\gevc$~region.
    The (black) points with error bars represent the data. 
    The total fit result is shown with the thick (red) solid line
    and the DPS component is shown with the thin (orange) solid line.
  }
  \label{fig:cmp:fits_my_psipsi_1}
\end{figure}

\begin{figure}[tb]
\begin{center}
\includegraphics[width=0.495\linewidth]{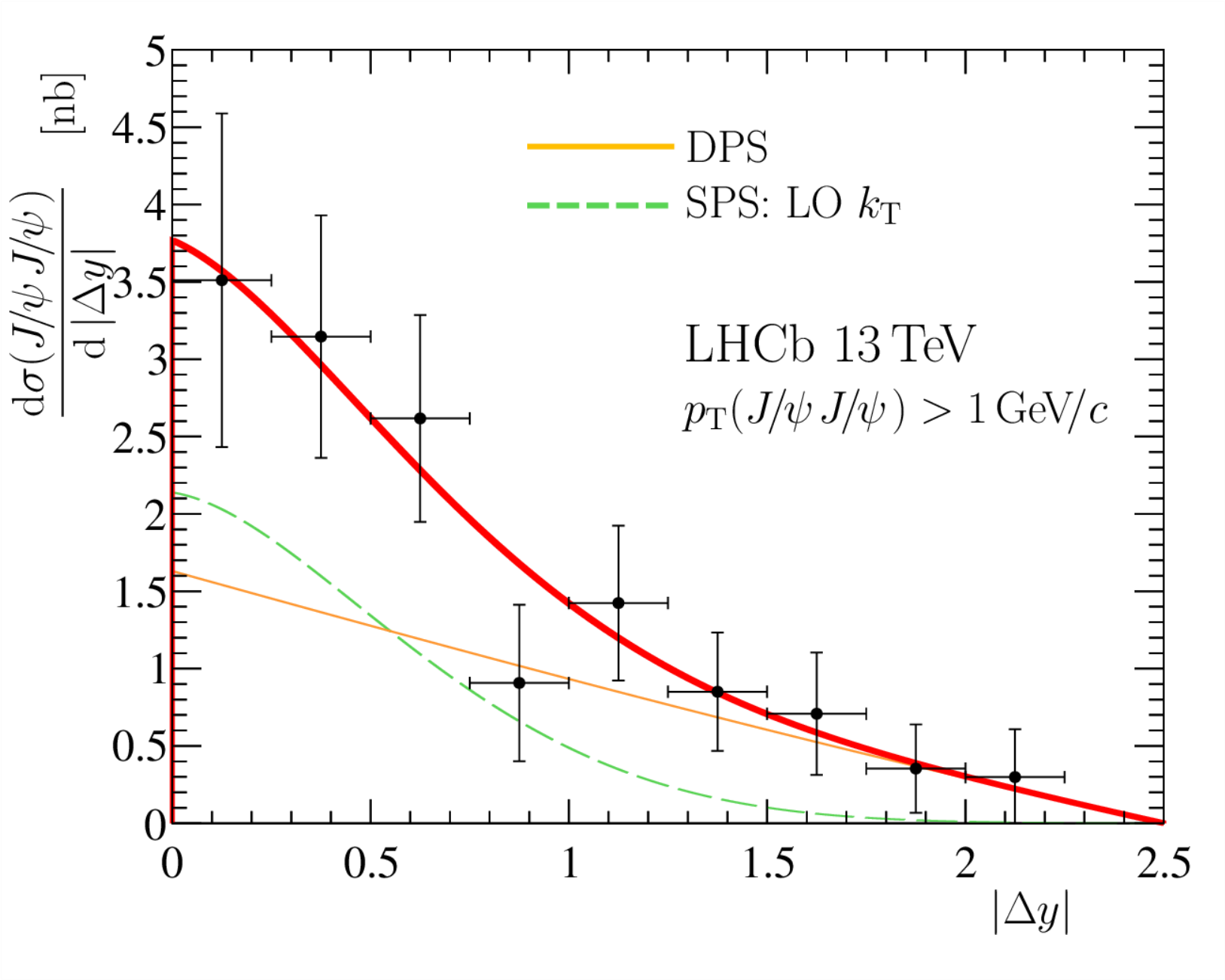}
\includegraphics[width=0.495\linewidth]{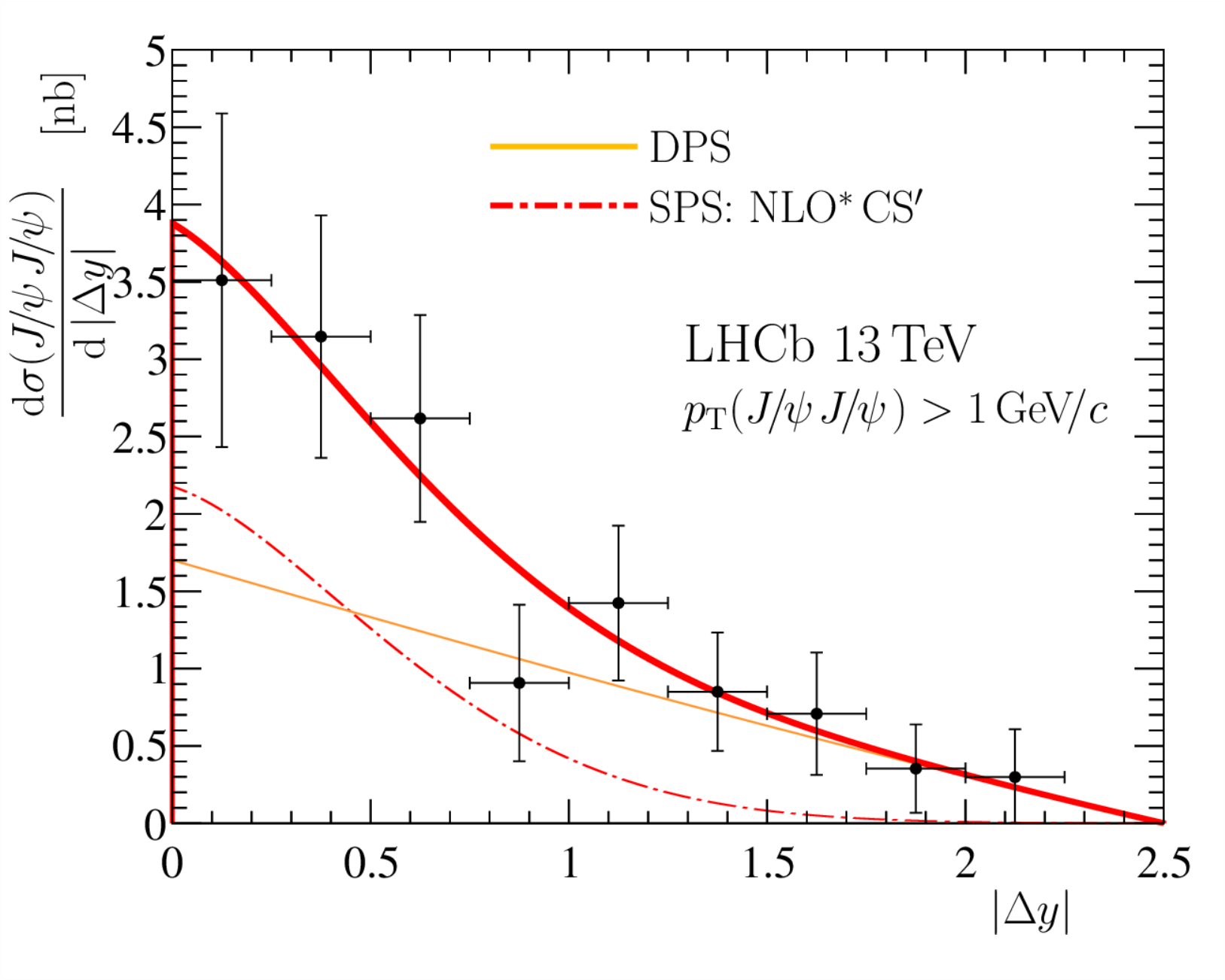}
\includegraphics[width=0.495\linewidth]{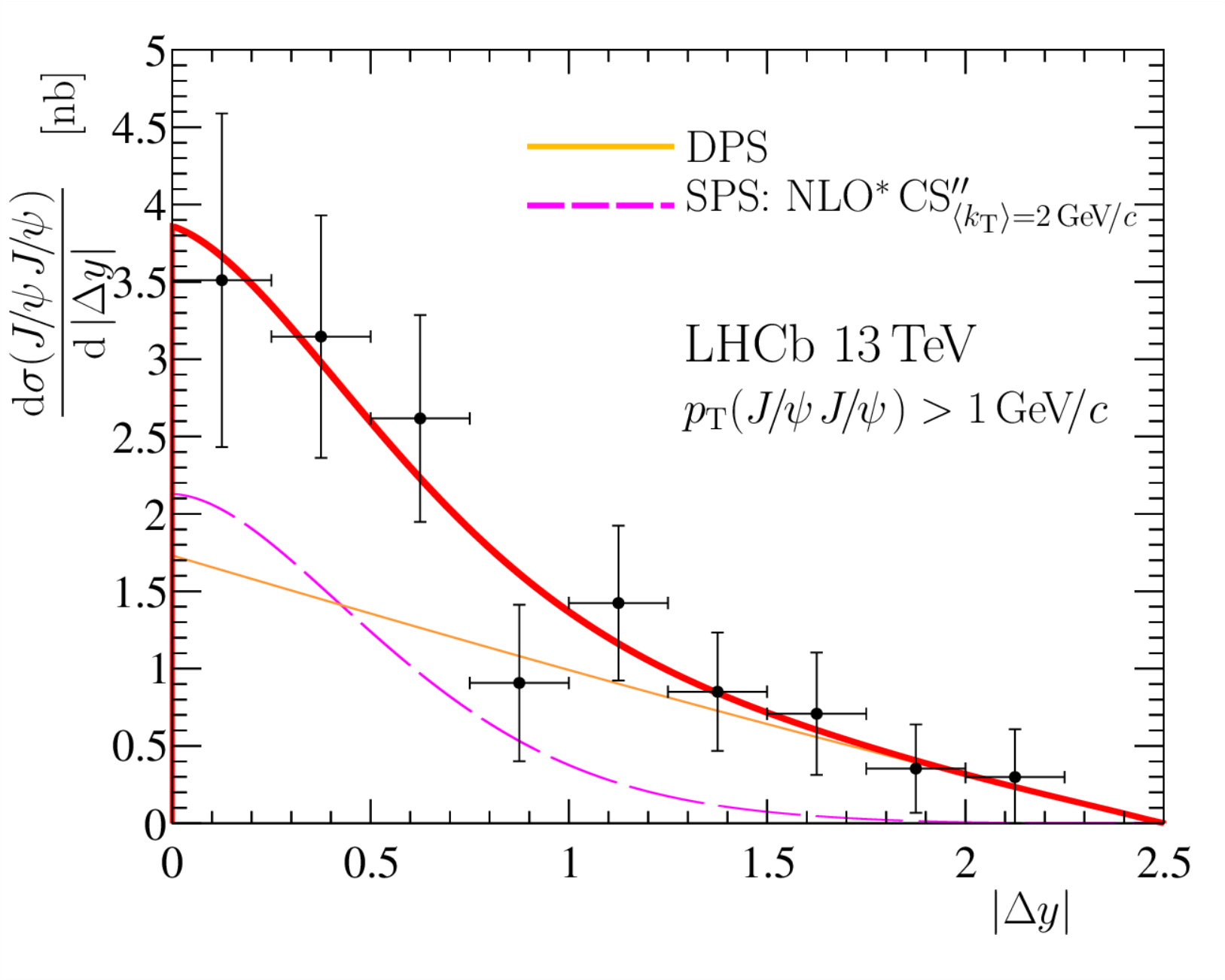}
\includegraphics[width=0.495\linewidth]{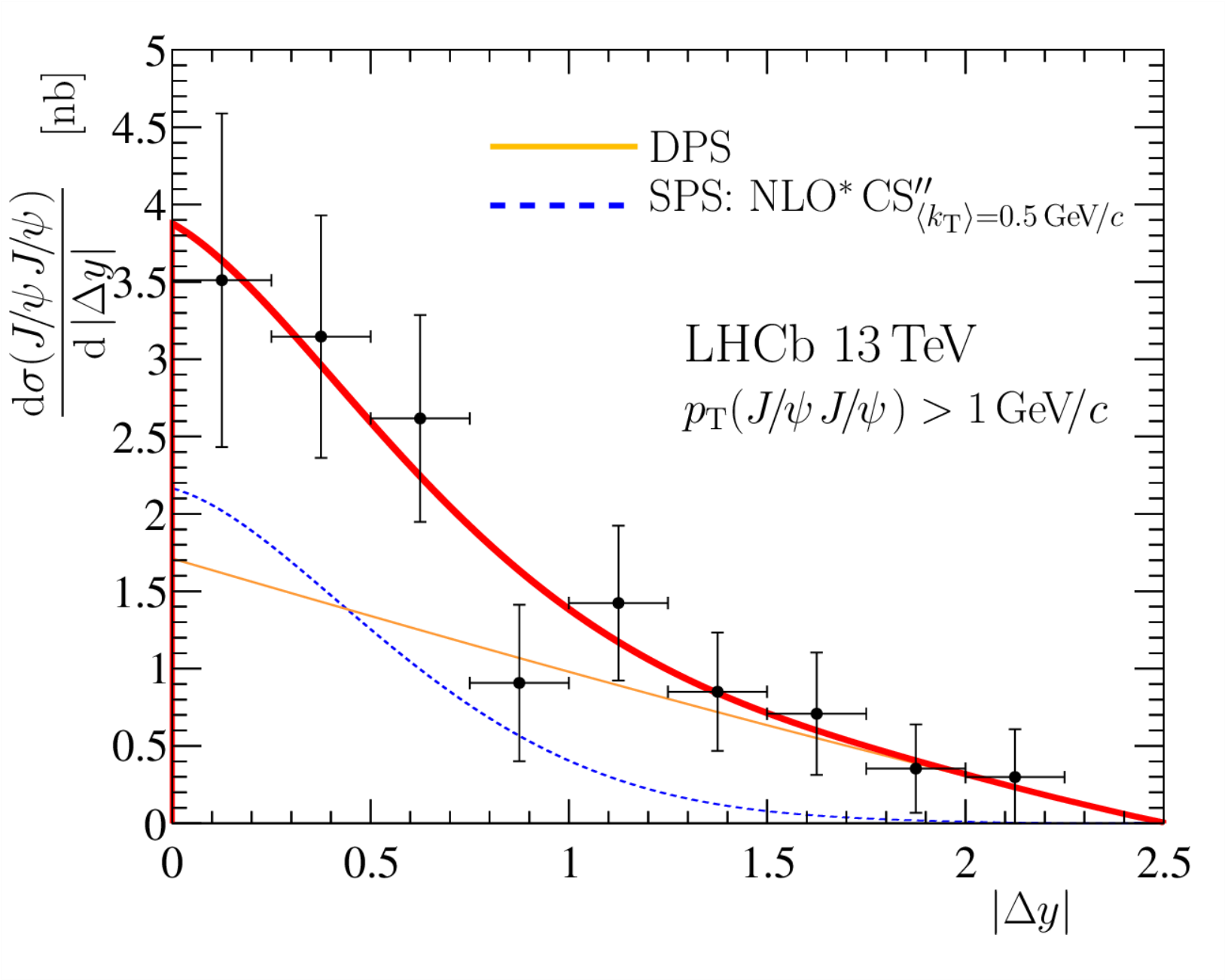}
\end{center}
  \caption { \small
    Result of templated DPS fit for 
    $\frac{ \mathrm{d}\sigma(\jpsi\jpsi)}{\mathrm{d} \left| \Delta y \right|}$ 
    for the $\pt(\jpsi\jpsi)>1\gevc$~region.
    The (black) points with error bars represent the data. 
    The total fit result is shown with the thick (red) solid line
    and the DPS component is shown with the thin (orange) solid line.
  }
  \label{fig:cmp:fits_dy_psipsi_1}
\end{figure}

\begin{figure}[tb]
\begin{center}
\includegraphics[width=0.495\linewidth]{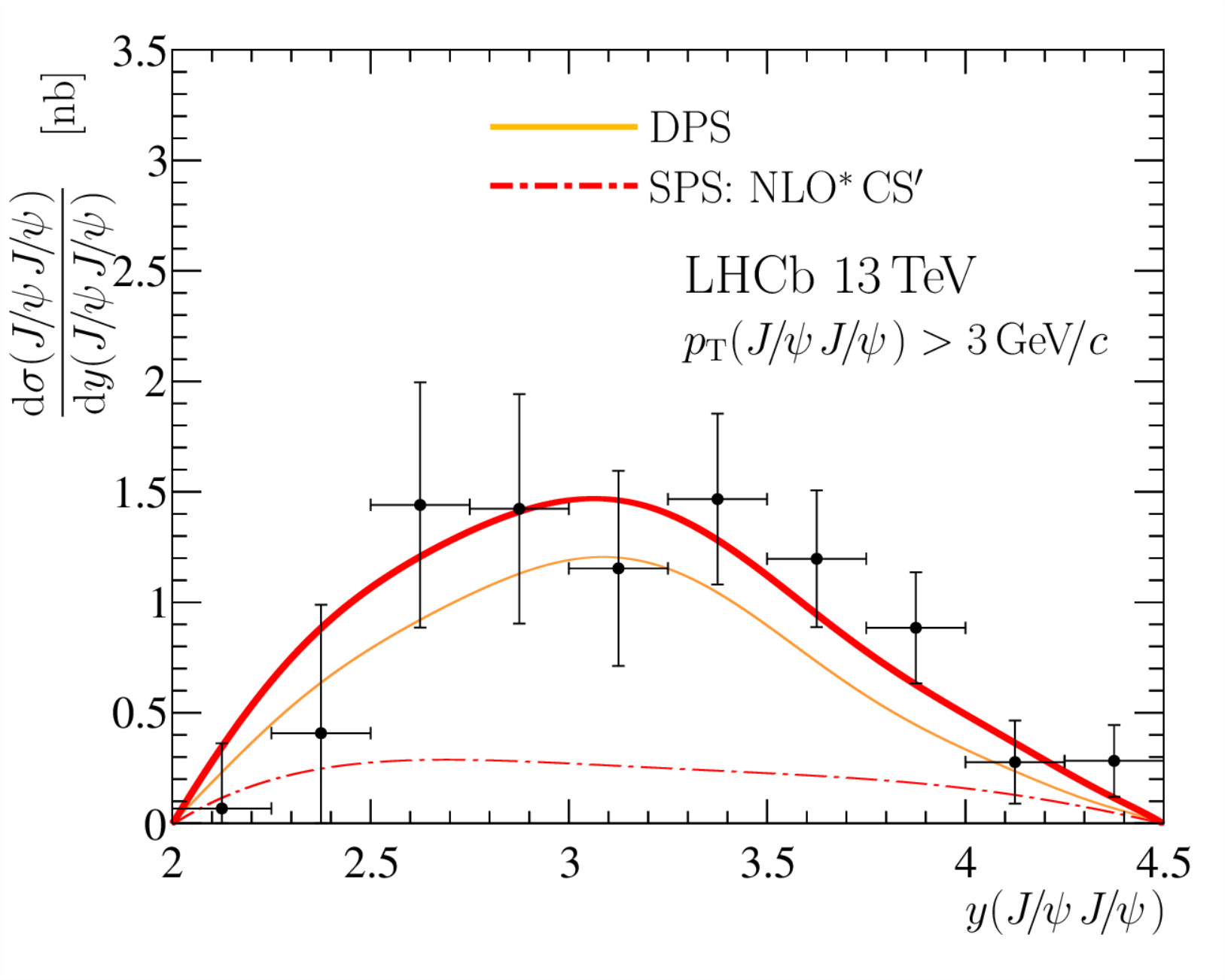}
\includegraphics[width=0.495\linewidth]{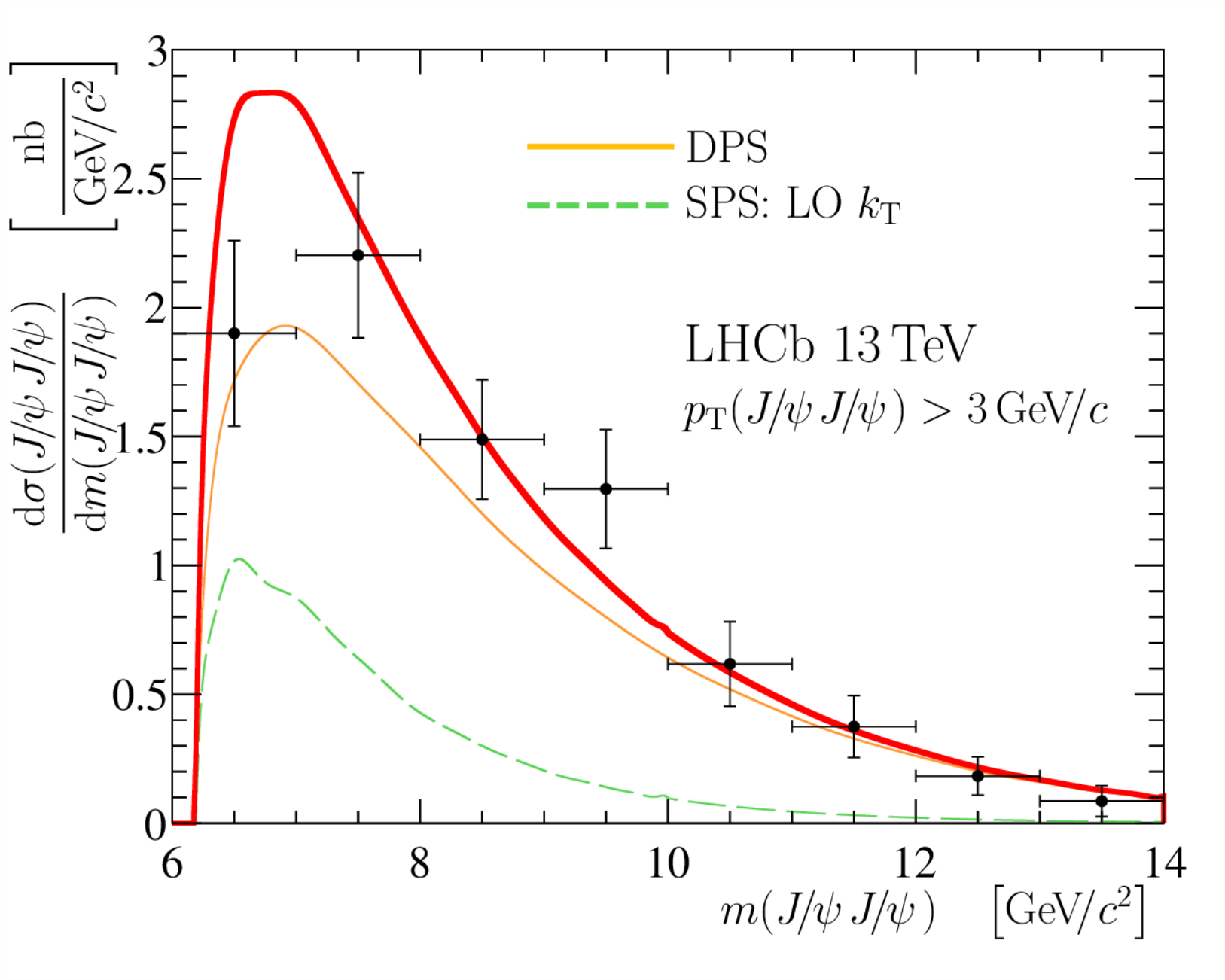}
\includegraphics[width=0.495\linewidth]{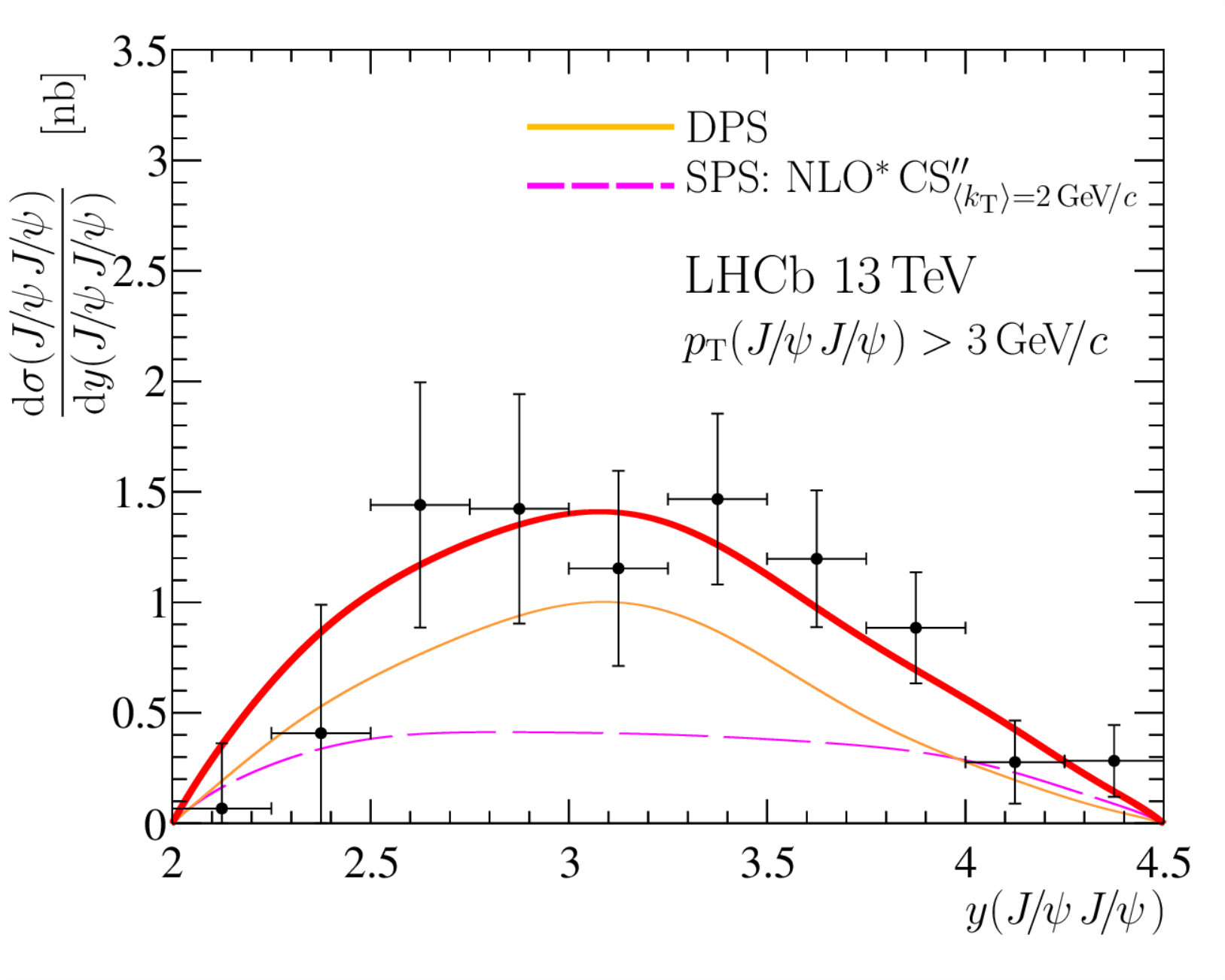}
\includegraphics[width=0.495\linewidth]{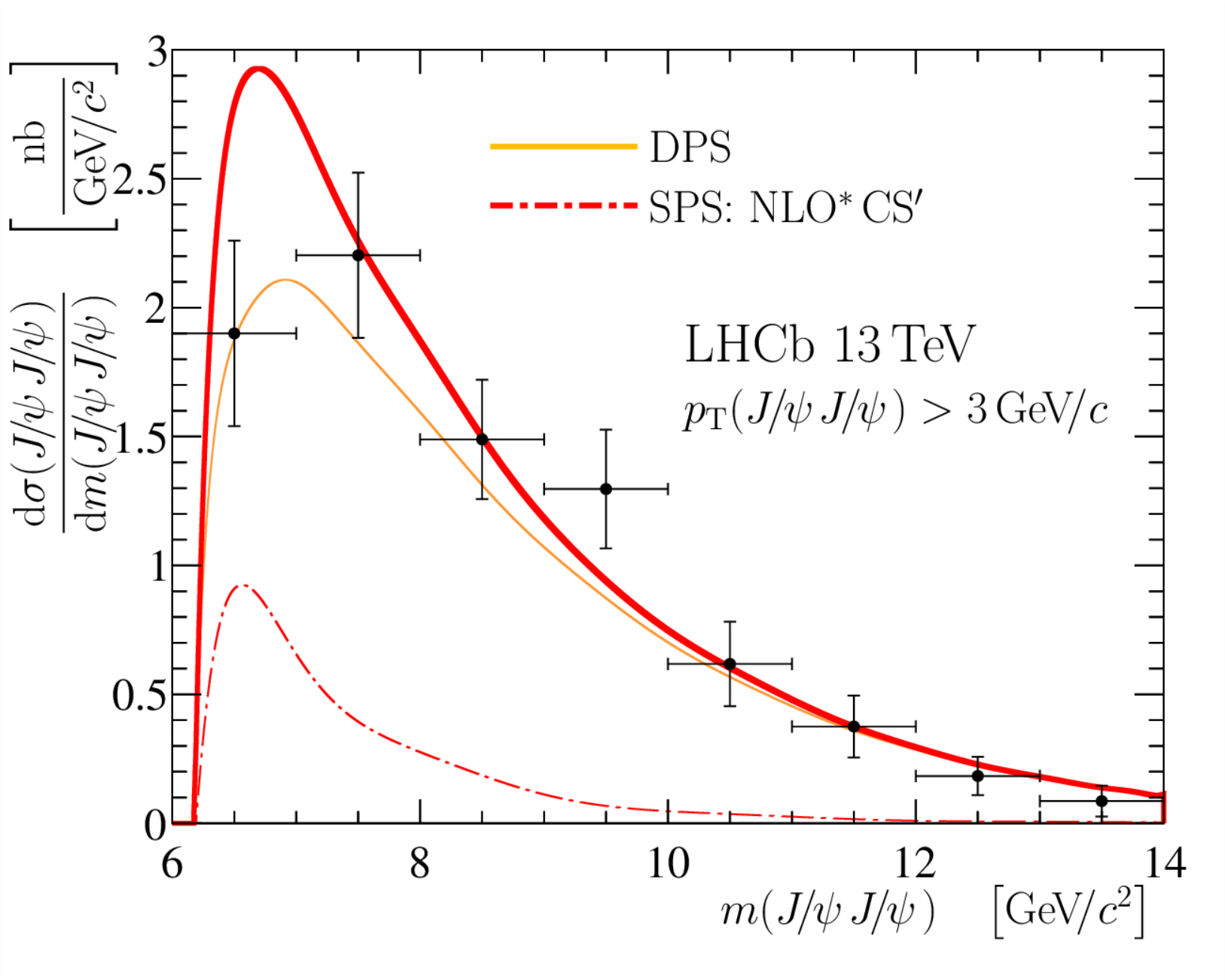}
\includegraphics[width=0.495\linewidth]{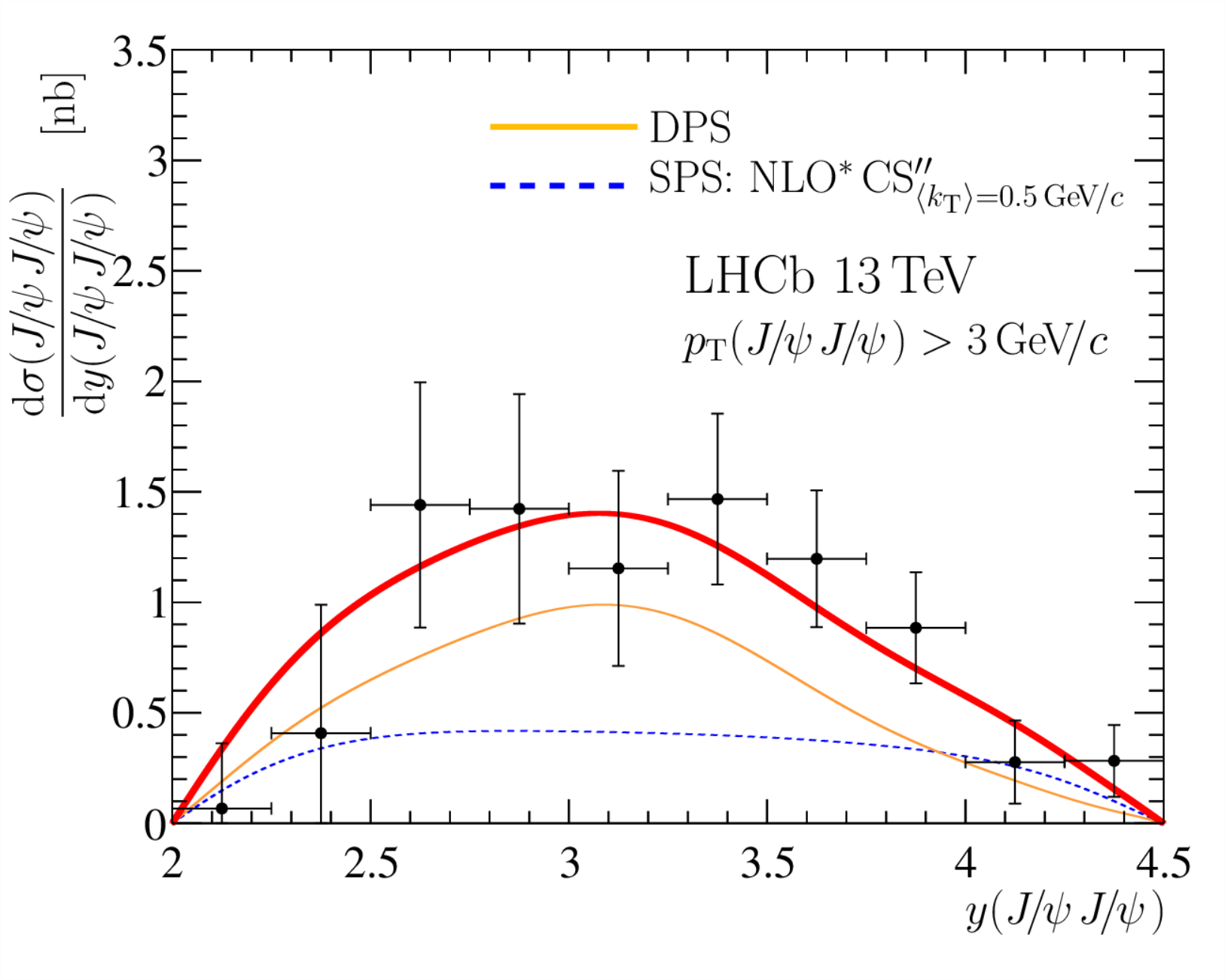}
\includegraphics[width=0.495\linewidth]{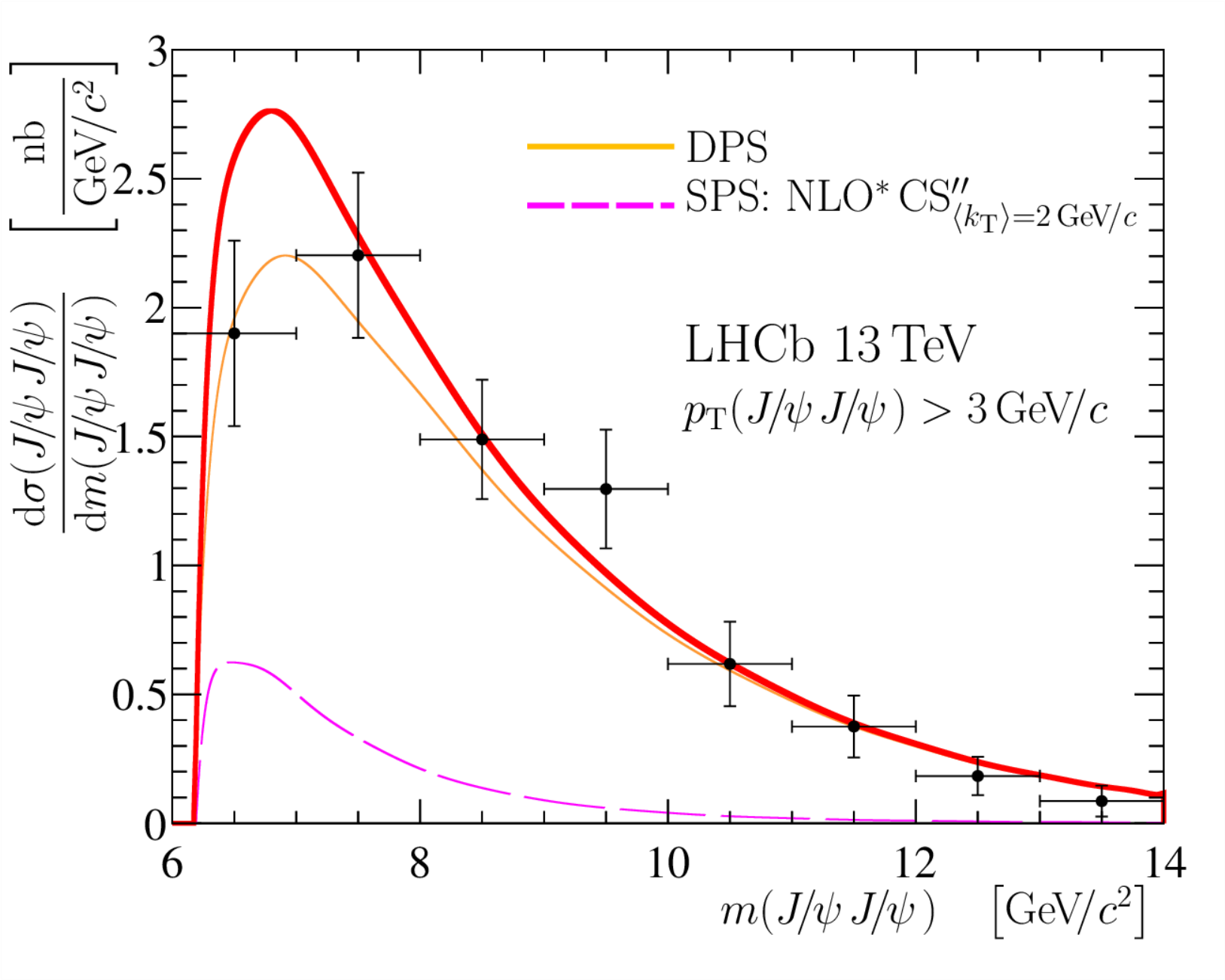}
\end{center}
  \caption { \small
    Result of templated DPS fit for 
    $\frac{ \mathrm{d}\sigma(\jpsi\jpsi)}{\mathrm{d} y(\jpsi\jpsi)}$ and
    $\frac{ \mathrm{d}\sigma(\jpsi\jpsi)}{\mathrm{d} m(\jpsi\jpsi)}$
    for the $\pt(\jpsi\jpsi)>3\gevc$~region.
    The (black) points with error bars represent the data. 
    The total fit result is shown with the thick (red) solid line
    and the DPS component is shown with the thin (orange) solid line.
  }
  \label{fig:cmp:fits_my_psipsi_3}
\end{figure}

\begin{figure}[tb]
\begin{center}
\includegraphics[width=0.495\linewidth]{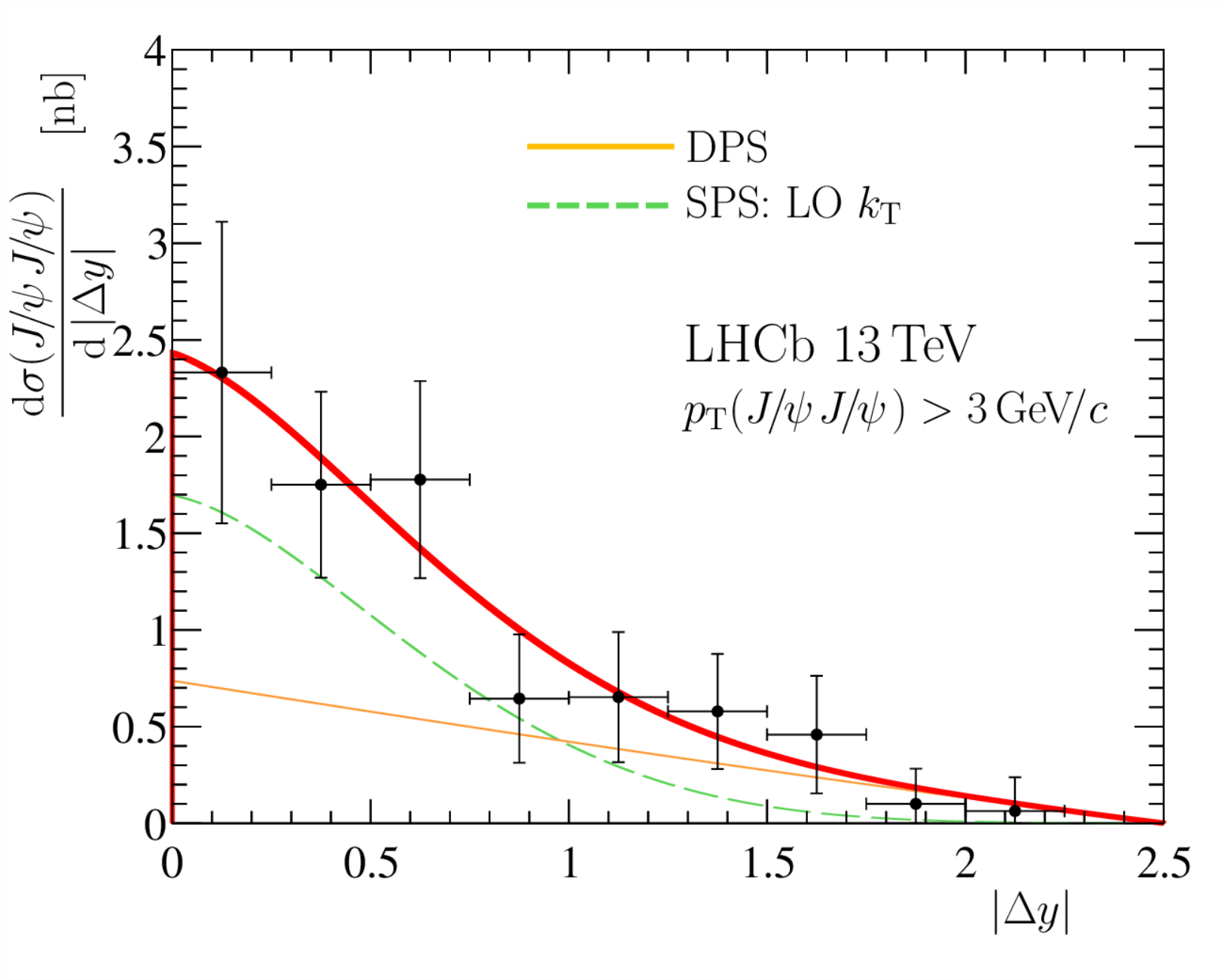}
\includegraphics[width=0.495\linewidth]{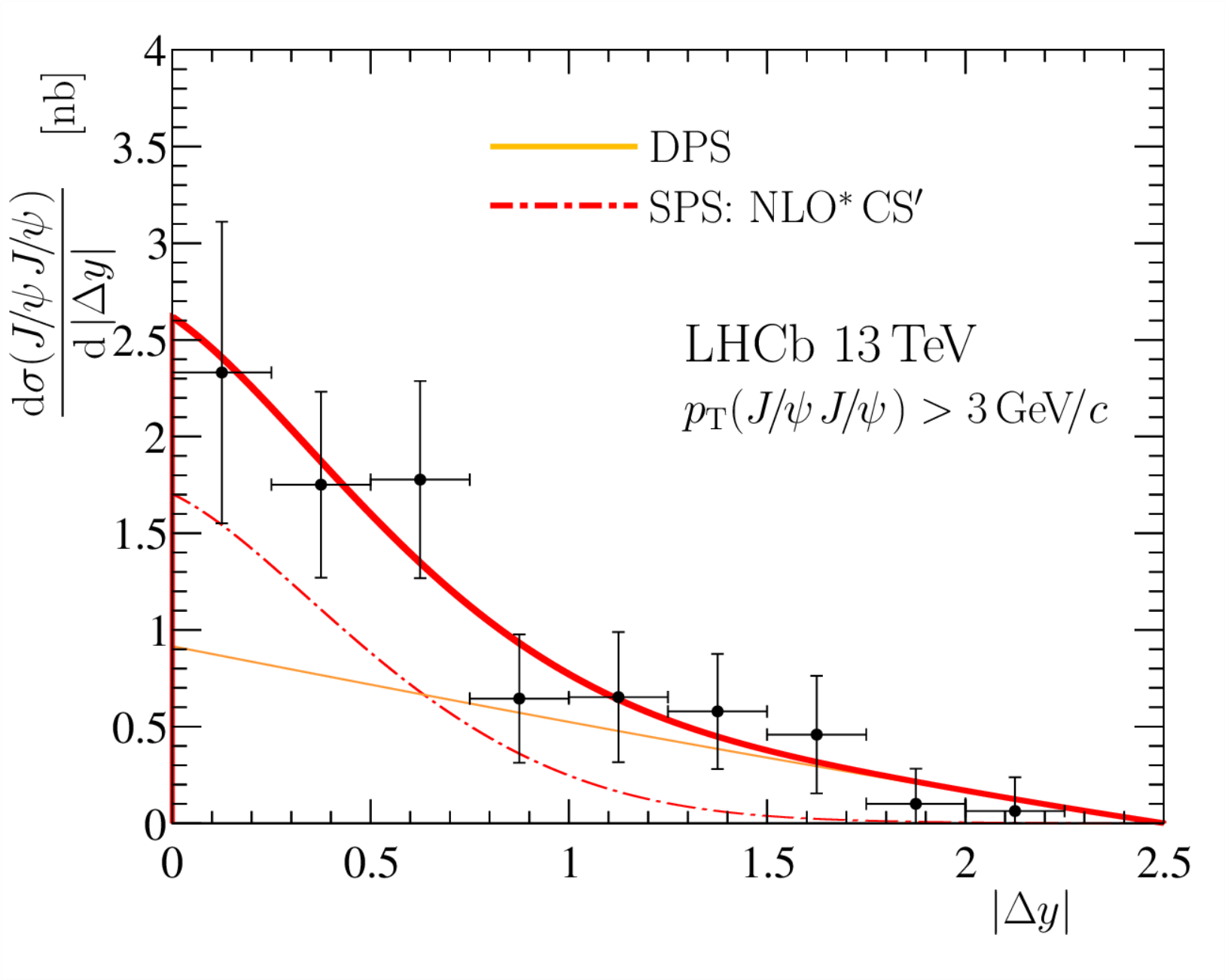}
\includegraphics[width=0.495\linewidth]{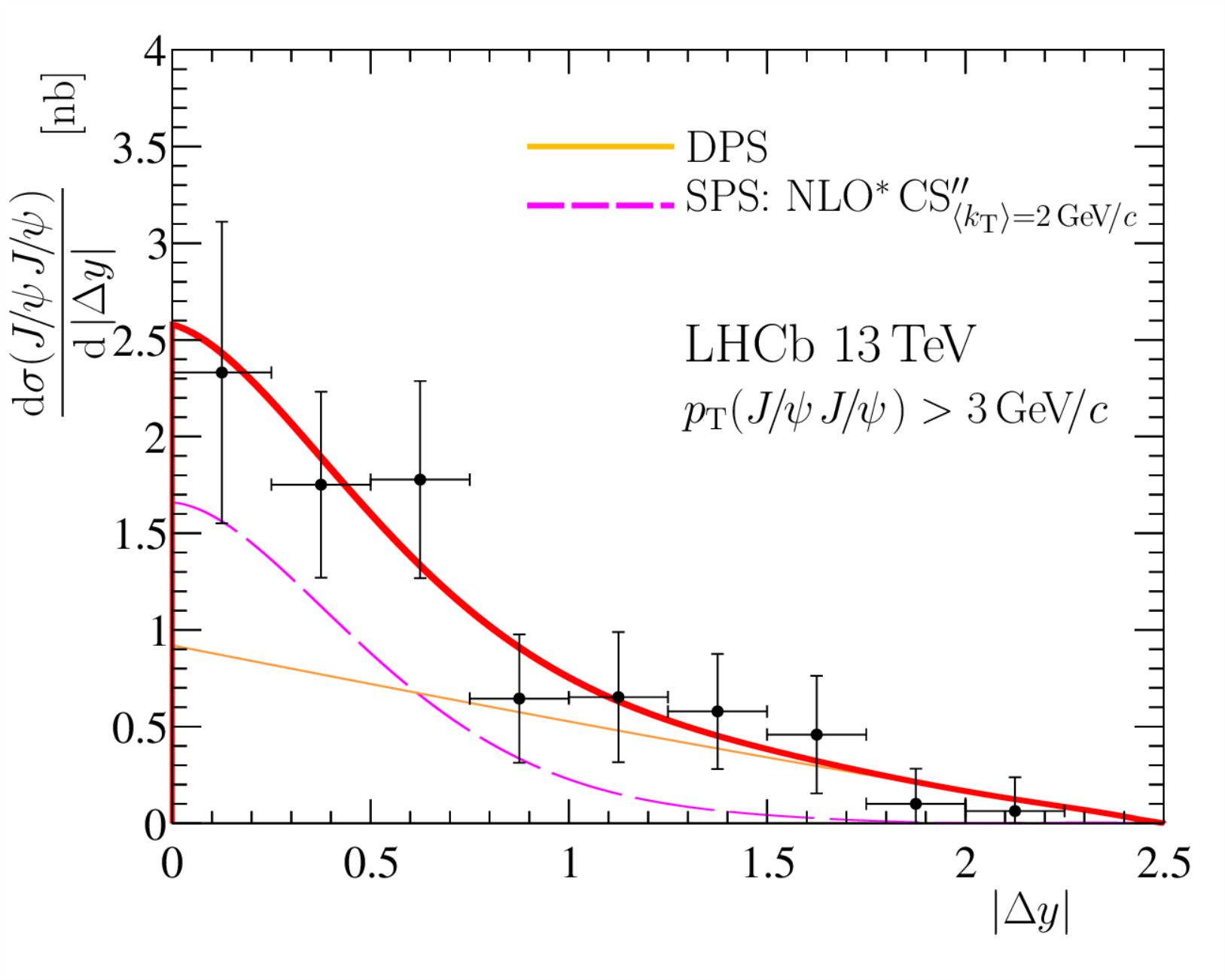}
\includegraphics[width=0.495\linewidth]{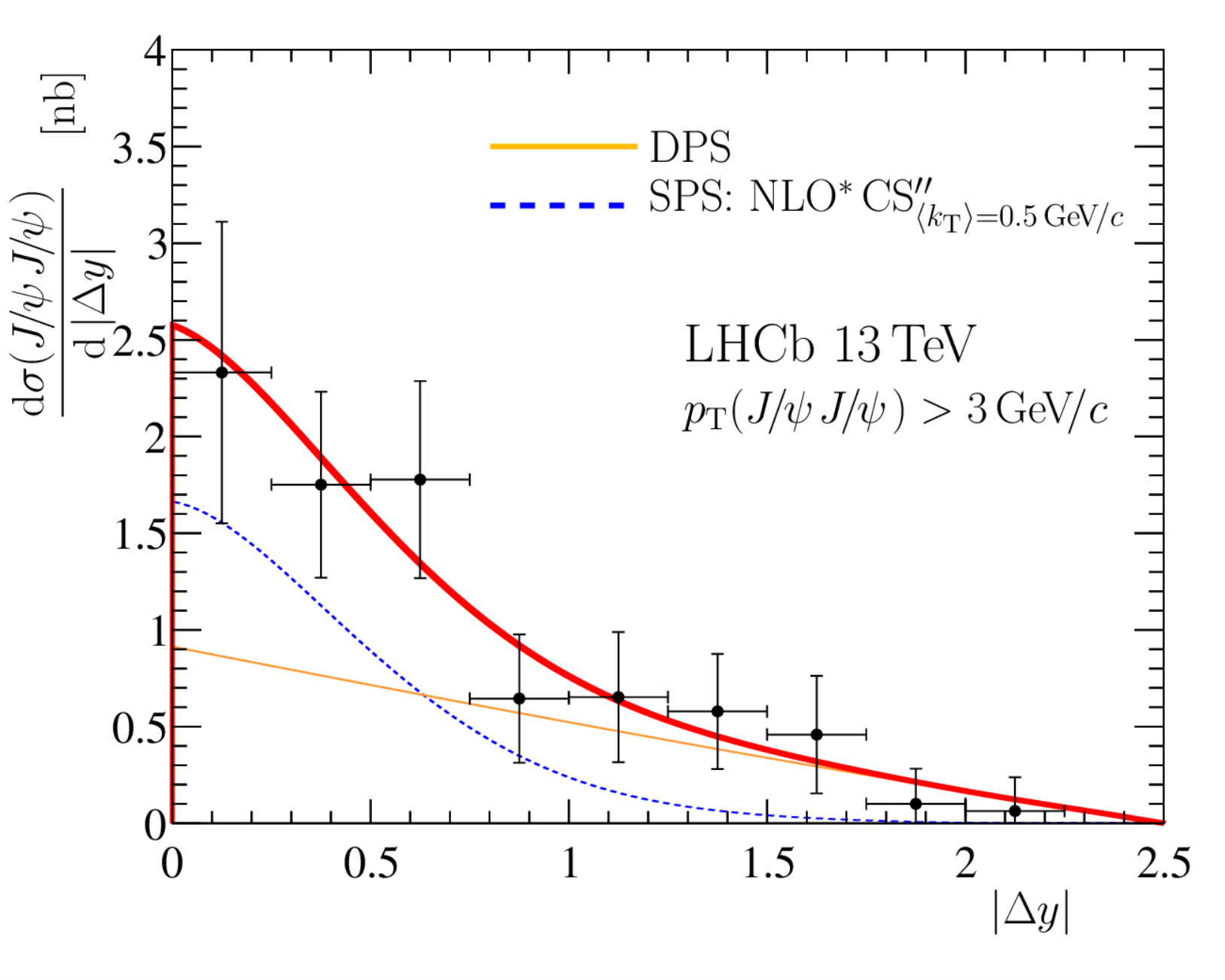}
\end{center}
  \caption { \small
    Result of templated DPS fit for 
    $\frac{ \mathrm{d}\sigma(\jpsi\jpsi)}{\mathrm{d} \left| \Delta y \right|}$ 
    for the $\pt(\jpsi\jpsi)>3\gevc$~region.
    The (black) points with error bars represent the data. 
    The total fit result is shown with the thick (red) solid line
    and the DPS component is shown with the thin (orange) solid line.
  }
  \label{fig:cmp:fits_dy_psipsi_3}
\end{figure}

The results of fits used for the determination of $\sigma_{\mathrm{eff}}$
are shown in Figs.~\ref{fig:cmp:fits_pty_psipsi},~\ref{fig:cmp:fits_m_psipsi} and~\ref{fig:cmp:fits_dy_psipsi}.
The fits used only for determination of $f_{\mathrm{DPS}}$ in 
\mbox{$\pt(\jpsi\jpsi)>1\gevc$}~and 
\mbox{$\pt(\jpsi\jpsi)>3\gevc$}~regions are shown in 
Figs.~\ref{fig:cmp:fits_my_psipsi_1},~\ref{fig:cmp:fits_dy_psipsi_1},~\ref{fig:cmp:fits_my_psipsi_3} and~\ref{fig:cmp:fits_dy_psipsi_3}.

\clearpage

\clearpage

\addcontentsline{toc}{section}{References}
\setboolean{inbibliography}{true}
\bibliographystyle{LHCb}
\bibliography{local,main,LHCb-PAPER,LHCb-CONF,LHCb-DP,LHCb-TDR}
 
\newpage
\centerline{\large\bf LHCb collaboration}
\begin{flushleft}
\small
R.~Aaij$^{40}$,
B.~Adeva$^{39}$,
M.~Adinolfi$^{48}$,
Z.~Ajaltouni$^{5}$,
S.~Akar$^{59}$,
J.~Albrecht$^{10}$,
F.~Alessio$^{40}$,
M.~Alexander$^{53}$,
S.~Ali$^{43}$,
G.~Alkhazov$^{31}$
% ...

{\footnotesize \it
% ...
$ ^{5}$Clermont Universit{\'e}, Universit{\'e} Blaise Pascal, CNRS/IN2P3, LPC, Clermont-Ferrand, France\\
% ...
$ ^{10}$Fakult{\"a}t Physik, Technische Universit{\"a}t Dortmund, Dortmund, Germany\\
% ...
$ ^{31}$Petersburg Nuclear Physics Institute (PNPI), Gatchina, Russia\\
% ...
$ ^{39}$Universidad de Santiago de Compostela, Santiago de Compostela, Spain\\
$ ^{40}$European Organization for Nuclear Research (CERN), Geneva, Switzerland\\
% ...
$ ^{43}$Nikhef National Institute for Subatomic Physics, Amsterdam, The Netherlands\\
% ...
$ ^{48}$H.H. Wills Physics Laboratory, University of Bristol, Bristol, United Kingdom\\
% ...
$ ^{53}$School of Physics and Astronomy, University of Glasgow, Glasgow, United Kingdom\\
% ...
$ ^{59}$University of Cincinnati, Cincinnati, OH, United States\\
% ...
}
\end{flushleft}

\end{document}